\documentclass[12pt]{article}

\usepackage[table]{xcolor}
\usepackage[normalem]{ulem}
\usepackage{booktabs}
\usepackage{longtable}
\usepackage{footmisc}
\usepackage{scicite}
\usepackage{ref_macros_sci}

\usepackage{hyperref}
\usepackage{comment}
\usepackage[sectionbib]{bibunits}
\defaultbibliography{ms} 

\defaultbibliographystyle{sn-nature} 

\usepackage{cancel}
\usepackage[normalem]{ulem}
\usepackage{graphicx}
\usepackage{setspace}
\usepackage{amsmath}
\usepackage{amssymb}
\usepackage{xspace}
\usepackage{multirow}
\usepackage{array}
\newcolumntype{C}[1]{>{\centering\arraybackslash}p{#1}}
\usepackage{rotating}

\usepackage{siunitx}

\usepackage{gensymb}

\newcommand{\nicer}{{\it NICER}\xspace}
\newcommand{\xmm}{{\it XMM-Newton}\xspace}
\newcommand{\target}{ASASSN-20qc\xspace}
\newcommand{\swift}{{\it Swift}\xspace}

\newcommand{\tess}{{\it TESS}\xspace}
\newcommand{\kms}{km~s$^{-1}$\xspace}
\newcommand{\ergs}{erg~s$^{-1}$\xspace}
\newcommand{\fluxunits}{erg~s$^{-1}$~cm$^{-2}$\xspace}
\newcommand{\farcs}{\mbox{$.\!\!^{\prime\prime}$}}%

\newcommand\T{\rule{0pt}{4.ex}}       % Top strut
\usepackage[final]{pdfpages}
\usepackage[left]{lineno}
\usepackage{times}
\usepackage{caption, subcaption, booktabs}

\usepackage{floatrow}
\DeclareFloatVCode{somespace}{\vspace{1.667\baselineskip}}
\newenvironment{sciabstract}{%
\begin{quote} \bf}
{\end{quote}}

\newcounter{lastnote}
\newenvironment{scilastnote}{%
\setcounter{lastnote}{\value{enumiv}}%
\addtocounter{lastnote}{+1}%
\begin{list}%
{\arabic{lastnote}.}
{\setlength{\leftmargin}{.22in}}
{\setlength{\labelsep}{.5em}}}
{\end{list}}

\title{A Case for a Binary Black Hole System Revealed via Quasi-Periodic Outflows}
\author{
Dheeraj R. Pasham$^{1}$, 
Francesco Tombesi$^{2,3,4,5,6}$,
Petra Sukov{\'a}$^{7}$, \\
Michal Zaja{\v{c}}ek$^{8}$,
Suvendu Rakshit$^{9}$,\\
Eric Coughlin$^{10}$,
Peter Kosec$^{1,23}$,
Vladim{\'\i}r Karas$^{7}$,\\
Megan Masterson$^{1}$,
Andrew Mummery$^{11}$,
Thomas W.-S. Holoien$^{12}$,
Muryel Guolo$^{13}$,\\
Jason Hinkle$^{14}$,
Bart Ripperda$^{15,16,17}$,
Vojt{\v{e}}ch Witzany$^{18}$,\\
Ben Shappee$^{14}$,
Erin Kara$^{1}$,
Assaf Horesh$^{19}$, \\
Sjoert van Velzen$^{20}$, 
Itai Sfaradi$^{19}$,
David L.\ Kaplan$^{21}$,\\
Noam Burger$^{19}$,
Tara Murphy$^{22}$,
Ronald Remillard$^{1}$,\\
James F. Steiner$^{23}$,
Thomas Wevers$^{24}$,
Riccardo Arcodia$^{1}$,\\
Johannes Buchner$^{25}$,
Andrea Merloni$^{25}$,
Adam Malyali$^{25}$,
Andy Fabian$^{26}$,\\
Michael Fausnaugh$^{1}$, 
Tansu Daylan$^{27}$, 
Diego Altamirano$^{28}$,\\
Anna Payne$^{14}$,
E.~C.~Ferrara$^{5,6,29}$.\\
{\small $^{\bf 1}$Kavli Institute for Astrophysics and Space Research,}\\ 
{\small Massachusetts Institute of Technology, MA, USA}\\
{\small $^{\bf 2}$Physics Department, Tor Vergata University of Rome, Via della Ricerca Scientifica 1, 00133 Rome, Italy}\\
{\small $^{\bf 3}$INAF – Astronomical Observatory of Rome, Via Frascati 33, 00040 Monte Porzio Catone, Italy}\\
{\small $^{\bf 4}$INFN - Rome Tor Vergata, Via della Ricerca Scientifica 1, 00133 Rome, Italy}\\ 
{\small $^{\bf 5}$Department of Astronomy, University of Maryland, College Park, MD 20742, USA}\\
{\small $^{\bf 6}$NASA Goddard Space Flight Center, Code 662, Greenbelt, MD 20771, USA}\\
{\small $^{\bf 7}$Astronomical Institute of the Czech Academy of Sciences, Prague, Czech Republic}\\
{\small $^{\bf 8}$Department of Theoretical Physics and Astrophysics, Masaryk University, Brno, Czech Republic}\\
{\small $^{\bf 9}$Aryabhatta  Research Institute of Observational Sciences (ARIES), Manora Peak, Nainital, 263002, India}\\ 
{\small $^{\bf 10}$Department of Physics, Syracuse University, Syracuse, NY, USA}\\
{\small $^{\bf 11}$Astrophysics, Department of Physics, University of Oxford, UK}\\
{\small $^{\bf 12}$The Observatories of the Carnegie Institution for Science, 813 Santa Barbara St., Pasadena, CA 91101, USA}\\
{\small $^{\bf 13}$Department of Physics and Astronomy, Johns Hopkins University, MD, USA}\\
{\small $^{\bf 14}$Institute for Astronomy, University of Hawaii, HI, USA}\\
{\small $^{\bf 15}$School of Natural Sciences, Institute for Advanced Study, 1 Einstein Drive, Princeton, NJ 08540, USA}\\
{\small $^{\bf 16}$NASA Hubble Fellowship Program, Einstein Fellow}\\
{\small $^{\bf 17}$Center for Computational Astrophysics, Flatiron Institute, 162 5th Avenue, New York, NY 10010, USA}\\
{\small $^{\bf 18}$Charles University, Prague, Czechia}\\
{\small $^{\bf 19}$Racah Institute of Physics, The Hebrew University of Jerusalem, Israel}\\
{\small $^{\bf 20}$Leiden Observatory, Leiden University, The Netherlands }\\
{\small $^{\bf 21}$University of Wisconsin Milwaukee, WI, USA}\\ \and
{\small $^{\bf 22}$University of Sydney, Australia}\\
{\small $^{\bf 23}$Center for Astrophysics \textbar\ Harvard \& Smithsonian, 60 Garden St, Cambridge, MA 02138, USA}\\
{\small $^{\bf 24}$European Southern Observatory, Chile}\\ 
{\small $^{\bf 25}$Max-Planck Institute for Extraterrestrial Physics, Germany}\\ 
{\small $^{\bf 26}$University of Cambridge, UK}\\
{\small $^{\bf 27}$Department of Physics and McDonnell Center for the Space Sciences, Washington University, St. Louis, MO 63130, USA}\\
{\small $^{\bf 28}$University of Southampton, UK}\\
{\small $^{\bf 29}$Center for Research and Exploration in Space Science \& Technology II (CRESST II),}\\
{\small NASA/GSFC, Greenbelt, MD 20771, USA}\\
\normalsize{$^\ast$To whom correspondence should be addressed; E-mail:  dheeraj@space.mit.edu}
}
\date{}
\begin{document} 
\baselineskip24pt
\maketitle 
\begin{bibunit}
\begin{sciabstract}
Binaries containing a compact object orbiting a supermassive black hole are thought to be precursors of gravitational wave events, but their identification has been extremely challenging. Here, we report quasi-periodic variability in X-ray absorption which we interpret as quasi-periodic outflows (QPOuts) from a previously low-luminosity active galactic nucleus after an outburst, likely caused by a stellar tidal disruption. We rule out several models based on observed properties and instead show using general relativistic magnetohydrodynamic simulations that QPOuts, separated by roughly 8.3 days, can be explained with an intermediate-mass black hole secondary on a mildly eccentric orbit at a mean distance of about 100 gravitational radii from the primary. Our work suggests that QPOuts could be a new way to identify intermediate/extreme-mass ratio binary candidates.
% Binaries containing a compact object orbiting a supermassive black hole are thought to be the precursors of gravitational wave events, but their identification has been extremely challenging. Here, we report the detection of quasi-periodic outflows (QPOuts) from a previously low-luminosity active galactic nucleus after an outburst, likely caused by a tidal disruption of a star. We rule out several models based on observed properties and instead show using general relativistic magnetohydrodynamic simulations that these QPOuts, separated by roughly 8.3 days, can be explained with an intermediate-mass black hole secondary on a mildly eccentric orbit at a mean distance of about 100 gravitational radii from the primary. Our work suggests that QPOuts could be a new way to identify intermediate/extreme-mass ratio binary candidates.
\end{sciabstract}

\section{Introduction}
ASASSN-20qc \cite{2020TNSTR3850....1S} is an astrophysical flare that originated from the nucleus of a galaxy at a redshift of 0.056 (luminosity distance of 260 Mpcs). It was discovered by the All-Sky Automated Survey for SuperNovae (ASAS-SN; \cite{asassn,benassasn}) on 20 December 2020. Throughout the paper, we reference times with respect to this discovery date of Modified Julian Date (MJD) 59203.27. A follow-up optical spectrum revealed the presence of several Hydrogen and Oxygen emission lines which facilitated the estimate of the redshift \cite{20qcclassification} (Methods section \ref{sisec:data}) and a supermassive black hole (SMBH) mass of log$(M_{\bullet}/M_{\odot}) = 7.5^{+0.7}_{-0.3}$ (Methods section \ref{subsec:bhmass} and \ref{sec:host}; Table \ref{tab:mbh}). A luminosity of 6$\times$10$^{40}$ erg s$^{-1}$ from archival {\it eROSITA} data (Fig.~\ref{fig:fig1}a and Methods section \ref{sisec:data}) indicates that prior to the outburst it was a low-luminosity Active Galactic Nucleus (AGN; see Methods section \ref{subsec:bpt}) accreting at $<$0.002\% of its Eddington limit.

\begin{sloppypar}

Roughly 52 days after \target's optical discovery the Neil Gehrels {\it Swift} observatory (\swift) observed it and detected X-rays. Following this detection, the Neutron star Interior Composition ExploreR (\nicer) started a high-cadence (1-2 visits per day) monitoring program (Fig.~\ref{fig:fig1}a and \ref{fig:fig1}c). We analyzed the \nicer soft X-ray (0.3-1.1 keV) energy spectra in the early phases of the outburst and found that the spectrum was thermal (accretion disk dominated) and contained systematic residuals reminiscent of a broad absorption trough (Fig.~\ref{fig:fig1}d). We also obtained an \xmm observation on 14 March 2021 (MJD 59287.34), roughly a month after the first \nicer exposure, noting the presence of broad absorption residuals. Subsequent \nicer spectra taken at various epochs of the outburst revealed that this absorption was variable throughout the outburst. A detailed photo-ionization modeling indicates that the dominant absorption feature is due to O~VIII transitions in the 0.75-1.00 keV observed energy band blue-shifted with a mildly-relativistic velocity of about 30\% of the speed of light. This evidence is indicative of an ultra-fast outflow (UFO) \cite{tombesi_ufo}. See Fig.~\ref{fig:xmmresiduals}, Methods section \ref{sisec:spectral_model} for a detailed discussion on X-ray spectral modeling.

\section{Results}
To probe the interplay between the variable outflow and the thermal continuum emission we calculated the ratio of the observed, background-subtracted count-rates in the energy band dominated by the outflow (0.75-1.00 keV) and the continuum emission (0.30-0.55 keV) (see shaded regions of Fig.~\ref{fig:fig1}d). This quantity, which we define as the Outflow Deficit Ratio (ODR), quantifies the amplitude of the outflow variability with respect to the continuum and it is shown in Fig.~\ref{fig:fig2}a. Surprisingly, the ODR curve showed repeating variations with a $\approx$8.5 day quasi-periodicity, which are not present in the unabsorbed continuum emission (SFs \ref{fig:lcs} and \ref{fig:hid}).
\end{sloppypar}

To quantify the variability and search for quasi-periodic signals in Fig.~\ref{fig:fig2}a, we computed the Lomb-Scargle Periodogram (LSP; \cite{scargle,lspnorm}) of the ODR curve (Fig.~\ref{fig:fig2}b). The highest power in the LSP is at $8.3\pm0.3$ days in multiple neighboring bins and is consistent with the time series in Fig.~\ref{fig:fig2}a. To estimate the false alarm probability that takes into account multiple bins, i.e., the chance probability of generating a signal as strong as the one observed from noise, we devised a detailed Monte Carlo method (see Fig. \ref{fig:whitetests}, Methods sections \ref{sisec:odrtiming} and \ref{sisec:noisenature}). The global false alarm probability of the observed $\approx$8.5 d quasi-periodicity is $<$2$\times$10$^{-5}$ ($>$4.2$\sigma$; see Fig.~\ref{fig:fig2}c). Global refers to a blind search for signal over all the frequencies sampled, i.e., $\sim$a day to 100 days (see Fig. \ref{fig:fig2}b).

\begin{sloppypar}
To further probe the nature of this quasi-periodicity, we extracted and fitted time-resolved \nicer X-ray spectra from individual maxima and minima in the ODR curve (see Methods section \ref{sisec:timeresolspecs}, Methods section \ref{sisec:nicercompspecs}, Table  \ref{tab: nicerxraydata}, and Fig.~\ref{fig:specplots}). Notably, the outflow has an order of magnitude higher column density ($N_H$) during the minima phases of the ODR curve with respect to the maxima: median value of $(12.6\pm 5.5)\times$10$^{21}$ cm$^{-2}$ and $(1.8\pm0.7)\times$10$^{21}$ cm$^{-2}$ for the minima and the maxima, respectively. The ionization parameter, defined as log$\xi$$=$$L/nr^2$, in units of erg~s$^{-1}$~cm, where $L$ is the ionizing luminosity between 1 Ryd and 1000 Ryd (1 Ryd $=$ 13.6 eV), $n$ is the number density of the material, and $r$ is the distance of the gas from the central source, is on average only slightly higher during the minima than during the maxima. Instead, the outflow bulk velocity is stable at around 0.35$c$, where $c$ is the speed of light (see Fig.~\ref{fig:specplots}).

% The outflow has a higher column density during the minima phases of the ODR, increasing roughly by a factor of ten from the median value of 1.8$\times$10$^{21}$ (standard deviation of 0.7$\times$10$^{21}$) cm$^{-2}$ for the maxima to 12.6$\times$10$^{21}$ (standard deviation of 5.5$\times$10$^{21}$) cm$^{-2}$ for the minima, respectively. Instead, the observed outflow velocity is stable at the mildly-relativistic value of about 35\% of the speed of light.}
\end{sloppypar}

Based on the above timing analysis and the time-resolved spectral modeling, we conclude that \target exhibits quasi-periodic outflows (QPOuts) about once every 8.5 days (precisely 8.3$\pm$0.3 days). By the term QPOuts we denote quasi-periodic variations of the outflowing material (see Fig. \ref{fig:fig2}).

\section{Discussion}
We considered several theoretical models to interpret the above observations including a precessing inner accretion disk, clumpy or slow outflow, X-ray reflection, accretion disk instabilities, quasi-periodic eruptions, repeating partial tidal disruption event (TDE), but disfavor them based on several independent lines of arguments. See Methods section \ref{sisec:spectral_model}, \ref{sisec:singleclump}, \ref{sisec:lateufo} and Methods section \ref{sisec:othermodels} for more details and Table~\ref{modelstable} for a summary of the strengths and the weaknesses of these various models. 

Instead, we propose a viable model with an orbiting, inclined perturber that repeatedly crosses the inner accretion flow. This scenario can explain the presence of QPOuts if the perturber is characterized by a sufficiently large influence radius at a given distance \cite{Sukova+2021:QPOutSims,1999PASJ...51..571S}. To further verify this model, we performed extensive 2D GRMHD simulations of an object in orbit around a SMBH using the HARMPI code \cite{tchekhovskoy2007,ressler2015} based on the original HARM code \cite{gammie2003,noble2006} (see Methods section \ref{msec:modeldiscussion} for details). Regardless of the specific setup, QPOuts are triggered by the passing perturber once per its orbit (see Fig.~\ref{fig_run1} and Table~\ref{tab:GRMHD_runs} for an overview). The simulations predict a persistent magnetized outflow from the inner flow with a roughly constant radial velocity profile, which is mass-loaded periodically when the secondary crosses the primary disk. This is consistent with the observation of a persistent outflow in the maxima which is boosted during the minima of the ODR. For all the cases, the perturber is highly inclined with respect to the equatorial plane of the accretion flow, which leads to the recurrent, periodic mildly-relativistic outflow regardless of the background accretion-flow state. An ordered and stable poloidal magnetic field in the funnel region accelerates the ejected matter to mildly-relativistic velocities. Furthermore, a mildly eccentric orbit with an eccentricity of $0.5-0.7$ can naturally induce departures from strict periodicity which is evident from the LSP peak FWHM of $\sim 1$ day as well as from the outflow-rate temporal profiles in the bottom panels in Fig.~\ref{fig_run1}. One caveat of the 2D GRMHD simulations is that while magnetorotational instability/MRI--which is responsible for accretion onto the SMBH--is active at the distance of the perturber, it decays after a few$\times$10,000 $M$ (or $\sim$100 days) in the inner regions of the accretion flow ($\sim$a few gravitational radii). Thus, making direct comparisons of simulations to data beyond 100 days becomes challenging. However, since the observed QPOuts span about 100 days, our simulations with active MRI were performed on similar timescales and they show that such a scenario provides a potential mechanism for producing QPOuts. Further work using 3D GRMHD simulations where MRI does not decay with time are needed to track such systems for extended periods (see SI section \ref{sisec:caveats} for more discussion).

The observed ratio of the outflow to the inflow rate of about 20\% during the ODR minima is consistent with a perturber influence radius of $\mathcal{R}\sim 3$ gravitational radii when compared to the analogous ratio derived from GRMHD simulations (see SI section \ref{sisec:naturepert} for details). Independent of the GRMHD simulations, simple analytical reasoning yields a similar estimate (see the second paragraph of SI section \ref{sisec:naturepert}). Taking into account that the ejected outflow clumps originate in the underlying flow, which can be treated as an advection-dominated accretion flow (ADAF: \cite{2014ARA&A..52..529Y}, \cite{Sukova+2021:QPOutSims}), and their sizes are comparable to $\mathcal{R}$, such a length-scale would be in agreement with the inferred column density of 10$^{22}$ cm$^{-2}$ of the spectroscopically detected UFO. Considering the tidal (Hill) length-scale of a massive perturber as well as the radius, within which the surrounding gas comoves with the perturber, we arrive at a rather broad range of the perturber masses $\sim 10^2-10^5\,M_{\odot}$. This broad range already includes the uncertainty in the primary SMBH mass (see SI section \ref{sisec:naturepert} for further discussion and EDFs \ref{fig_influence_radius}, \ref{fig_Hill_syn_radius}). %The required influence radius is inconsistent with an orbiting star with a fast stellar outflow as well as with a strongly magnetized neutron star. Both these would produce a bow shock with a  length-scale given by the ram pressure balance with the stellar-wind or the electromagnetic pressure. Such a value is smaller than one gravitational radius for both standard thin disk \cite{1973A&A....24..337S} and ADAF (see EDFs  \ref{fig_stag_radius_wb} and \ref{fig_perturber_regime}).

\begin{figure}[!ht]
\begin{center}
%\hspace{-0.35in}
\includegraphics[width=\textwidth, angle=0]{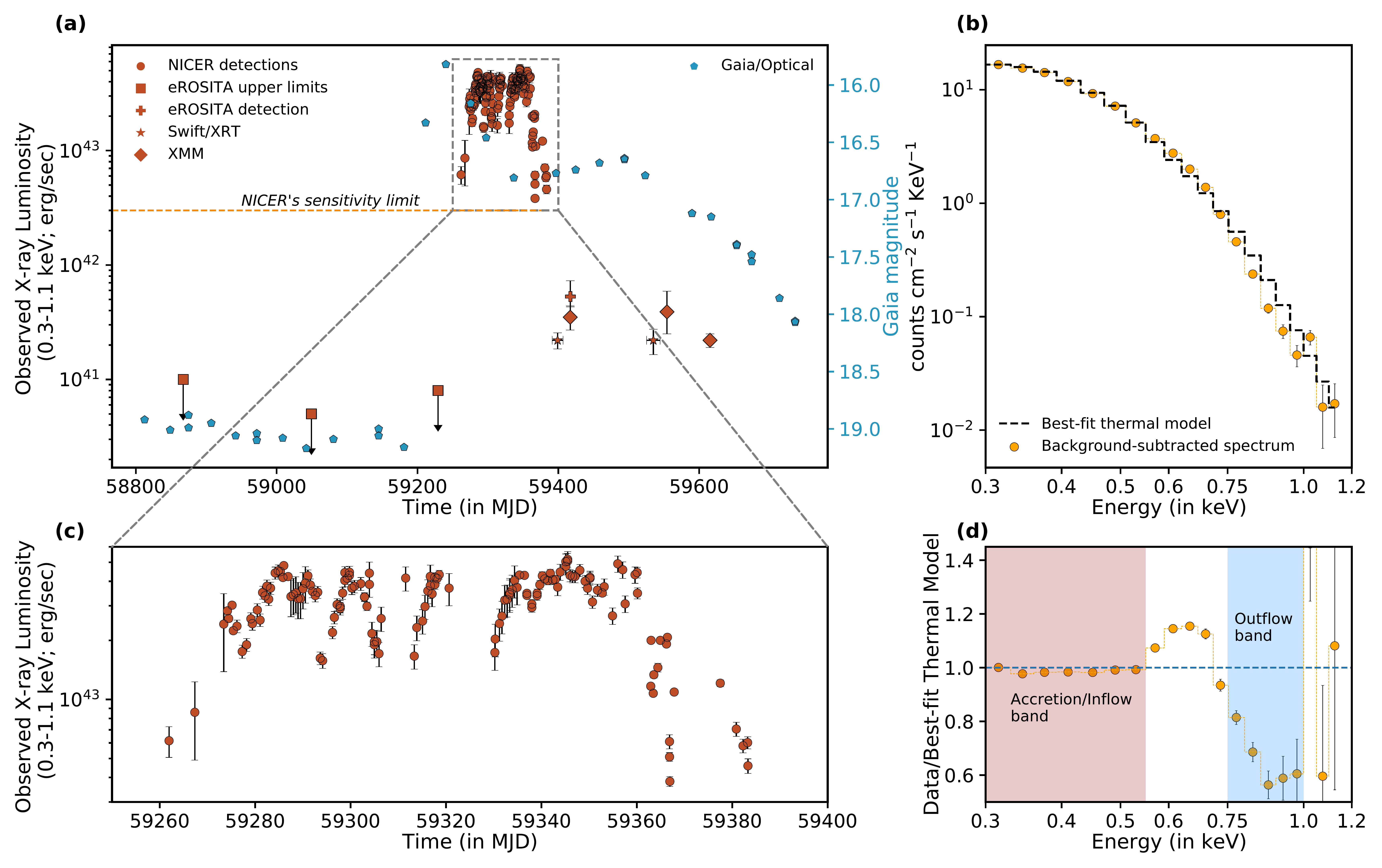}
\end{center}
%\vspace{-.35cm} 
\caption{{\bf \target's long-term evolution and a sample X-ray spectrum highlighting the outflow. (a) \target's observed X-ray and optical evolution.} Orange data represents X-ray (0.3-1.1 keV) data acquired by various instruments. The blue data shows the {\it Gaia} magnitude. The horizontal (dashed) line represents \nicer's sensitivity limit of 3$\times$10$^{42}$ \ergs for a source at redshift, $z,=0.056$. {\bf (b) The combined X-ray spectrum using all \nicer data} acquired over epochs of high absorption (yellow) and the best-fit emission model (black histogram). {\bf (c) Zoom-in of the outburst near the X-ray peak}. {\bf (d) Ratio of the average energy spectrum} using all \nicer data acquired over epochs of minima in outflow deficit ratio (ODR) and the best-fit thermal model. The outflow band is defined as the 0.75-1.00 keV band while the inflow/accretion band is defined as the bandpass where the ratio is near 1, i.e., 0.30-0.55 keV band.}
\label{fig:fig1}
\end{figure}

\begin{figure}[!ht]
\begin{center}
%\hspace{-1.35cm}
\includegraphics[width=\textwidth, angle=0]{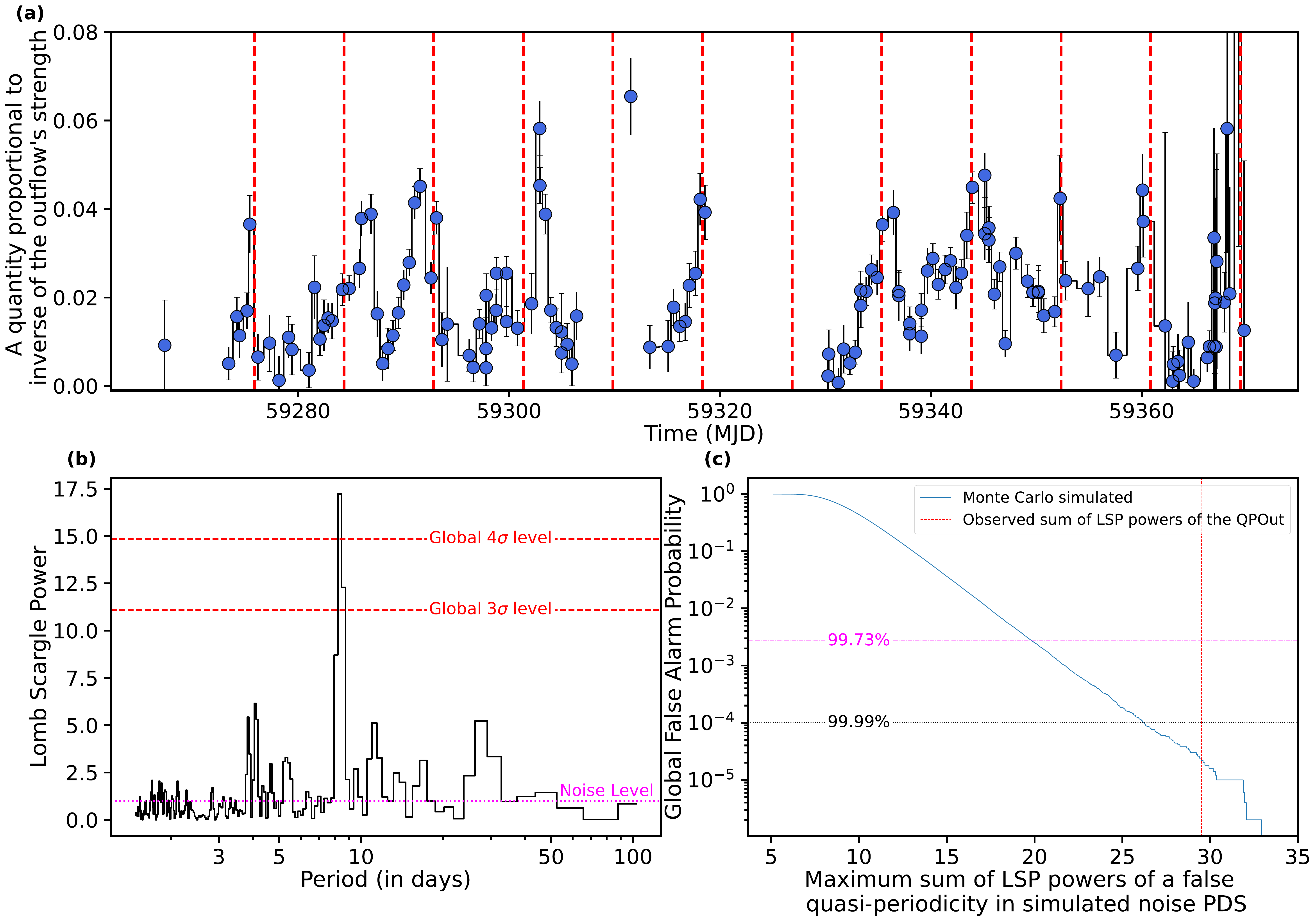}
\end{center}
%\vspace{-.35cm} 
\caption{ {\bf Summary of \target's timing analysis. (a) \target's outflow deficit ratio (ODR) versus time.} ODR is defined as the ratio of background-subtracted count rates in 0.75-1.00 keV (outflow) and 0.3-0.55 keV (continuum) bands. A lower ODR value implies a stronger outflow and vice versa. The dashed vertical red lines are uniformly separated by 8.5 days. {\bf (b) Lomb Scargle periodogram (LSP) of the ODR.} The strongest signal is near 8.5 days. The horizontal dashed red lines show the 3 and 4$\sigma$ global false alarm probabilities as per \cite{scargle}. The noise in the periodogram is consistent with white with a mean LSP power value of 1 (see Methods section \ref{sisec:noisenature}). {\bf (c) Global (trials-accounted) false alarm probability.} This curve was generated using extensive Monte Carlo simulations (see Methods section \ref{sisec:odrtiming}). The global statistical significance of the 8.5 d quasi-periodicity is $>$ 4.2$\sigma$.}
\label{fig:fig2}
\end{figure}

\begin{figure}[!htp]
    \centering
    \includegraphics[width=\textwidth]{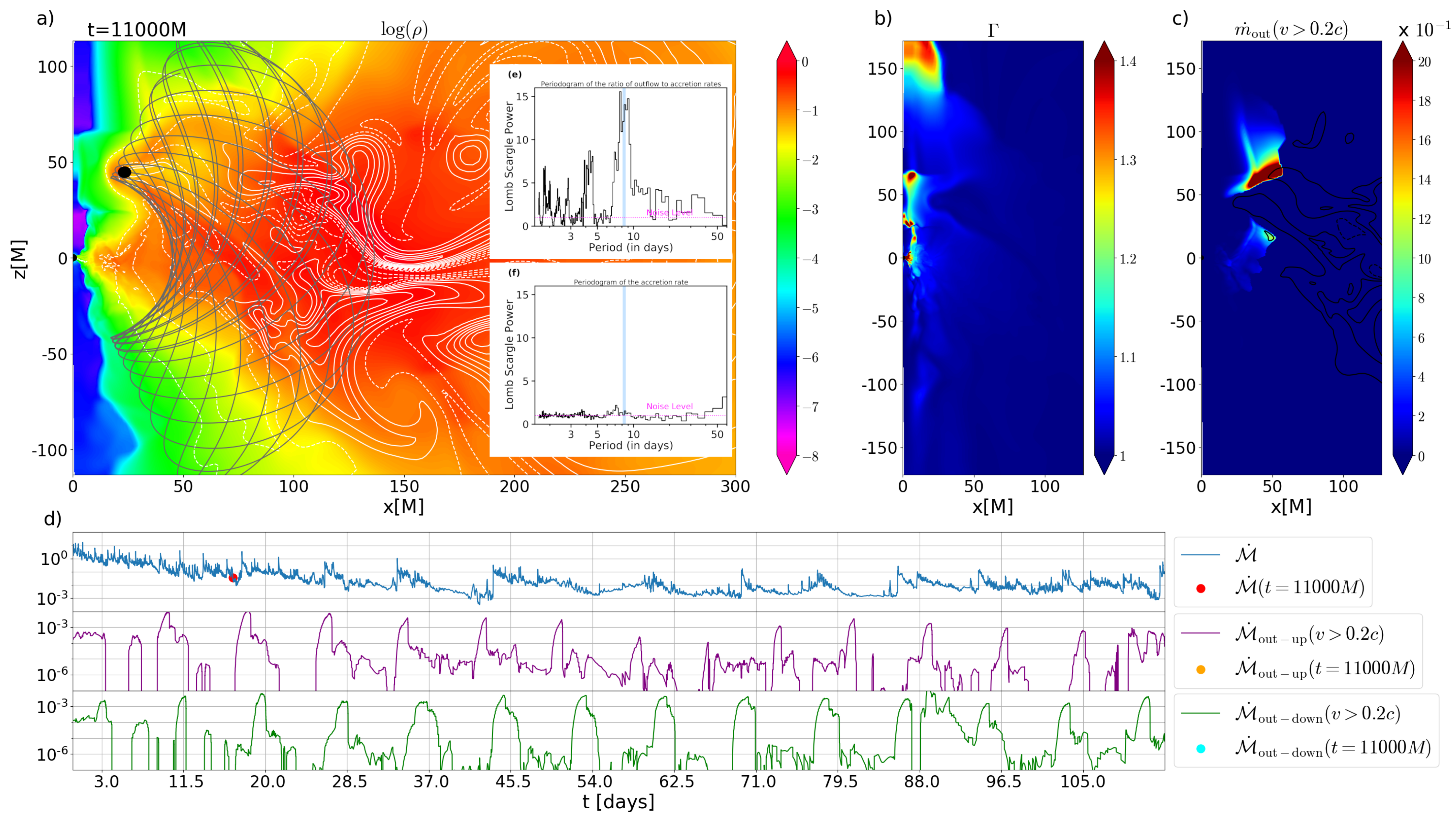}
    \caption{{\bf A sample snapshot from our GRMHD simulation (2D HARM, Run 14 from Table \ref{tab:GRMHD_runs})}. For this case, the SMBH mass was set to 10$^{7.4}$ M$_{\odot}$ and the perturbing companion is in an elliptical orbit (eccentricity, e=0.5) with an observed orbital period of 8.5 days and has an influence radius of 3 gravitational radii ($1M=GM_{\bullet}/c^2=0.25(M_{\bullet}/10^{7.4}\,M_{\odot})\,{\rm AU}$). {\bf (a)} Spatial distribution of the logarithm of mass density expressed in arbitrary units. The horizontal and the vertical axes are spatial coordinates expressed in gravitational radii (units of $M$). The white contours indicate the magnetic field configuration. The position and size of the perturber is shown by the black circle, while the grey line displays its trajectory in the 2D slice.  {\bf (b) Spatial distribution of the Lorentz factor of the gas bulk motion.  (c) Spatial distribution of the mass-outflow rate with $v>0.2c$.} The outflow rate is colour-coded using arbitrary units according to the colour-bar to the right. {\bf (d) Temporal profiles of the inflow rate (blue), the outflow rate through the upper funnel (purple), and the outflow rate through the lower funnel (green).} The inflow and the outflow rates are expressed in arbitrary units. The time is expressed in days in the observed frame. The coloured points/dots indicate the time of the snapshot. Vertical lines are uniformly separated by 8.5 days. \textbf{(e, f) Lomb Scargle Periodogram of the ratio of the outflow to the inflow rates (e) and the accretion rate (f) from run 14 sampled exactly as the real data.} The peak signal in panel (e) is broad with a value of 8.5$^{+0.7}_{-1.1}$ d and is consistent with the observed value of 8.3$\pm$0.3 d (shaded blue band) while no such signal is present in the accretion rate periodogram (f), i.e., an elliptical binary can reproduce the observed quasi-periodicity in the outflow strength without similar variations in the continuum. }
       \label{fig_run1}
\end{figure}

\begin{figure}[!t]
    \centering
    \includegraphics[width=0.81\textwidth]{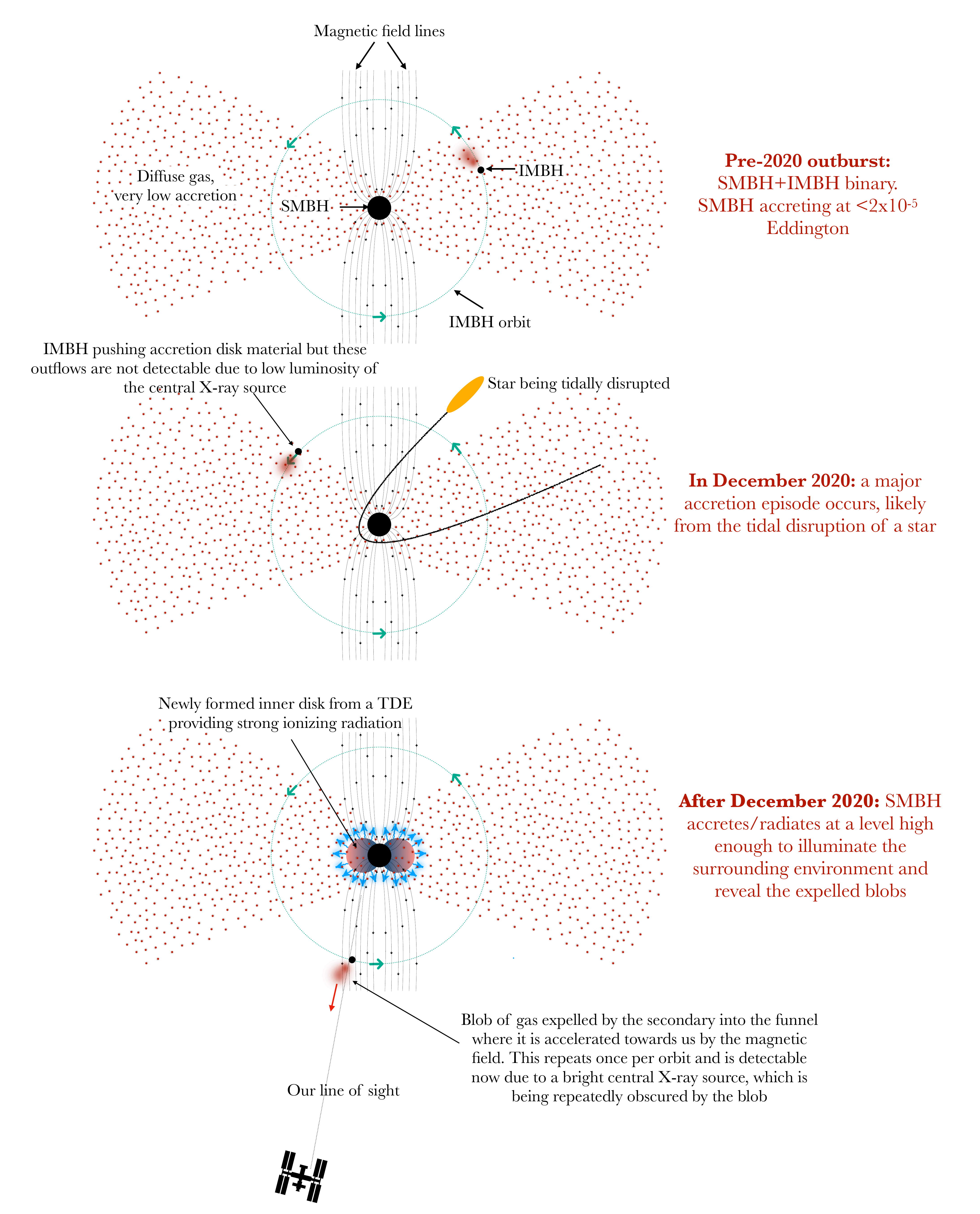}
    \caption{{\bf Schematic of a potential model for \target}. A gravitationally bound (pre-existing) IMBH located at roughly 100 R$_{g}$ from the central SMBH can explain the repeated outflows seen here. The overall outburst could have been triggered by a tidal disruption of a passing star by the SMBH which creates a compact accretion disk which naturally enhances the X-ray emission and consequently illuminates the surrounding environment and the presence of the IMBH secondary. Secondary plunges through the pre-existing (non-TDE) accretion flow, modulating the outflow on the orbital period. Relative sizes are not to scale.}
    \label{fig:schematic}
\end{figure}

Distinct from the QPOuts, the optical light curve shown in the Fig.~ \ref{fig:uvotevol} exhibits a smooth rise, peak, and decay on a timescale of $\sim$ 150 days. This timescale is broadly consistent with the canonical fallback time of the debris from a TDE (e.g., \cite{Rees88}) with a black hole mass of $M_{\bullet} \sim 10^{7}M_{\odot}$ and a solar-like star. The evolution of the optical/UV temperature and photosphere radius during the outburst is also very similar to those of known TDEs (compare the Fig.~\ref{fig:uvotevol} with Figure 8 of \cite{van_Velzen_21} and Figure 1 of \cite{jason}). The time delay between the X-ray and the optical outbursts of a few months (see Fig. \ref{fig:fig1}a) has also been seen in several TDEs (e.g., \cite{2017ApJ...837L..30P,2017ApJ...851L..47G}). Finally, the soft X-ray spectrum is also strikingly similar to thermal X-ray TDEs. Therefore, a reasonable interpretation is that the overall outburst in the optical, UV and X-rays was induced by a TDE, which produces a bright inner accretion disk, i.e., a soft X-ray source, which is quasi-periodically obscured by the blobs driven by the orbiting perturber (see Fig. \ref{fig:schematic}).

Attributing the outburst to a TDE we can further constrain the mass of the secondary based on the argument that the gravitational wave inspiral time should be greater than the typical time for a stellar disruption in a galaxy. Using a TDE rate of 10$^{-4}$ yr$^{-1}$ (e.g., see Fig. 10 in \cite{tderates}) would require the SMBH--perturber system have a merger timescale of $\gtrsim 10^4$ years. This limits the perturber mass to the range of  $10^2$--$10^4\,M_{\odot}$, i.e., to the intermediate-mass black-hole (IMBH) range (see bottom panel of Fig.~\ref{fig_Hill_syn_radius}).
For such mass and distance of the secondary, the gravitational radiation is weak and the period of the system will not evolve significantly in the next decade, making the signal lay outside the frequency range of the upcoming space-based gravitational-wave observatory LISA.
The unique combination of a SMBH-IMBH pair experiencing the TDE makes such observation rather rare, though not entirely implausible. Within the cosmological volume inside $z\sim 0.06$, we estimate $N_{\rm pair,TDE}=0.07-5.3$ TDEs per year in hosts with tight SMBH-IMBH pairs (out of $\sim 2.5$ million galaxies; see SI section \ref{subsec_formation_channels} for further discussion on the estimated event rate and the detectability of the system in gravitational waves). 

In summary, our work highlights the new astrophysical phenomenon of QPOuts and the importance of high cadence optical and X-ray monitoring observations to potentially uncover electromagnetic signatures of tight binary black hole systems. The identification of such SMBH--IMBH binaries, i.e., intermediate/extreme mass ratio inspirals (I/EMRIs), has fundamental implications for multi-messenger astrophysics and for our understanding of black hole growth and evolution.

% FIG+++FIG+++FIG+++FIG+++FIG+++FIG+++FIG+++FIG+++FIG+++FIG+++FIG+++FIG++++++FIG+++FIG+++
%                   Figure 1: LIGHT CURVE AND SPECTRUM
% FIG+++FIG+++FIG+++FIG+++FIG+++FIG+++FIG+++FIG+++FIG+++FIG+++FIG+++FIG++++++FIG+++FIG+++

\clearpage
\vfill\eject
% FIG+++FIG+++FIG+++FIG+++FIG+++FIG+++FIG+++FIG+++FIG+++FIG+++FIG+++FIG++++++FIG+++FIG+++
% FIG+++FIG+++FIG+++FIG+++FIG+++FIG+++FIG+++FIG+++FIG+++FIG+++FIG+++FIG++++++FIG+++FIG+++
%                   Figure 2: ODR CURVE AND LSP/FAP
% FIG+++FIG+++FIG+++FIG+++FIG+++FIG+++FIG+++FIG+++FIG+++FIG+++FIG+++FIG++++++FIG+++FIG+++
\newpage
\vfill\eject

% FIG+++FIG+++FIG+++FIG+++FIG+++FIG+++FIG+++FIG+++FIG+++FIG+++FIG+++FIG++++++FIG+++FIG+++
%                   FIGURE 3: SNAPSHOT FROM THE GRMHD SIMULATION
% FIG+++FIG+++FIG+++FIG+++FIG+++FIG+++FIG+++FIG+++FIG+++FIG+++FIG+++FIG++++++FIG+++FIG+++
\clearpage
\vfill\eject

% FIG+++FIG+++FIG+++FIG+++FIG+++FIG+++FIG+++FIG+++FIG+++FIG+++FIG+++FIG++++++FIG+++FIG+++
%                   FIGURE 3: SNAPSHOT FROM THE GRMHD SIMULATION
% FIG+++FIG+++FIG+++FIG+++FIG+++FIG+++FIG+++FIG+++FIG+++FIG+++FIG+++FIG++++++FIG+++FIG+++

\clearpage

% FIG+++FIG+++FIG+++FIG+++FIG+++FIG+++FIG+++FIG+++FIG+++FIG+++FIG+++FIG++++++FIG+++FIG+++
%                   TABLE 1: PROS AND CONS OF ALL MODELS CONSIDERED IN THIS WORK
% FIG+++FIG+++FIG+++FIG+++FIG+++FIG+++FIG+++FIG+++FIG+++FIG+++FIG+++FIG++++++FIG+++FIG+++
% \clearpage
% \vspace{-0.8in}
\begin{table}[ht]
\tiny
    \centering
    \caption{A summary of the strengths and the weaknesses of various models considered in this work to explain \target's observed X-ray spectro-timing variability. See Methods sections \ref{sisec:spectral_model}, \ref{sisec:singleclump}, \ref{sisec:lateufo} and SI for more details.}
    \label{modelstable}
    \begin{tabular}{|C{0.15\textwidth}|C{0.25\textwidth}|C{0.455\textwidth}|C{0.15\textwidth}|}
        \hline 
        &&& \\
        \textbf{Model/class of models} & \textbf{Strengths} & \textbf{Weaknesses} & \textbf{Notes} \T \\ \hline \T 
         Inner disk precession & Thought to be commonly seen in stellar-mass black hole binaries \cite{Ingram:LFQPOs} & The lack of strong continuum modulation and the observed changes from high column, high ionization parameter to low column, low ionization along with constant outflow speed and more importantly the lack of a strong quasi-periodicity in the 0.3-0.55 keV continuum variations are inconsistent with precession with all known types of outflows \cite{Keigo10, keigo14, nomura16, elvis00, kingpoundsufos} & Disfavoured based on physical reasoning \\ \T
         Clumpy outflow & - & The outflow geometry would need to be fine tuned to have uniformly separated clumps. The probability of formation of such clumps by chance is less than 1 in 50,000 & Disfavoured due to low likelihood  \\ \T
         Slow outflow & Slow outflow can, in principle, produce similar spectral signatures & - The \xmm/RGS and EPIC/pn spectrum rule out a slow outflow that can produce such a broad feature \newline - A typical slow outflow is  distant from the SMBH and cannot produce a rapid ($\sim$ week timescale) quasi-periodic variability seen here & Disfavoured  based on physical reasoning (see Methods section \ref{sisec:noslow})  \\ \T
         X-ray reflection by a corona & Seen in several highly accreting AGN with an X-ray corona \cite{xrayreflection} & Lack of a Compotonizing corona/powerlaw component in the X-ray spectrum & Disfavoured based on lack of evidence in data (see Methods section \ref{sisec:spectral_model})  \\ \T
        X-ray reflection by a disk & Argued to operate at least in one changing-look AGN \cite{Masterson22} & - Lack of a geometrically thick surface for reflection, would require a fine-tuned disk geometry \newline - Unphysically large fraction of reflected emission compared to the primary thermal emission & Disfavoured based on statistical argument and physical grounds (see Methods section \ref{sisec:spectral_model} for more discussion)  \\ \T
         Magnetically arrested accretion disk & Preliminary work by \cite{bart} suggests that outflows can be produced through repeated magnetic reconnection events & - Based on state-of-the-art high-resolution simulations it is unclear if such outflows would be quasi-periodic in nature \newline - Such regular outflows are not seen in lower-resolution simulations \newline - Lack of strong quasi-periodicity in the continuum variations & viable but no clear indication in the state-of-the-art simulations (but see Fig. 8 of \cite{bart})\\ \T
         Quasi-periodic eruptions (QPEs) & Seen in a small sample of AGN & - QPEs manifest as large amplitude flux bursts as opposed to changes in ODR. \newline - Variable outflows have not been reported in known QPE sources. & Disfavoured because the observed signal is distinct (see Methods section \ref{supsec:qpemodel})\\ \T
         Repeating partial stellar tidal disruption & Argued to operate in at least 3 systems \cite{Payne21,fykrepeatingtde,zhu22} & - The expected orbital period would be orders of magnitude longer than what is seen here \cite{Cufari22} \newline - No evidence for a similar variability in the optical light curve \newline - A stellar core's influence radius would be too small to produce the observed outflow & Disfavoured based on physical reasoning  \\ \T
         Stellar debris stream &  Could provide obscuration when highly inclined  & -Stellar debris would be tidally spread along the whole orbit, turning off the periodicity; the material would need to be continuously replenished (see partial TDE above) & Disfavoured  based on physical reasoning \\ \T
         Radiation pressure driven outflows & Observed in a sample of accreting stellar-mass black holes \cite{2000ApJ...542L..33J} & - The persistence of the outflow over a factor of $>$200 change in X-ray flux suggests negligible radiation driving  \newline - Fine tuning of the disk properties for obtaining short-enough instability period \cite{2023A&A...672A..19S} \newline - No evidence for a similar variability in the soft X-ray continuum & Disfavoured based on the need for fine tuning  \\ \T
         A scaled-up version of quasi-periodic oscillations  & Occurring in stellar-mass black-hole binaries (e.g., \cite{ronclin}) & The lack of a strong quasi-periodicity in the thermal continuum (0.3-0.55 keV band) & Disfavoured due to lack of a precendent\\ \T
         An orbiting object repeatedly perturbing the SMBH accretion disk & - can explain QPOuts \newline - ultrafast outflow production supported by GRMHD simulations \newline - Consistent with TDE statistics and production rates of SMBH--IMBH binaries \cite{smbh_imbhrates}& - An IMBH distance of $\sim 100\,r_{\rm g}$ makes full 3D GRMHD simulations computationally expensive \newline - For 2D simulations, magnetorotational instability enabling accretion tends to stop operating after sometime (within a few$\times$10,000 $M$ which corresponds to $\sim$100 days), which makes comparison with data limited at later epochs & Viable but no precedent (see SI section \ref{msec:modeldiscussion} and \ref{sisec:grmhdsims}; see section \ref{sisec:caveats} for caveats) \\ \T
         &&& \\ \hline
    \end{tabular}
\end{table}
\vfill\eject

\clearpage
\section*{{\Huge Material and Methods.}}
% \begin{bibunit}
\singlespace
\begin{scilastnote}
\begin{sloppypar}
\item[] {\bf Acknowledgments.}\\ 
The authors declare that they have no competing interests.\\

DP and FT would like to thank Dr. Jelle Kaastra for insight into cross-calibration issues between the EPIC/pn and RGS spectra. 

FT acknowledges funding from the European Union - Next Generation EU, PRIN/MUR 2022 (2022K9N5B4). 

PS has been supported by the fellowship Lumina Quaeruntur No.\ LQ100032102 of the Czech Academy of Sciences. Computational resources used for this work were funded also by the Ministry of Education, Youth and Sports of the Czech Republic through the e-INFRA CZ (ID:90140). 

ERC acknowledges support from the National Science Foundation through grant AST-2006684

PK acknowledges support by NASA through the NASA Hubble Fellowship grant HST-HF2-51534.001-A awarded by the Space Telescope Science Institute, which is operated by the Association of Universities for Research in Astronomy, Incorporated, under NASA contract NAS5-26555.

MZ acknowledges the GA\v{C}R JUNIOR STAR grant No. GM24-10599M for financial support. 

VK acknowledges the OPUS-LAP/GA\v{C}R-LA bilateral grant (2021/43/I/ST9/01352/OPUS 22 and GF23-04053L).

PyXspec was developed by Craig Gordon and Keith Arnaud HEASARC Software Development, Astrophysics Science Division, Code 660.1, NASA/GSFC, Greenbelt MD 20771.

Support for TW-SH was provided by NASA through the NASA Hubble Fellowship grant HST-HF2-51458.001-A awarded by the Space Telescope Science Institute (STScI), which is operated by the Association of Universities for Research in Astronomy, Inc., for NASA, under contract NAS5-26555.

BR is at the Flatiron Institute which is supported by the Simons Foundation. Support for this work was provided by NASA through the NASA Hubble Fellowship grant HST-HF2-51518.001-A awarded by the Space Telescope Science Institute, which is operated by the Association of Universities for Research in Astronomy, Incorporated, under NASA contract NAS5-26555.

TD acknowledges support from the McDonnell Center for the Space Sciences at Washington University in St. Louis.

SR acknowledges the support of the Start-up Research Grant (SRG) from the Science and Engineering Research Board (SERB), a statutory body of the Department of Science \& Technology, Government of India, through SRG grant number SRG/2021/001334.

ECF is supported by NASA under award number 80GSFC21M0002.\\

DLK is supported by NSF grant AST-1816492.

A.H. is grateful for the support by the the United States-Israel Binational Science Foundation (BSF grant 2020203) and by the  Sir Zelman Cowen Universities Fund. This research was supported by the ISRAEL SCIENCE FOUNDATION (grant No. 1679/23). 

All data needed to evaluate the conclusions in the paper are present in the paper and/or the Supplementary Materials. The time-resolved energy spectra from \nicer and the ultrafast outflow XSTAR table model are made available on zenodo and can be accessed here: \url{https://zenodo.org/records/10069726}. The X-ray and optical light curves can be provided by DRP  pending scientific review and a completed material transfer agreement. Requests for the data in this paper should be submitted to: p.dheerajreddy@gmail.com.

\end{sloppypar}
\end{scilastnote}

\newpage
%%%%%%%%%%%%%%%%%%%%%%%%%%%%%%%%%%%%%%%%%%%%%%%%%%%%%%%%%%%%%%%
%
%                   SUPPLEMENT SECTION
%
%%%%%%%%%%%%%%%%%%%%%%%%%%%%%%%%%%%%%%%%%%%%%%%%%%%%%%%%%%%%%%%%
% \setcounter{page}{1}
\renewcommand{\theequation}{S\arabic{equation}}
\setcounter{figure}{0}
\setcounter{table}{0}

%++++++++++++++++++++++++++++++++++++++++++++++++++++++++++++++++++++++++++++
% ------------------------------- DATA INTRODUCTION ------------------------
%++++++++++++++++++++++++++++++++++++++++++++++++++++++++++++++++++++++++++++

%%%%%%%%%%%%%%%%%%%%%%%%%%%%%%%%%%%%%%%%%%%%%%%%%%%
%%%%%%%%%%%%%%%%%%%%%%%%%%%%%%%%%%%%%%%%%%%%%%%%%%%
%%%%%%%%%%%%%%%%%%%%%%%%%%%%%%%%%%%%%%%%%%%%%%%%%%%

%%%%%%%%%%%%%%%%%%%%%%%%%%%%%%%%%%%%%%%%%%%%%%%%%%%%%%%%%%%%%%%%%%%%%%%%%%%%%%%%%%%%%%%%%%%%%%%%%%%%%%%%
%%%%%%%%%%%%%%%%%%%%%%%%%%%%%%%%%%%%%%%%%%%%%%%%%%%%%%%%%%%%%%%%%%%%%%%%%%%%%%%%%%%%%%%%%%%%%%%%%%%%%%%%
%%%%%%%%%%%%%%%%%%%%%%%%%%%%%%%%%%%%%% SUPPLEMENTARY INFORMATION %%%%%%%%%%%%%%%%%%%%%%%%%%%%%%%%%%%%%%%
%%%%%%%%%%%%%%%%%%%%%%%%%%%%%%%%%%%%%%%%%%%%%%%%%%%%%%%%%%%%%%%%%%%%%%%%%%%%%%%%%%%%%%%%%%%%%%%%%%%%%%%%
%%%%%%%%%%%%%%%%%%%%%%%%%%%%%%%%%%%%%%%%%%%%%%%%%%%%%%%%%%%%%%%%%%%%%%%%%%%%%%%%%%%%%%%%%%%%%%%%%%%%%%%%

\clearpage
\setcounter{section}{0}
\setcounter{figure}{0}
\setcounter{table}{0}

% \noindent{\LARGE{\textbf{}}}

\section{Data and Reduction:}\label{sisec:data}
% \section{\Large{\bf Data and Reduction}}\label{supsec:data}
For this work we acquired/used multiwavelength data in the X-ray, optical, UV, and the radio bands. Data reduction for each of the telescopes/instruments are described below. Throughout this paper, we adopt a standard $\Lambda$CDM cosmology with H$_{0}$ = 67.4 km~s$^{-1}$~Mpc$^{-1}$, $\Omega_{m}$ = 0.315 and $\Omega_{\Lambda}$ = 1 - $\Omega_{m}$ = 0.685 \cite{planck}. Using the Cosmology calculator of \cite{Wright2006} \target's redshift of 0.056 corresponds to a luminosity distance of 259.5 Mpcs.
%++++++++++++++++++++++++++++++++++++++++++++++++++++++++++++++++++++++++++++
% ----------------------- NICER DATA REDUCTION ------------------------------
%++++++++++++++++++++++++++++++++++++++++++++++++++++++++++++++++++++++++++++
% ^o^o^o^o^o^o^o^o^o^o^o^o^o^o^o^o^o^o^o^o^o^o^o^o^o^o^o^o^o^o^o^o^o^o^o^o^o^o
% ************************************* X-ray *******************************
% ^o^o^o^o^o^o^o^o^o^o^o^o^o^o^o^o^o^o^o^o^o^o^o^o^o^o^o^o^o^o^o^o^o^o^o^o^o^o
\subsection{X-ray}\label{supsec:xray}
\target's X-ray data used in this work were acquired by six different instruments: \nicer's X-ray Timing Instrument (XTI; \cite{nicer}), \xmm's European Photon Imaging Camera's (EPIC) pn \cite{epicpn} and MOS \cite{epicmos} detectors, \xmm's Reflection Grating Spectrometer (RGS; \cite{rgs}), \swift's X-Ray Telescope (XRT; \cite{swift, xrt}), and the eROSITA instrument \cite{erosita} on-board the Russian/German Spectrum-Roentgen Gamma (SRG) mission. \nicer provided high-cadence monitoring data of the majority of the outburst while \xmm performed five exposures: one near the peak of the outburst (MJD 59287.34) and four after its luminosity decreased by a factor of $\gtrsim$200 compared to the peak (on MJDs 59416.76, 59552.55, 59556.75, 59615.36; See Fig. \ref{fig:fig1}). A few \swift exposures were taken early in the outburst and two sets of high-cadence monitoring--with one exposure per day lasting 1-2 ks--were performed for 15 days and 20 days after the source faded in X-rays, i.e., between MJD 59391.19-59406.72 and 59525.39-59544.44, respectively (see Fig.~\ref{fig:fig1}a). eROSITA provided limits on X-ray flux from prior to the optical outburst and a detection during the decline phase. 

\subsubsection{\nicer's XTI}\label{sec:nicerdatareduc}
The \nicer X-ray observatory has been operating on board the the International Space Station (ISS) since July of 2017. Its primary instrument is the XTI, which is made up of 56 co-aligned X-Ray Concentrators (XRCs) which focus X-rays into apertures of Focal Plane Modules (FPMs). Each FPM consists of a single-pixel (non-imaging) Silicon Drift Detector (SDD; \cite{sdd}) with a field of view area of roughly 30 arcmin$^{2}$. At the beginning of science operations 52 out of 56 FPMs were active. The combination of these detectors provides a nominal bandpass of 0.3-12 keV with a peak effective area of $\sim$1900 cm$^{2}$ near 1.5 keV. This large effective area in the soft X-rays, good spectral resolution (E/$\Delta$E $\sim$ a few 10s (see \url{https://heasarc.gsfc.nasa.gov/docs/nicer/mission_guide/}) combined with rapid maneuvering capability makes \nicer an ideal facility to perform spectral monitoring studies of variable soft X-ray phenomena like TDEs. 

\nicer started monitoring \target on 13 February 2021 as part of an approved guest observer program (PI: Pasham, program number: 3139) performing multiple visits per day when possible. In this work we include 162 observation IDs (obsIDs) totalling $\approx$300 ks of exposure time spread across 1921 Good Time Intervals (GTIs) before any data screening was applied.  

We started \nicer data reduction with the raw data, i.e., unfiltered ({\it uf}) event files, publicly available on the High Energy Astrophysics Science Archive Research Center (HEASARC)'s archive: \url{https://heasarc.gsfc.nasa.gov/cgi-bin/W3Browse/w3browse.pl}. These data were reduced/cleaned using the \nicer data reduction tools packaged as {\tt NICERDAS} which itself is part of the High Energy Astrophysics Software (HEASoft). We used HEASoft version 6.29c (released on 1 September 2021) with the latest \nicer calibration files {\it xti20210707} (20 July 2021). \nicer version {\it 2021-08-31\_V008c} was used.  

\begin{sloppypar}
\nicer data is organized in the form of obsIDs where often each obsID contains multiple exposures taken over a period of one day. The initial data reduction to produce the unfiltered but calibrated event files ({\it ufa}), cleaned event files ({\it cl}) and Good Time Intervals (GTIs) was done on a per obsID basis using the standard {\tt nicerl2} tool. We used the following filters to extract the GTIs: {\it nicersaafilt=YES}, {\it saafilt=NO}, {\it trackfilt=YES}, {\it ang\_dist=0.015}, {\it st\_valid=YES}, {\it elv=15}, {\it br\_earth=30}, {\it cor\_range=``-''}, {\it min\_fpm=38}, {\it underonly\_range=``*-*''}, {\it overonly\_range=``*-*''}, {\it overonly\_expr=``NONE''}. Except for the {\it underonly\_range}, {\it overonly\_range}, and {\it overonly\_expr} parameters the rest are the default values as recommended by the \nicer data analysis guide: \url{https://heasarc.gsfc.nasa.gov/lheasoft/ftools/headas/nimaketime.html}. Instead of screening GTIs based on the {\it underonly\_range}, {\it overonly\_range}, and {\it overonly\_expr values}--which are proxies for screening out epochs of optical light leak and high particle background--we chose to screen them based on background-subtracted rates in the so-called S0-band (0.2-0.3 keV) and the HBG band (13-15 keV) as suggested by \cite{3c50}. Screening this way at a later stage, i.e, after computing the background spectrum, minimizes the total amount of data loss.

After extracting the unfiltered but calibrated ({\it ufa}) event files, calibrated ({\it cl}) event files and the GTIs, we performed further analysis on a per GTI basis. First, we identify all the so-called ``hot'' detectors in each GTI, i.e., those affected by optical light leak and produce spuriously large amounts of charge. This is done by first estimating the mean count rate in the 0.0--0.2 keV band for each of the active FPMs in a given GTI. This array of 52 values is sigma-clipped, and detectors with values more than 4$\sigma$ above the median of the sigma-clipped values are marked as ``hot'' for a given GTI. This information is also used further down the analysis pipeline while extracting time-resolved energy spectra (see section \ref{sisec:timeresolspecs}). Using the 3c50 background model \cite{3c50}, we estimated a background for each GTI by taking care to exclude the ``hot'' detectors. As per the recommendation given by \cite{3c50}, a given GTI is considered valid only if the following two conditions are met: 1) absolute value of the background subtracted count rate in S0-band, i.e., 0.2-0.3 keV, is less than 10 cps, and 2) absolute value of background-subtracted count rate in HGB band, i.e., 13-15 keV, is less than 0.1 cps. Finally, we also require that the observed 15-18 keV rate in a given GTI be within one standard deviation of the distribution of all observed 15-18 keV rates to exclude false flares. GTIs that do not satisfy these conditions are discarded and not included in further analysis. After the data screening we were left with 239 ks of exposure spread over 364 GTIs. 

As recommended by the \nicer data analysis guide (https://heasarc.gsfc.nasa.gov/docs/nicer/analysis\_threads/cal-recommend/), we impose a conservative systematic uncertainty of 1.5\%, i.e., {\it systematic 0.015} in {\it XSPEC}, during all spectral modeling.

%++++++++++++++++++++++++++++++++++++++++++++++++++++++++++++++++++++++++++++
% ------------------- XMM/EPIC DATA REDUCTION ------------------------------
%++++++++++++++++++++++++++++++++++++++++++++++++++++++++++++++++++++++++++++
\subsubsection{\xmm EPIC}\label{sec:epicdatared}
For all the \xmm observations (obsIDs: 0852600301, 0891800101, 0891803701, 0891803801, 0893810701; PI: Pasham; see the Table \ref{tab:xmmxraydata}) we started our data reduction with their raw Observation Data Files (ODF). Using \xmm's science analysis software (xmmsas version 19.1.0) we reprocessed the EPIC-pn and MOS data using the standard tools {\tt epproc} and {\tt emproc}, respectively. During the first observation (obsID: 0852600301/XMM\#1), because the MOS data were taken in the {\it small window} mode there was no source-free area on the CCD to extract a background from. Because of this reason we decided to exclude MOS data from obsID 0852600301. The rest of the observations were taken in the full window mode with ample area to estimate a background. To enhance the signal to noise of the resulting spectra we used both the pn and MOS datasets from the rest of the observations, i.e., 0891800101 (XMM\#2), 0891803701, 0891803801 (XMM\#3), and 0893810701 (XMM\#4). 

After producing the cleaned event files we extracted GTIs without background flares (non-flare GTIs) using the 10-12 keV light curve as outlined in the \xmm data analysis guide. For obsID 0852600301, we also extracted the instrumental GTIs for pn. By combining these two sets of GTIs (instrumental and non-flare) we extracted a set of GTIs without any background flares and when pn was actively operating. For the other 4 datasets we extracted GTIs when background flaring was low and when both the pn and the MOS detectors were operating.

\end{sloppypar}
The source spectra and event files were estimated using a circular aperture centred on the optical position of (ra, dec) = (04:13:02.450, -53:04:21.72) (J2000.0 epoch) and a radius of $\ang{;;33}$. This radius corresponds to roughly 90\% of the light from a point source as estimated by the fractional encircled energy of the EPIC-pn instrument. For the 4 datasets where the source decreased by more than two orders of magnitude, we used a smaller circular extraction region of $\ang{;;25}$ to minimize background contamination. Background spectra and events were extracted from two nearby circular regions, away from any point sources, each with radii of $\ang{;;45}$. While extracting the spectra we imposed additional filters of {\it \#XMMEA\_EP \&\& (FLAG==0) \&\& (PATTERN$<$=4)} to only include the high quality events for pn. For MOS we used {\it \#XMMEA\_EM \&\& (FLAG==0) \&\& (PATTERN$<$=12)}. 

The final spectra from obsID 0852600301 were grouped using the xmmsas tool {\tt specgroup} to ensure a minimum of 20 counts per bin and an oversampling of 3. $\chi^2$ statistics were used for fitting spectral models. For the case of obsIDs 0891800101, 0891803701, 0891803801 and 0893810701, due to low counts, we used a minimum of 1 count per bin with an oversampling of 3, and used the Cash statistic while spectral modeling. ObsIDs 0891803701 and 0891803801 were taken a few days apart so we modelled them together to improve the signal-to-noise. 

%++++++++++++++++++++++++++++++++++++++++++++++++++++++++++++++++++++++++++++
% ------------------------ RGS DATA REDUCTION ------------------------------
%++++++++++++++++++++++++++++++++++++++++++++++++++++++++++++++++++++++++++++
\subsubsection{\xmm RGS}\label{sec:rgsdatareduc}
\target was detected by the RGS only during the first observation (obsID: 0852600301). We use the latest pipeline RGS data products, which include the source and background spectral files, together with the instrument response files. We consider only the first order spectra, which provide the highest signal-to-noise. In order to improve the signal-to-noise we first stacked the RGS 1 and RGS 2 spectra. Then using the {\tt ftgrouppha} task we binned the spectrum using the optimal scheme described by \cite{optmin} with an additional requirement of at least 1 count per spectral bin. We then fit using the Cash statistics, in order to exploit the high-energy resolution of the instruments. We focused the analysis in the observer-frame energy band of 0.35-0.75 keV, which is found to be clearly dominated by the source counts. 

%++++++++++++++++++++++++++++++++++++++++++++++++++++++++++++++++++++++++++++
% ------------------- SWIFT/XRT DATA REDUCTION ------------------------------
%++++++++++++++++++++++++++++++++++++++++++++++++++++++++++++++++++++++++++++
\subsubsection{\swift XRT}\label{sec:xrtdatareduc}
\swift monitored \target between 20 February 2021 and 26 November 2021. Between 20 February and 15 April the source was observed once every 3-5 days (proposer: Hinkle) while high-cadence (one visit per day) observations were made during 26 June to 11 July and 7 November to 26 November (Proposer: Pasham). The duration of individual visits/exposures varied between 1000 and 2000 seconds.
 
We started our XRT data analysis with the raw data from the HEASARC public archives and reprocessed them using the standard HEASoft tool {\tt xrtpipeline}. All XRT data were taken in the so-called Photon Counting (PC) data mode. We only used events with grades between 0 and 12 as recommended by the data analysis guide. Source events were extracted from an aperture of $\ang{;;30}$. Background events were extracted in an annulus with inner and outer radii of $\ang{;;60}$ and $\ang{;;180}$, respectively. We ensured that there weren't any point sources within this background annulus. 

With a mean background-subtracted 0.3-1.1 keV count rate of $\approx$1.6$\times$10$^{-3}$ counts s$^{-1}$, \target was barely detected in the individual exposures during the two high cadence campaigns. Therefore, we combined the data from these epochs to extract one average flux measurement per campaign (see Fig. \ref{fig:fig1}a). 

%++++++++++++++++++++++++++++++++++++++++++++++++++++++++++++++++++++++++++++
% ------------------- eROSITA DATA REDUCTION ------------------------------
%++++++++++++++++++++++++++++++++++++++++++++++++++++++++++++++++++++++++++++
\subsubsection{SRG/eROSITA}\label{sec:erositareduc}
eROSITA \cite{erosita}, the soft X-ray instrument aboard the Spectrum-Roentgen-Gamma mission \cite{Sunyaev+2021:srg}, started the first of eight X-ray all-sky surveys (eRASS1-8, each completed in six months) on 13 December 2019. It has since scanned over the coordinates of \target in the first four eRASS1-4, although no source was detected with significance until eRASS4. Data were processed using eROSITA Science Analysis Software v946 (eSASS; \cite{Brunner+2021:esass}). Photons were extracted around the source coordinates within a circular aperture of radius $\ang{;;30}$, while background counts were extracted within an offset source-free circle of radius $\ang{;;156}$.

In particular, in eRASS1 the telescope passed several times over \target between 2020-01-16 (03:42:27 UTC) and 2020-01-22 (11:42:41) without detecting it (net exposure of $\approx890\,$s). Assuming the spectral model obtained by \nicer, a 3-sigma upper limit of the observed flux in rest-frame $0.3-1.1\,$keV can be inferred at $\lesssim1.5\times10^{-14}\,$ \fluxunits. The same for eRASS2 between 2020-07-16 (00:47:24) and 2020-07-22 (16:47:42, net exposure of $912\,$s), with an inferred 3-sigma upper limit of $\lesssim6.6\times10^{-15}\,$ \fluxunits, and for eRASS3 between 2021-01-12 (22:42:27) and 2021-01-17 (10:42:42, net exposure of $538\,$s), with $\lesssim1.2\times10^{-14}\,$\fluxunits. Data products were also extracted from the cumulative image combining all the first three eRASS scans, namely a net exposure of $2339$\,s taken from 2020-01-16 to 2021-01-17: the stacked signal on the cumulative image allows the source to be as bright as $\lesssim1.3\times10^{-14}\,$ \fluxunits (at 3$\sigma$). During eRASS4 the telescope scanned over \target between 2021-07-19 (19:47:27 UTC) and 2021-07-24 (07:47:42). The source was detected with a total number of 14 counts in the $0.2-2.3\,$keV band in 663\,s of net exposure. Fitting the spectrum with a {\it diskbb} model results in a median (and related 16th and 84th percentiles) observed flux of $7.2^{+2.3}_{-1.9}\times10^{-14}\,$ \fluxunits between rest frame $0.3-1.1\,$keV.

% ^o^o^o^o^o^o^o^o^o^o^o^o^o^o^o^o^o^o^o^o^o^o^o^o^o^o^o^o^o^o^o^o^o^o^o^o^o^o
% ***************************** Optical and UV *******************************
% ^o^o^o^o^o^o^o^o^o^o^o^o^o^o^o^o^o^o^o^o^o^o^o^o^o^o^o^o^o^o^o^o^o^o^o^o^o^o
\subsubsection{Optical and UV}
\target's optical and ultraviolet (UV) data used here were obtained by \swift's UV Optical Telescope (UVOT; \cite{uvot}), the All-Sky Automated Search for SuperNovae (ASAS-SN; \cite{asassn, benassasn}), and the Transiting Exoplanet Survey Satellite (\tess; \cite{tess}). To study the host galaxy we also used archival optical, UV and infrared data prior to 2020, i.e., before the optical and the X-ray outbursts (see section \ref{sec:host}). Two optical spectra were obtained by the FLOYDS spectrograph: one near the peak of the optical light curve \cite{firstspec} and another after the source faded in X-rays. A high signal-to-noise optical spectrum was also obtained using the MagE spectrograph on the Magellan telescope \cite{mage} on 19 Aug 2021, i.e., after the X-ray outburst ended. Four more high SNR optical spectra were obtained with LDSS-3 on Magellan. A description of the reduction procedures for these datasets is described below.

%++++++++++++++++++++++++++++++++++++++++++++++++++++++++++++++++++++++++++++
% ------------------- UVOT DATA REDUCTION -----------------------------------
%++++++++++++++++++++++++++++++++++++++++++++++++++++++++++++++++++++++++++++
\subsubsection{\swift UVOT and archival data}\label{sec:uvotdatareduc}
\swift~ UVOT \cite{Roming_2005} images were taken simultaneously with XRT observations (section \ref{sec:xrtdatareduc}). We reduce the observations using the {\tt uvotsource} task in HEAsoft using a $\ang{;;5}$ aperture. 

To estimate the host galaxy properties (see section \ref{sec:host})  and to subtract its contribution to the \swift UVOT photometry, we compile the host galaxy spectral energy distribution (SED) using archival observations in the UV through IR bands. In the mid-IR we use WISE \cite{Cutri_13} W1, W2, and W3 magnitudes. We also use DES \cite{Abbott_18} Kron magnitudes in  g, r, i, z, and Y optical bands, while for the UV, we performed aperture photometry on the GALEX \cite{Bianchi_11} NUV and FUV images with {\tt gPhoton} package \cite{Million_16} using a $\ang{;;5}$ aperture. 

%++++++++++++++++++++++++++++++++++++++++++++++++++++++++++++++++++++++++++++
% ------------------- ASASSN DATA REDUCTION ---------------------------------
%++++++++++++++++++++++++++++++++++++++++++++++++++++++++++++++++++++++++++++
\subsubsection{ASASSN}
ASAS-SN began surveying the sky in 2013 with the goal of identifying bright transients across the whole sky with an un-targeted survey. From 2013 - 2017 ASAS-SN expanded from two to eight telescopes with $V$-band filters mounted on two mounts at two stations: Haleakala Observatory (Hawaii) and Cerro Tololo International Obervatory (CITO, Chile). In late 2017 we added 12 additional telescopes on three additional mounts at one at McDonald Observatory (Texas), one at South African Astrophysical Observatory (SAAO, South Africa), and an second station at CTIO.  Our stations are hosted by the Las Cumbres Observatory Global Telescope Network (LCOGT; \cite{LCO}). Finally, in late 2018 we switched the 8 original telescopes from $V$-band to $g$-band and we scan the entire visible sky down to $g\sim18.5$ mag nightly.   

ASAS-SN units use  FLI ProLine Cooled $2k\times2k$ CCD cameras with 14-cm aperture Nikon telephoto lenses. The units' field-of-view is 4.5 degrees on a side (20 degrees$^2$) with pixel size of 8.0 arcsec. Ideally, each observation epoch consists of three dithered 90 second exposures, though we are currently averaging 2.7 exposures per epoch due to scheduling and weather events. Furthermore, our observations are split between those taken with legacy $V$-band filters and $g$-band filters which we plan to use going forward. The limiting $V$- and $g$-band magnitudes are $m \sim 17.5$ and $m\sim18.5$, respectively. The original CTIO and Hawaii stations used $V$-band filters for observations up until the spring of 2019 when they were switched to $g$-band. The latter three stations have been using $g$-band filters since beginning of their operations. 

%++++++++++++++++++++++++++++++++++++++++++++++++++++++++++++++++++++++++++++
% ------------------- TESS DATA REDUCTION -----------------------------------
%++++++++++++++++++++++++++++++++++++++++++++++++++++++++++++++++++++++++++++
\subsubsection{TESS}
Fortuitously, \tess~ captured the rise of the outburst in the optical band at an unprecedented cadence of one exposure every 30 minutes. We extracted a light curve following the procedures in \cite{faustesslc}.  Briefly, we use the ISIS image subtraction software \cite{aladlupton98, alard_2000} to subtract a median ``reference'' image from individual TESS Full Frame Images (FFIs) after convolving with a spatially variable kernel.  This provides a correction for instrumental systematic errors due to pointing jitter, pointing drift from velocity aberration, and intrapixel sensitivity variations.  Combined with some additional post-processing steps to remove scattered light from the Earth/moon and non-uniform pixel sensitivity due to CCD ''straps'', difference imaging has been shown to perform well in the background dominated regime for TESS data (see, for example, \cite{vallely_2019}). We then perform forced photometry at the location of the transient in the differenced TESS images using a model of the instruments Pixel Response Function at that location. 

Previously, \tess captured the rise of a stellar TDE ASASSN-19bt and it was found that the optical brightness rose as t$^{\alpha}$, where $\alpha = 2.10 \pm 0.12$ \cite{holoien19a}. This value is similar to the ``fireball'' model commonly used to the fit the early rises of supernovae \cite{riess99}. Two other normal TDEs have had their rise slopes measured, albeit not with high cadence TESS data. These are ASASSN-19dj, with a rise slope of $\alpha = 1.90^{+0.42}_{-0.36}$ measured from ASAS-SN $g$-band data \cite{hinkle21a}, and AT2019qiz, with a rise slope of $\alpha = 1.99 \pm 0.01$ measured from the bolometric light curve \cite{nicholl20}. Additionally, several other nuclear transients have had rise slopes measured with \tess, with flatter rises than these TDEs. These include the repeating TDE ASASSN-14ko, with a rise slope of $\alpha = 1.01 \pm 0.07$ \cite{Payne21} for the first flare observed by TESS when assuming a single power-law model, and $\alpha = 1.10 \pm 0.04$ and $\alpha = 1.50 \pm 0.10$ for the first and the second flares observed by TESS, respectively, when assuming a curved power-law model \cite{payne2022b}, and the ANT ASASSN-20hx, with a rise slope of $\alpha = 1.05 \pm 0.06$ \cite{hinkle22}.

We modeled \target's \tess~ light curve with a function of the form flux $\propto$ (t - t$_{0}$)$^{\alpha}$ excluding data after MJD 59199, when the \tess background flux began to dominate the signal. We find a best-fit t$_{0}$ of MJD = $59189.5 \pm 0.3$ and power-law index $\alpha$ of $1.35 \pm 0.09$, flatter than the three TDEs with measured rise slopes but steeper than either ASASSN-14ko or ASASSN-20hx.

%++++++++++++++++++++++++++++++++++++++++++++++++++++++++++++++++++++++++++++
% ------------------- FLOYDS DATA REDUCTION ---------------------------------
%++++++++++++++++++++++++++++++++++++++++++++++++++++++++++++++++++++++++++++
\subsubsection{FLOYDS optical spectra}\label{sec:floydsreduc}
Two spectra were taken by Las Cumbres Observatory \cite{LCO}, using the FLOYDS spectrograph on the 2.0m Faulkes Telescope South. Spectra cover a wavelength range of 3500–10000 \r{A} at a resolution R$\approx$300–600. Data were reduced using \texttt{floyds\_pipeline}: \url{https://github.com/lcogt/floyds_pipeline}, which performs cosmic ray removal, spectrum extraction, and wavelength and flux calibration using standard IRAF/PyRAF routines as described in \cite{Valenti14}.

%++++++++++++++++++++++++++++++++++++++++++++++++++++++++++++++++++++++++++++
% ------------------- MagE DATA REDUCTION -----------------------------------
%++++++++++++++++++++++++++++++++++++++++++++++++++++++++++++++++++++++++++++
\subsubsection{Magellan/MagE optical spectrum}\label{sec:magereduc}
ASASSN--20qc was observed on 19 August 2021 with the Magellan Echellete spectrograph (MagE), mounted on the Magellan Baade telescope located at Las Campanas Observatory, Chile. The observation was 3600 seconds long, and we used a 0.7 arcsec slit, which delivers a FWHM spectral resolution of 50 km s$^{-1}$ at 4000 \AA. The spectrum was reduced using the dedicated MagE data reduction pipeline \cite{Kelson2000, Kelson2003}. The flux calibration was performed using a spectrophotometric standard star Feige 110 observed during the night.

\subsubsection{Magellan/LDSS-3 optical spectrum}\label{sec:ldssreduc}
We obtained 4 spectra in 2021 and 2022 (9 November 2021, 25 January 2022, 10 March 2022 and 16 August 2022) using the Low-Dispersion Survey Spectrograph 3 (LDSS-3) on the 6.5-m Magellan Clay telescope. Each set of observations included 4 1200s exposures of the target using a 0\farcs{9} slit and the VPH-All grism and was taken at parallactic angle. We used {\sc Iraf} to reduce our LDSS-3 spectra following standard procedures, including bias subtraction, flat-fielding, one-dimensional spectral extraction, wavelength calibration using a comparison lamp spectrum, and median combination of the individual exposures into a single final spectrum. We flux-calibrated our spectra using observations of spectrophotometric standard stars obtained on the same nights as our science spectra.

% ^o^o^o^o^o^o^o^o^o^o^o^o^o^o^o^o^o^o^o^o^o^o^o^o^o^o^o^o^o^o^o^o^o^o^o^o^o^o
% ************************************* Radio ********************************
% ^o^o^o^o^o^o^o^o^o^o^o^o^o^o^o^o^o^o^o^o^o^o^o^o^o^o^o^o^o^o^o^o^o^o^o^o^o^o
\subsubsection{Radio}
The position of \target was observed by the Australian SKA Pathfinder (ASKAP) Telescope as part of the Rapid ASKAP Continuum Survey (RACS; \cite{askap2020,askaphale2021}) and the ASKAP Variables and Slow Transients survey (VAST, \cite{vast2013,vast2021}). Overall, there are 11 observing epochs of \target with ASKAP (1 with RACS and 10 with VAST). All of the observations were conducted at a central frequency of 887.5 MHz with a bandwidth of 288 MHz. The data were reduced using the VAST pipeline \cite{vastpipeline} and the full set of measurements is presented in the Table  \ref{tab:Radio_Observations}.

The source is detected in the first ASKAP epoch (RACS data) on 2019 May 4. It then seems to fluctuate (note also that the image rms level is also fluctuating between epochs) until the last ASKAP observation undertaken more than two years later on 2021 August 22. However, given the large flux density errors, it is not possible to determine statistically whether these observed fluctuations originate from variability of the source of are merely statistical fluctuations. There is also no significant change in the radio flux density in the two observing epochs following \target's optical discovery. The observed mean flux density of the source is 1.13\,mJy which translates to a luminosity of $3.7 \times 10^{37}\,{\rm erg s}^{-1}$. 

%++++++++++++++++++++++++++++++++++++++++++++++++++++++++++++++++++++++++++++
% ------------------- MODELING OPTICAL SPECTRA: BLACK HOLE MASS -------------
%++++++++++++++++++++++++++++++++++++++++++++++++++++++++++++++++++++++++++++
\section{Optical spectral modeling and black hole mass}\label{subsec:bhmass}
A fundamental parameter of probing the underlying  physics is the black hole mass. We estimated this from the optical spectra. First, we re-scaled all the spectra based on the photometric magnitude obtained from ASAS-SN automated pipeline. Then we performed multi-component spectral decomposition using \textsc{PYQSOFIT} developed by \cite{2018ascl.soft09008G} to measure the spectral information. A detailed description of the spectral decomposition method is given in \cite{2020ApJS..249...17R}. In brief, first, we corrected the spectrum for Galactic extinction using the Milky Way extinction law of \cite{1989ApJ...345..245C} with Rv=3.1 and the \cite{1998ApJ...500..525S} map. Then the spectrum was transformed to the rest-frame using a redshift of 0.056. 

The continuum was modeled using a combination of AGN power-law ($f_{\lambda}=A\lambda^{\alpha_{\lambda}}$) and optical Fe II template from \cite{1992ApJS...80..109B} to represent various blended Fe II emission lines. As stellar absorption features were not visible in the spectra, decomposition of the host galaxy contribution was not attempted. During the continuum fitting, all the strong Balmer emission lines were masked.  The best-fit continuum model (PL + FeII) was subtracted resulting in a pure emission line spectrum, which was modeled using multiple Gaussian components.  

Emission lines were decomposed into broad and narrow components where each narrow component was modeled using single Gaussian with a maximum Full Width at Half Maximum (FWHM) of 900 \kms to separate Type 1 AGN from Type 2 AGN following previous studies (e.g., \cite{2020ApJS..249...17R}), while the broad components were modeled using two Gaussians each having FWHM larger than 900 \kms.  The velocity and width of the narrow components were tied together within an emission line complex. The broad H$\beta$ and H$\alpha$ components were modeled using two Gaussians and [O III]$\lambda\lambda$5007,4959 doublets were modeled using two Gaussians (one for the core and another for the wing). During the fit, the flux ratio of [O III] and [N II] doublets were fixed at the theoretical values i.e. F(5007)/F(4959) = 3 and F(6585)/F(6549) = 3. All the emission lines in a given line complex was fitted together. The emission line information from  spectral decomposition is given in Table \ref{tab:mbh}.

The FWHM of H$\beta$ and H$\alpha$ were measured to be 2108$\pm$183 \kms and $2654\pm441$ \kms, respectively at the epoch of 11 January 2021 when the source was in the high state with monochromatic luminosity at 5100 \r{A} ($\log L_{5100}$) of $43.97\pm0.01$ erg/s. The AGN continuum was very blue. However, the source became 
fainter by 30 July 2021 with $\log L_{5100} ($\ergs$) = 43.84\pm0.01$. A strong, very broad component in H$\alpha$ was found. The He I 5876 \r{A}, which was undetectable in January 2021 also became stronger. A very broad component (of FWHM $\sim 10,000$ \kms) in H$\alpha$ is clearly visible in the August and the November spectra. Compared to the January spectrum, the later spectra show stronger H$\beta$ and H$\alpha$ emission lines. The R4570, defined as the flux ratio between Fe II emission in the wavelength range of $4435-4685$ \r{A} to the H$\beta$ broad component, is found to be $\approx$0.6. This value is typical for narrow-line Seyfert 1 galaxies \cite{2017ApJS..229...39R}.     

The black hole mass was estimated from the monochromatic luminosity at $L_{5100}$ and the width of the H$\beta$ broad component using virial relation given by \cite{2015ApJ...801...38W}. The black hole mass estimates are found to be consistent in all epochs with an average value of $M_{\bullet}=10^{7.5} M_{\odot}$. The reported errorbars on the black hole mass in Table \ref{tab:mbh} only include measurement uncertainties. They do not include the uncertainty ($>$0.4 dex) associated with the virial relation due to the systematics involved in the calibration, unknown geometry, and the kinematics of the broad line region (BLR).  

The virial mass measurements have several caveats and biases, e.g., the virial assumption evidence of which has been found in several AGNs with multiple emission lines and from the velocity resolve reverberation mapping, host-galaxy subtraction, unknown geometry and kinematics, radiation pressure effect, and the use of different line width indicators: FWHM vs. line dispersion (a detailed discussion can be found in \cite{2013BASI...41...61S}). The validity of the virial assumption can be tested if for an increase in the luminosity of the source, the line width decreases given enough response time. However, the limited dynamic range in the variability, and the measurement errors in the spectral parameters, especially in the line widths, make this a challenging task for \target.

%++++++++++++++++++++++++++++++++++++++++++++++++++++++++++++++++++++++++++++
% ------------------- BPT DIAGRAM: AGN CONFIRMATION -------------------------
%++++++++++++++++++++++++++++++++++++++++++++++++++++++++++++++++++++++++++++

\section{\target's location in the BPT and the WHAN diagrams suggests it is an AGN}\label{subsec:bpt}
The BPT \cite{bpt} and the WHAN diagrams \cite{Cid2011} are commonly used tools to classify different class of emission line objects based on the narrow line fluxes of [N II]6584/$H\alpha$, [OIII]5007/H$\beta$ and H$\alpha$ equivalent width (see Table \ref{tab:mbh}).

For the BPT diagram, we took the error weighted average of all the epochs if measurements are reliable better than 1-$\sigma$ uncertainty. We note that due to strong blending of the narrow and the broad components in H$\alpha$ and H$\beta$ regions, estimation of the narrow components flux is challenging and the uncertainty in the flux measurement is large.  By overplotted the Kewley \cite{Kewley2001} extreme starburst curve, Kauffmann \cite{Kauffmann2003} empirical relation and Schawinski \cite{Schawinski2007} separation line of LINER and AGNs we find that \target clearly falls in the AGN region. 

For the WHAN diagram, we use the three measurements where the errorbars are reasonable. Similar to the BPT diagram, the WHAN diagram also suggests that \target is an AGN.

%++++++++++++++++++++++++++++++++++++++++++++++++++++++++++++++++++++++++++++
% ---------------------- HOST GALAXY PROPERTIES -----------------------------
%++++++++++++++++++++++++++++++++++++++++++++++++++++++++++++++++++++++++++++
\section{\target's host-galaxy properties and black hole mass}\label{sec:host}
To estimate the host properties we model the pre-flare Spectral Energy Distribution (SED; Table \ref{tab:hostgalphot}) using the flexible stellar population synthesis module (FSPS: \cite{Conroy_09}). We also included a non-stellar power-law continuum, available on FSPS, that represents an AGN contribution to the SED prior to \target. We use the {\tt Prospector} \cite{Johnson_21} software to run a Markov Chain Monte Carlo (MCMC) sampler \cite{Foreman-Mackey_13}. In the {\tt Prospector} fitting we assume an exponentially decaying star formation history (SFH), and a flat prior on the six free model parameters: stellar mass ($M_{\star}$), stellar metallicity ($Z$), color excess due to dust extinction E(B-V), assuming the extinction law by \cite{Calzetti_2000}, the stellar population age ($t_{age}$), the e-folding time of the exponential decay of the SFH ($\tau_{\rm{sfh}}$), and the fraction of the total light that is produce by the AGN ($f_{AGN}$).

% TABLE_TABLE_TABLE_TABLE_TABLE_TABLE_TABLE_TABLE_TABLE_TABLE_TABLE_TABLE
%               Extended Data Table 2: Host Galaxy Properties
% TABLE_TABLE_TABLE_TABLE_TABLE_TABLE_TABLE_TABLE_TABLE_TABLE_TABLE_TABLE

From  the  best  fit  template  spectrum  we  derive: log($M_{\star}/M_{\odot})=10.13^{+0.02}_{-0.01}$, log$(Z/Z_{\odot})=-0.55^{+0.02}_{-0.04}$, $E(B-V)=0.01^{+0.01}_{-0.01}$ mag, $t_{\rm age} = 3.23^{+0.22}_{-0.28}$ Gyr, $\tau_{\rm{sfh}} =0.43^{+0.04}_{-0.05}$ Gyr and $f_{AGN} = 0.05^{+0.01}_{-0.01}$. The color excess is in complete agreement with the Galactic value $E(B-V)=0.0137$ mag \cite{Schlafly_2011} requiring no additional extinction from the host galaxy. We estimate the host galaxy fluxes in the UVOT bands from the posterior distribution of the population synthesis models. The host contribution was then subtracted from the UVOT measured photometry (Table \ref{tab:hostgalphot}). The uncertainty on the host galaxy model was propagated into our measurements of the host-subtracted fluxes.

We also estimate $M_{\bullet}$ from the host galaxy mass by applying the \cite{Greene_2020} relation:  log $M_{\bullet}/M_{\odot} = 7.56 + 1.39[{\rm log}(M_{\star}/M_{\odot}) - 10.48]$. This results in log $M_{\bullet}/M_{\odot} = 7.06^{+0.02}_{-0.01} \pm 0.79$, where 0.79 dex is the intrinsic scatter of the relation. This values agrees with the virial mass measurements in section \ref{subsec:bhmass} and Fig.~\ref{tab:mbh}.

%++++++++++++++++++++++++++++++++++++++++++++++++++++++++++++++++++++++++++++
% ---------------------- SHOW Swift/XRT IMAGE WITH ONE POINT SOURCE ---------
%++++++++++++++++++++++++++++++++++++++++++++++++++++++++++++++++++++++++++++
\subsection{\target~ dominates the X-ray emission in \nicer/XTI's field of view}
\nicer/XTI is a single-pixel (non-imaging) instrument with a field of view of roughly 30 arcmin$^{2}$ in area. Therefore, to rule out a contaminating point source we extracted an image from the combined \swift/XRT images which shows a single point source coincident with the optical coordinates. This demonstrates that \target dominated the X-ray emission in \nicer's field of view and contamination by other sources was negligible. 
%%%%%%%%%%%%%%%%%%%%%%%%%%%%%%%%%%%%%%%%%%%%%%%%%%%%%%%%%%%%%%%%%%%%%%%%%%%%%%%%%%%%%%%%%%
%%%%%%%%%%%%%%%%%%%%%% Swift/XRT image %%%%%%%%%%%%%%%%%%%%%%%%%%%%%%%%%%%%%%%%%%%%%%%%%%%
%%%%%%%%%%%%%%%%%%%%%%%%%%%%%%%%%%%%%%%%%%%%%%%%%%%%%%%%%%%%%%%%%%%%%%%%%%%%%%%%%%%%%%%%%%

%++++++++++++++++++++++++++++++++++++++++++++++++++++++++++++++++++++++++++++
% ------------------- X-RAY ENERGY SPECTRAL ANALYSIS   ----------------------
%++++++++++++++++++++++++++++++++++++++++++++++++++++++++++++++++++++++++++++
\section{X-ray energy spectral modeling: \nicer and \xmm detect an ultrafast outflow}\label{sisec:spectral_model}
We started our X-ray energy spectral analysis with an average spectrum derived from the first few weeks of \nicer data. This spectrum was soft with essentially no source X-ray events above $\approx$ 1.1 keV. Following this revelation early in the outburst we requested for a 50 ks \xmm observation to get a deep X-ray snapshot of \target (XMM\#1). For this we triggered an approved \xmm guest observer program 085260 (PI: Pasham).

We then turned our focus to the \xmm dataset for a detailed spectral study. Similar to the earlier \nicer spectrum, \xmm/EPIC-pn spectrum was also soft with the background becoming comparable to the source beyond roughly 1.5 keV. To avoid uncertainties from background estimation and to match with \nicer's bandpass we only considered the energy range of 0.3-1.1 keV for further analysis. For spectral modeling we used the {\it XSPEC} spectral fitting package \cite{xspec} and a {\tt Python} interface to {\it XSPEC} known as {\tt PyXspec}.

We started by modeling the \xmm/EPIC-pn spectrum from XMM\#1 with simple phenomenological models: a thermal accretion disk modified by MilkyWay's neutral absorbing column and a power-law modified by MilkyWay's neutral absorbing column of 1.2$\times$10$^{20}$ cm$^{-2}$. The MilkyWay column along the direction of \target was estimated using the HEASARC nH calculator: \url{https://heasarc.gsfc.nasa.gov/cgi-bin/Tools/w3nh/w3nh.pl} \cite{heasarcnh}. These two models were defined as {\it tbabs*zashift(diskbb)} and {\it tbabs*zashift(pow)} in XSPEC. The {\it zashift} component accounts for the host galaxy redshift of 0.056. The former model resulted in a $\chi^2$ of 187.5 with 19 degrees of freedom (dof) while the latter yielded a $\chi^2$ of 2026.8 with 19 dof. In both cases, strong systematic residuals were evident (see panels (a) and (b) of Fig.~\ref{fig:xmmresiduals}). In the case of the power-law model the best-fit photon index was roughly 6. Adding a Gaussian to the power-law model improves the fit resulting in a $\chi^2$/dof of 13.4/15 (Fig.~\ref{fig:xmmresiduals}c). However, again the best-fit power-law index is extreme with a value of 8.5$\pm$0.2. Typically, AGN have a power-law index value of $\approx$1.8 with extreme values up to 3\cite{dadina08}. An index value of 8.5 is unphysical because when extrapolated to lower energies would imply an unrealistically high intrinsic luminosity. Also, such a steep index can be explained by the fact that in the narrow bandpass of 0.3-1.1 keV we are fitting the Wien's portion of the black body emission, which naturally leads to a steep index when modeled with a power-law. Using a thermal disk plus a power-law model, i.e., {\it tbabs*zashift(diskbb+pow)}, does not improve the fit significantly when compared with the disk only model (Fig.~\ref{fig:xmmresiduals}d). In fact, the best-fit power-law normalization value is pushed to a value close to zero. Given the soft nature of the spectrum we proceed with the thermal model for the continuum, i.e., {\it tbabs*zashift(diskbb)}, which gives an inner disk temperature of roughly 90 eV.

The residuals show a systematic behavior with an excess near 0.65 keV and a deficit near 0.85 keV (Fig.~\ref{fig:xmmresiduals}a). These are remarkably similar to the residuals seen in the early X-ray spectra of the TDE ASASSN-14li which has been interpreted as a newly launched ultra-fast outflow \cite{14liufo}. Similar residuals at energies between 0.3-10 keV have been seen in X-ray spectra of numerous AGN, and these are also interpreted as ultra-fast outflows (see e.g., \cite{pcygni1, pcyg2, tombesi_ufo}). Motivated by these previous studies we fit the residuals by first adding an absorption line. As customary in X-ray spectral analysis in {\it XSPEC}, we model the absorption feature adding an inverted (negative intensity) Gaussian line. This is mathematically equivalent of including a multiplicative Gaussian absorption line. The overall $\chi^2$/dof improved from 187.5/19 with {\it tbabs*zashift(diskbb)} to 43.1/16 with {\it tbabs*zashift(diskbb+gauss\_absorption)} (Fig.~\ref{fig:xmmresiduals}e). Except for the redshift of the host galaxy and the MilkyWay column, all the other model parameters were allowed to be free for the above fits. Because there are still systematic deviations near 0.65 keV (Fig.~\ref{fig:xmmresiduals}e), we added a Gaussian emission line which improves the fit to 11.8/13 (Fig.~\ref{fig:xmmresiduals}g). We also fitted by adding the Gaussian emission line first which yielded a $\chi^2$/dof of 50.7/16 (Fig.~\ref{fig:xmmresiduals}f). We also experimented with considering a multiplicative Gaussian component for the absorption line, modeled as ``gabs'' in {\it XSPEC}, {\it tbabs*zashift(gabs*diskbb+gauss)} which yielded a similar good fit with $\chi^2$/dof of 11.8/13 (Fig.~ \ref{fig:xmmresiduals}h).

Encouraged by the above Gaussian fits we implemented in XSPEC a physically-motivated XSTAR \cite{xstar} table model consisting of ionized gas between the illuminating central X-ray source and us, the observer (see details of the XSTAR model in section \ref{sisec:xstar}). This model gives a good fit with $\chi^2$/dof of 19.9/16, and the best-fit parameters imply the presence of a ultra-fast outflow (UFO) moving towards us at $\approx$0.33$c$, where $c$ is the speed of light (Fig.~\ref{fig:xmmresiduals}i).

To rule out that the feature is not an artifact of limited bandpass we also fit the 0.3-2.0 keV bandpass of EPIC-pn spectrum. This gave results consistent with the above parameters. We also tested if a powerlaw component maybe present after adding the UFO. Adding the powerlaw improved the $\chi^{2}$ by 5 with 2 additional degrees of freedom. Based on the Akaike information criterion we conclude that a powerlaw in not statistically required by the data.

We also analyzed the combined RGS1 and RGS2 spectrum (0.35-0.75 keV) from XMM\#1, which has roughly a factor of 30 higher resolving power than EPIC-pn. We find clear evidence for two narrow outflow components in the RGS spectrum. These can be modeled with an XSTAR table model with a velocity broadening of 100 \kms. The derived column density and ionization parameter values are [(4.6$^{+4.5}_{-3.2}$)$\times$10$^{21}$ cm$^{-2}$, 2.7$^{+0.5}_{-0.4}$ erg~s$^{-1}$~cm],  and [(2.0$^{+7.9}_{-1.2}$)$\times$10$^{21}$ cm$^{-2}$, 1.5$^{+0.5}_{-0.9}$ erg~s$^{-1}$~cm] for the two outflows. The velocity shift is consistent with an outflow velocity of $\sim$1000 \kms for the first one and with zero for the second one. These parameters are consistent with warm absorbers (WAs) typically detected in local Seyfert galaxies (e.g., \cite{2017Natur.549..488R}) and are discussed in detail in a separate paper \cite{2023ApJ...954..170K}.

The \xmm dataset does not allow us to reliably perform a joint RGS and EPIC-pn fit for several reasons. (i) Because the spectrum is extremely soft, this limits the RGS band to 0.35-0.75 keV and the EPIC-pn to 0.3-1.1 keV. In this overlapping, very soft band the two instruments are known to have significant cross-calibration uncertainties (e.g., \cite{2010A&A...516A..61D}). (ii) The very soft source spectrum and limited energy band does not allow us to employ the typical method of using the RGS data below 1.5-2 keV and the EPIC-pn from 1.5-2 keV up to 10 keV. (iii) If we perform a joint fit using the RGS between 0.35-0.75 keV and the EPIC-pn between 0.7-1.1 keV, the source continuum would not simply extend from the RGS to the EPIC-pn band. This is because the broad absorption feature would significantly affect the continuum shape and intensity in the 0.7-1.1 keV band. Consequently, we cannot use a simple cross-normalization constant between the two instruments. Therefore, for the aforementioned reasons, we performed separated fits to the EPIC-pn and the RGS data.

\subsection{The broad absorption residuals cannot be explained with slow outflows}\label{sisec:noslow}
While a detailed study of these WAs will be presented in a separate work \cite{2023ApJ...954..170K}, we address three specific questions here. First, can the slow-moving outflows found in RGS data explain the residuals seen in the low-resolution EPIC-pn and \nicer spectra? To answer this, we fit a model consisting of thermal emission modified by two slow outflows to the pn spectrum. The parameters of the outflows were constrained to be within 99\% of the best-fit values from RGS modeling. The exact model we used was {\it tbabs*WA1*WA2*zashift(diskbb)}, where WA1 and WA2 are the two warm absorbers. This model gives a very poor best-fit $\chi^{2}$/dof of 79.6/15 with similar residuals as without WA1 and WA2 (Fig.~\ref{fig:xmmresiduals}j). From this we conclude that the X-ray spectral residuals seen Fig.~\ref{fig:xmmresiduals}a cannot be explained by the two slow outflows seen in the RGS data.

Second, can we explain the residuals in Fig.~\ref{fig:xmmresiduals}a with a 3$^{rd}$ warm absorber? Adding a 3$^{rd}$ slow outflow to the EPIC-pn data improves the fit and results in a $\chi^2$/dof of 19.2/13, which is however still worse than the case of a single mildly-relativistic outflow. The best-fit column density and ionization parameter of this 3$^{rd}$ WA are 1.2$^{+0.6}_{-0.4}\times$10$^{23}$ cm$^{-2}$ and 4.1$^{+0.5}_{-0.6}$ erg~s$^{-1}$~cm, respectively. Because it is a slow outflow by definition and EPIC-pn spectrum does not have the sufficient spectral resolution in the soft X-ray band to discriminate velocity shifts lower than $\sim$10,000 km~s$^{-1}$, its velocity shift was fixed to zero. Now, the presence of such a putative very-high column third WA component in the EPIC-pn should lead to intense narrow absorption lines and ionization edges in the RGS data in 0.35-0.75 keV band. To test for this, we modeled the RGS data with 3 WAs, with the third WA having the same parameters inferred from the EPIC-pn data. Adding the 3rd WA to the RGS data provides a much worse fit (C-stat/dof of 677.7/337) with respect to just two WAs components (C-stat/dof of 523.9/337). Therefore, we conclude that the presence of a 3rd very-high column WA is excluded by the RGS data and that the broad residual feature in Fig.~\ref{fig:xmmresiduals}a is better interpreted as broad OVIII resonant absorption, with a blue-shift of $\sim$0.3$c$ and broadening of $\sim$30,000 \kms, instead of OVII-VIII edges with a low velocity shift and broadening of $\sim$100 \kms.

Finally, we also address the question: how does the inclusion of the two RGS WA components in the EPIC-pn spectrum alter the inferred properties of the UFO? To answer this, we fit two models, one consisting of a thermal component modified by two slow outflows and one UFO, i.e., {\it tbabs*WA1*WA2*UFO*zashift(diskbb)} in XSPEC, and another with a thermal component modified by the UFO alone, i.e., {\it tbabs*UFO*zashift(diskbb)} in XSPEC. In the former case, we constrained the parameters of the two WAs to be within 99\% of the best-fit RGS values. The best-fit column, ionization parameter, and the line of sight velocity values of the UFO with and without the slow components are consistent with each other within the 90\% uncertainties. Thus, we conclude that the two warm absorbers detected in the RGS spectrum do not affect the properties of the UFO, and we do not include them in further modeling. 

% A detailed study of these WAs will be presented in a separate work (Kosec et al., in prep.).

\subsection{A spectral model with two thermal components akin to quasi-periodic eruptions is ruled out}\label{supsec:qpemodel}
We also tested if two slow outflows and two thermal components can fully describe the X-ray spectrum. The motivation for two thermal components comes from studies of quasi-periodic eruptions \cite{Miniutti+2019:GSN069,Giustini+2020:RXQPEs}. For this we fit {\it tbabs*zashift(WA1*WA2*(diskbb+diskbb))} to the pn spectrum. Similar to above analysis, the parameters of WA1 and WA2 were constrained to be within the 99\% uncertainty of the best-fit values from RGS. This resulted in a best-fit $\chi^{2}$/dof of 69.8/13 with systematic residuals between 0.55-1 keV (Fig.~\ref{fig:xmmresiduals}k). We also experimented with {\it bbody} for the second thermal component but that did not improve the fit (Fig.~\ref{fig:xmmresiduals}l). From this analysis we concluded that two thermal components plus two slow outflows cannot explain \target's X-ray spectrum.

\subsection{Relativistic reflection is disfavored}
X-ray reflection in the inner regions of the accretion flow can, in principle, produce residuals similar in shape to those seen in Fig.~\ref{fig:xmmresiduals}a. The typical picture in the AGN context is that there is a compact corona that emits a non-thermal (powerlaw) X-ray spectrum. Part of this coronal emission reflects off the inner accretion disk, where general relativistic effects are strong, to produce relativistically broadened emission features (see for example \cite{garcia14} and references therein). This scenario is disfavored for \target due to the lack of an obvious power-law continuum emission from a putative compact X-ray corona around the black hole, required to effectively illuminate the accretion disk.

Alternatively, it has been argued that X-ray reflection can also occur in the absence of a compact corona \cite{Masterson22}. For example, ASASSN-18el is a nuclear outburst lasting for over 3 years \cite{Ricci2021, Masterson22}. Its X-ray spectra during the early phases of the outburst were soft with negligible emission beyond 3 keV, i.e., a weak corona. When fit with a thermal model these spectra result in broad residuals between 0.7-2 keV (see Fig. 2 of \cite{Masterson22} and Fig. 2 of  \cite{Ricci2021}). \cite{Masterson22} have modeled ASASSN-18el's residuals with relativistically broadened reflection. The underlying picture in their model is that the inner accretion disk produces the overall thermal continuum, and because the system is accreting near the Eddington limit, a powerful outflow is launched off the disk. Thus, the total emission reaching us comprises of two components: direct disk/thermal emission and thermal emission reflected off the outflow. Because the outflow is launched from very close to the black hole the reflected emission is subject to relativistic effects. For this scenario, \cite{Masterson22} developed {\it xillverTDE}, a reflection model in which the incident spectrum is a soft thermal continuum instead of a powerlaw/non-thermal emission from a corona. We also considered xillverTDE for \target and as a starting point applied a model similar to ASASSN-18el: {\it tbabs*ztbabs*(zashift(diskbb) + relconv(xillverTDE))}. Here {\it diskbb} and {\it xillverTDE} are the direct and reflected emission, respectively. {\it relconv} accounts for relativistic broadening which is necessary given the broad residuals. Considering all the spectra corresponding to the phases where the residuals are strong, i.e., the so-called Min phases in ODR curve (see Methods section \ref{sisec:odrtiming}) this model results in a combined $\chi^2$/dof of 99.7/72. Here we allowed Fe abundance, inner disk inclination, density and ionization parameter of the material facilitating reflection to be free across all spectra. The redshift of the xillverTDE was fixed at the redshift of the host galaxy. Taken at face value, this reflection model implies an improvement in $\chi^2$ of 11.6 at a expense of 36 additional degrees of freedom when compared with the UFO model (Table \ref{tab: nicerxraydata}). Tying the Fe abundance across all the Min spectra results in a $\chi^2$/dof of 96/81. Tying the density of the reflecting material or the ionization parameter across all the spectra results is a worse fit (reduced $\chi^2$ $>$ 2.5). 

Following our in-depth spectral modeling with {\tt xillverTDE} we disfavor the reflection model for this source for the following reasons: 

\begin{itemize}
    \item In ASASSN-18el \cite{Masterson22} suggested that a clumpy outflow could provide a reflecting medium. However, in the case of \target the best-fit reflection model does not require an outflow, i.e., the redshift of {\tt xillverTDE} component is fixed at the host galaxy value of 0.056 while modeling. Allowing it to be free results in positive (red-shifted) values which would imply material falling into the black hole and is therefore inconsistent with the reflection scenario.

    \item Photons emitted from an accretion disk may be gravitationally bent over the black hole and subsequently illuminate the 'far side' of the accretion flow. It is possible therefore that the disk's thermal emission could itself be the source of a reflection spectral component. However, only the photons emitted very close to the black hole undergo sufficient ray bending to illuminate the far side of the accretion disk, and for a Schwarzschild black hole the fraction of the liberated energy that is then reabsorbed is on the order of 1\% \cite{2000ApJ...528..161A}. For higher black hole spins this fraction increases, but is still limited to $\sim 10\%$, even for the most rapidly rotating ($a = 0.99$) black holes (see Fig. 2b of \cite{2000ApJ...528..161A}). From fitting the \xmm and the \nicer spectra with relativistic reflection, i.e., {\it tbabs*zashift(diskbb + relconv*xillverTDE)}  in {\it XSPEC}, we find that the reflected component dominates the observed flux by a factor of few to up to 10 over the direct thermal component. This is inconsistent with reflection in the  gravitational light bending and disk illumination scenario for a standard disk. To produce such high fluxes in the reflected component would require a fine-tuned disk geometry, which seem contrived.  
    
    \item Lack of a suitable interpretation for the observed variability. For instance, a putative disk precession was already disfavored. Furthermore, there is no clear separation between the best-fit parameters (disk inclination, column, etc) between the min and max spectra.
        
    \item Finally, from a statistical point of view, the reflection model improves the $\chi^2$ only marginally for a large number of additional parameters. Therefore, the reflection model is not statistically superior to the UFO absorption model for this source.
\end{itemize}

%%%%%%%%%%%%%%

\subsection{The outflow is not an artifact of averaging data}
The absorption feature near 0.85 keV is present in the majority of the  phase-resolved \nicer spectra. Because these spectra are obtained by combining data over a certain period (a few days in some cases), it is, in principle, possible that the 0.85 keV could be a artifact of varying spectral properties (blackbody temperature and normalization). However, the presence of the same residuals in a few hours of \xmm snapshot affirms that the feature near 0.85 keV is not an artifact of a varying spectrum.  

\subsection{On the non-detection of an emission line}
We note that the ratios of the data with respect to the continuum may in principle be reminiscent of a P-Cygni profile, where the emission appears red-shifted and the absorption blue-shifted (e.g., \cite{tombesi_ufo, pcygni1}). However, we do not find a statistically significant requirement for an emission component after including the XSTAR absorption table. 

There can be two reasons why the data do not require an additional emission feature associated with a putative P-Cygni profile. One reason is that the variable outflow observed along the line of sight is in the form of a cloud or its physical extent is limited. Therefore, its emission would be expected to be intrinsically weak. 

A second reason may be that the outflow could be geometrically broad and extended, but the emission line arising from a $\sim$0.3c outflow would be so broadened due to Doppler effects (with a width of up to $\sim$1 keV) resulting in a very marginal contribution over the continuum. The narrow energy band (E=0.3-1.1 keV) of the source spectrum and the limited S/N of the observations would make the detection of such a very broad emission feature currently impossible. 
The detection of such a broad and faint emission feature would require X-ray spectrometers with a much higher effective area and energy resolution than currently available, consistent with those proposed for the Athena and the Lynx X-ray observatories.

%++++++++++++++++++++++++++++++++++++++++++++++++++++++++++++++++++++++++++++
% ------------------- COUNT RATE TO LUMINOSITY CURVE -----------------
%++++++++++++++++++++++++++++++++++++++++++++++++++++++++++++++++++++++++++++
\subsection{Computing the observed luminosity vs time curve}
The individual \nicer GTIs do not have enough counts to compute the observed luminosity and other parameters of the outflow. Therefore, we use the mean count rates and the observed flux measurements from time-resolved spectra from Table \ref{tab: nicerxraydata} and section \ref{sisec:timeresolspecs}. We compute the 0.3-1.1 keV observed luminosity of a GTI by scaling the 0.3-1.1 keV count rate to the value of luminosity in a given epoch. The resulting curve is shown in panel (a) of Fig.~\ref{fig:fig1}.

%++++++++++++++++++++++++++++++++++++++++++++++++++++++++++++++++++++++++++++
% ------------------- X-RAY TEMPERATURE -----------------
%++++++++++++++++++++++++++++++++++++++++++++++++++++++++++++++++++++++++++++
\section{The temperature of the \target blackbody continuum emission}\label{sisec:temp}
The X-ray spectrum of \target is very well modeled with a single blackbody continuum component with a temperature of $T \simeq 90$ eV. This is consistent with what is usually found for X-ray spectra of TDEs (See the right panel of Figure 2 of \cite{suvi2021}). This phenomenological modeling is required in order to characterize the continuum shape and normalization, but we do not derive physical conclusions from it. The phenomenological blackbody emission is only used as the input ionizing continuum in the XSTAR photoionization code to calculate the absorption tables (see Methods section \ref{sisec:xstar} for more details on table models).   

From a phenomenological point of view, we note that an hybrid accretion disk solution combining an ADAF-type hot flow and a standard thin disk is often suggested as a description of the emission for low-luminosity AGN (e.g., \cite{2014MNRAS.438.2804N}). Depending on the thin disk truncation radius and the temperature, geometry, and extent of the inner ADAF, it is plausible that blackbody optical/UV diskphotons are up-scattered by the hot ADAF gas, similarly to the putative X-ray corona in more luminous AGN. In the case of \target, being the disk quite limited in spatial extent, the resultant spectrum would most likely be approximated with a single blackbody with increased temperature. 

Moreover, classical estimates based on steady state accretion theory do not apply to a disk system undergoing a large amplitude outburst like \target, where the disk is not in inflow equilibrium. A disk system out of the steady state can have a higher surface density in its innermost regions, leading to a higher temperature at the inner edge of the disk. This is particularly true for TDEs around higher mass black hole's where the incoming star's tidal radius approaches the black hole's ISCO.

To further demonstrate that the blackbody temperature inferred for \target is consistent with a TDE-disk system we simulate mock 0.3-1.1 keV X-ray spectra for time-dependent and fully relativistic accretion disk systems, using the techniques described in \cite{MummeryBalbus2020, Mummery2021}. These mock spectra were produced using full photon ray-tracing calculations, and therefore include all leading order relativistic effects. By fitting these mock X-ray spectra with a phenomenological blackbody profile, a temperature of the spectrum can be extracted. We compute the fitted temperature of the X-ray spectrum, produced for a black hole mass $M = 10^7 M_\odot$ and disk mass $M_d = 0.5 M_\odot$, for a range of black hole spins and disk-observer inclination angles. Note that this fitted temperature will differ from the physical temperature of the inner disk primarily due to the effects of Doppler and gravitational shifts, and the colour correction of the disk emission  (as discussed in \cite{Mummery2021}). We only consider inclination angles consistent with the obscuring UFO scenario ($\theta_{\rm inc} < 22^\circ$). Each X-ray spectrum was produced at a time corresponding to the peak of the disk’s bolometric light curve. We find that a temperature of 85 eV is within an acceptable parameter space.

\section{A single clumpy outflow is disfavored}\label{sisec:singleclump}
A steady and clumpy outflow launched at the onset of the X-ray outburst is disfavored due to the presence of the quasi-periodicity. This is because in order to produce the observed modulations in the ODR curve the clumps would need to be arranged in a preferred manner around the central black hole. This is highly unlikely for any intrinsically random distribution of clumps. The timing analysis in section \ref{sisec:odrtiming} already computes the odds of this happening to be less than 1 in 50,000.

%++++++++++++++++++++++++++++++++++++++++++++++++++++++++++++++++++++++++++++
% -------- OUTFLOW IS PRESENT EVEN AFTER THE OUTBURST ENDED -----------------
%++++++++++++++++++++++++++++++++++++++++++++++++++++++++++++++++++++++++++++
\section{The outflow is present even at 200 times lower X-ray luminosity}\label{sisec:lateufo}
To test the strength of the outflow as a function of observed luminosity we also obtained \xmm exposures after the initial outburst ended. While the first few \xmm exposures were too short, i.e., low signal-to-noise, we detect the same UFO signature in XMM\#3. The C-stat/dof without the UFO was found to be 125.8/89. Including the outflow improved the C-stat/dof to 97.9/86, i.e.,  $\Delta$C-stat of 27.9 for 3 additional dofs which corresponds to a confidence level of 99.99\%. The unfolded spectrum along with the residuals is shown in Fig.~\ref{fig:xmmlateufo}.

\section{Extracting Composite Spectra from \nicer data}\label{sisec:nicercompspecs}
To study the spectral properties during the epochs of ODR maxima and minima we combined exposures and obtained composite energy spectra. While combining the data we remove the detectors marked as ``hot'' based on the 0.0-0.2 keV count rate as described above. The main steps for extracting time-resolved \nicer energy spectra are as follows:
\begin{enumerate}
    \item First, we extract the combined ufa and cl event files using the start and the end times of all GTIs within a given epoch.
    \item Then, we use the 3c50 model on these combined ufa and cl files to estimate the average background and source spectra. All the detectors marked as ``hot'' at least once in any of the individual GTIs are excluded.
    \item using the tools {\tt nicerarf} and {\tt nicerrmf} we extract an arf and rmf for each epoch.
    \item Then, we group the spectra using the optimal binning criterion described by \cite{optmin} also ensuring that each bin have at least 25 counts.
    
\end{enumerate}

\section{\nicer time-resolved energy spectral analysis shows the same strong--weak outflow oscillatory pattern}\label{sisec:timeresolspecs}

We modelled the energy spectra of the individual maxima and minima in the ODR curve using the ionized outflow model. The results are shown in Table \ref{tab: nicerxraydata}. It is evident that both the absorbing column and the ionizing fraction are more than an order of magnitude higher during the epochs of ODR minima than during the maxima (Fig.~\ref{fig:specplots}). 

Some of the spectra, during some maxima, did not require an outflow component. In these spectra the $\chi^{2}$/dof was close to 1 with a thermal component alone. These are marked by shaded orange regions in Fig.~\ref{fig:specplots}. 

\section{Outflow energetics}\label{sec:energetics}
As conventionally done in the literature, we conservatively estimate the outflow launching radius to be the distance at which the observed velocity is equivalent to the escape velocity from the SMBH (e.g. \cite{2012ApJ...753...75C, 2013MNRAS.430.1102T, 2015MNRAS.451.4169G, 2021ApJ...920...24C}): $r = 2 G M_{\bullet}/v_{out}^2$. This can be written also in units of the gravitational radius $r_g = GM_{\bullet}/c^2$ as: $r = 2 (v_{out}/c)^{-2} r_g$. Considering a black hole mass of log$(M_{\bullet}/M_{\odot}) = 7.4$ we estimate a gravitational radius of $r_g = 3.7\times 10^{12}$ cm. The mass outflow rate can be estimated using the equation: $\dot{M}_{out} = 4\pi C_f r N_H \mu m_p v_{out}$. Where $N_H$ is the column density, $\mu = 1.4$ is the mean atomic mass per proton, $m_p$ is the proton mass, and $C_f$ is the global covering fraction typically assumed to be 0.5 for AGN disk outflows (e.g., \cite{2013MNRAS.430.1102T, 2015MNRAS.451.4169G, 2012ApJ...753...75C}). Then, the kinetic power of the outflow can be estimated using the formula: $\dot{E}_{out} = 1/2 \dot{M}_{out} v_{out}^2$. We also calculated the ratio between the outflow kinetic power and the unabsorbed luminosity in the 0.3--1.1 keV band, $\dot{E}_{out}/L$. From this, we estimated also the ratio between the mass outflow rate and the mass accretion rate $\dot{M}_{out}/\dot{M}_{acc}$, considering $\dot{M}_{acc} = L/\eta c^2$ and a typical radiative efficiency $\eta = 0.1$. 

Using the best-fit parameters reported in Table \ref{tab:xmmxraydata} and Table \ref{tab: nicerxraydata}, we show the estimates for the outflows detected in the time-resolved NICER analysis and in the XMM-Newton spectra in Table~\ref{tab:energetics}. We note that we are not reporting error bars in our calculations, as they are considered as order-of-magnitude estimates. However, the important point here is not the absolute value of each parameter, but the difference between the average parameters in the Min and Max phases of the ODR, which is independent on the model assumptions and overall uncertainties. 
For the NICER Min phases, we derive the following average quantities: launching radius $r \simeq 18 r_g$, mass outflow rate $\dot{M}_{out} \simeq 0.002 M_{\odot}/yr$, kinetic power  $\dot{E}_{out} \simeq 6\times 10^{42}$ erg/s, a ratio $\dot{E}_{out}/L \simeq 10$\%, and a ratio $\dot{M}_{out}/\dot{M}_{acc} \simeq 18$\%. For the NICER Max phases, we derive the following average quantities: launching radius $r \simeq 18 r_g$, mass outflow rate $\dot{M}_{out} \simeq 0.0003 M_{\odot}/yr$, kinetic power  $\dot{E}_{out} \simeq 0.9\times 10^{42}$ erg/s, a ratio $\dot{E}_{out}/L \simeq 1$\%, and a ratio $\dot{M}_{out}/\dot{M}_{acc} \simeq 2$\%. Comparing the estimates in the NICER Min and Max phases we see that the outflow launching radius is consistent, but the overall mass flux and energetics are one order of magnitude lower for the latter. 

From these estimates, we can infer some information regarding the potential impact of the outflow on the accretion flow and on its host galaxy feedback. The outflow is launched from the disk along the boundary of the accretion flow and the much less dense, but highly magnetized funnel. Therefore, simulations show that it will not significantly interfere with the accretion flow. This is supported by the estimate of the instantaneous mass flux, reported in Table~\ref{tab:energetics}, which is limited to about 20\% of the mass accretion rate. On the other hand, given its mildly-relativistic velocity, the outflow is found to have a power reaching up to about 10\% of the peak luminosity. This relatively high power suggests that it could temporarily drive feedback into the host galaxy \cite{2005Natur.433..604D}.

Regarding the XMM-Newton spectra, they are not exactly placed in the Min and Max phases, so their values are not directly comparable to the NICER time-resolved analysis. From Table~\ref{tab:energetics} we see that the outflow was statistically detected in two out of four XMM-Newton phases. In XMM1 the outflow has values comparable to the NICER Max phases. Instead, for XMM3, which was performed much later, during the low-luminosity state of the source, the values of the outflow mass-flux and energetics seem comparable to the NICER Min phases. However, we note that the high values of the ratios $\dot{E}_{out}/L$ and $\dot{M}_{out}/\dot{M}_{acc}$ in XMM3 may indicate that the outflow could likely be magnetically accelerated and that the disk radiative efficiency may be lower than the typical value assumed for the high-luminosity state of the source.

%++++++++++++++++++++++++++++++++++++++++++++++++++++++++++++++++++++++++++++
% ------------------- QP-OUTS ----------------------
%++++++++++++++++++++++++++++++++++++++++++++++++++++++++++++++++++++++++++++

\section{Outflow Deficit Ratio Timing Analysis}\label{sisec:odrtiming}
It is evident from \nicer's soft X-ray light curve that the source underwent a major outburst increasing by a factor of $>$600 and thereafter decreasing by a factor of roughly 200. Also, near the peak, i.e., between MJD 59260 and 59370, the source is variable. With an unprecedented high-cadence soft X-ray coverage \nicer data provides a unique opportunity to study the co-evolution, if any, of the UFO with the accretion (thermal continuum). Therefore, to track the evolution of the outflow  with respect to thermal continuum, we extracted a hardness ratio defined as the ratio of the countrate in the outflow band, 0.75-1.0 keV, over the continuum band, 0.3-0.55 keV and refer to it as the outflow deficit ratio or ODR. Surprisingly, the ODR vs time plot shows repeated flares which appear to  recur roughly once every 8.5 days (see Fig. \ref{fig:fig2}a). {\it As the ODR is inversely proportional to the strength of the outflow, a lower value would imply a stronger outflow and vice versa. } To verify that the LSP signal near 8.5 days in Fig.~\ref{fig:fig2}b is robust against the choice of the period-finding algorithm, we also implemented the phase dispersion minimization algorithm \cite{1978ApJ...224..953S} and the weighted wavelet Z-transform \cite{1996AJ....112.1709F,Pyleoclim}, {\bf see Figs.~\ref{fig_pdm} and \ref{fig_wwz}, respectively}. They both found the signal at the same frequency as the LSP, confirming the signal's  robustness against algorithm selection. For all further timing analysis we use the LSP throughout the rest of the paper. To test the statistical significance of the peak near 8.5 days in the LSP we first establish that the power values in the LSP are consistent with white noise.

\subsection{Values in the LSP are consistent with white noise}\label{sisec:noisenature}
To test for the presence of a quasi-periodicity in the ODR curve we computed its Lomb Scargle Periodogram (LSP; \cite{scargle, lspnorm}). The LSP was sampled at $N_{i}$ \emph{independent} frequencies as per Eq. 13 of \cite{lspnorm}. Consistent with the ODR curve, the highest peak in the LSP is near 8.5 d (see panel (b) of Fig.~\ref{fig:fig2}). To assess the global statistical significance (false alarm probability) of this LSP excess near 8.5 days we perform more analyses. We first turn our focus to understand the nature of the underlying noise in the LSP because an accurate characterization of the noise in the LSP is of utmost importance for estimating the statistical significance.

The ODR tracks the outflow's relative strength compared to the thermal continuum. By construction, because we are dividing by 0.3-0.55 keV flux, i.e., the band dominated by accretion-driven fluctuations, we expect to suppress any red noise present in the continuum. By eye, the ODR values between the flares appears to be roughly constant. To verify this more rigorously, we performed additional statistical tests. 

First, we normalize the LSP to have a mean value of 1 by dividing the LSP with the mean of all power values excluding bins near 8.5 d. We then compute the empirical distribution (EDF) and the probability density functions (PDF) of these LSP power values and compare them with the expected 1-$e^{-z}$ distribution expected for LSP if the power values were derived from white noise \cite{scargle}. Here, $z$ is a variable representing LSP powers. The EDF and PDF are shown in panels (a) and (b) of Fig.~\ref{fig:whitetests} are qualitatively consistent with the expected exponential distribution. 

Next, we investigate the nature of the distribution of LSP powers quantitatively. We performed the Kolmogorov-Smirnov (K-S) and the Anderson-Darling goodness-of-fit tests under the null hypothesis that the LSP powers are white, i.e., their values between $\approx$1.5 and 100 d, except for bins near 8.5 d, are exponentially distributed. The underlying principle behind these statistics is that they measure the maximum deviation between the EDF of the data and that of a comparison distribution. Therefore, the better the distribution fits the data, the smaller these statistic values will be. 

We computed the K-S statistic using the EDF of LSP powers and the expected 1-$e^{-z}$ distribution for white noise. To evaluate whether this value can be used to reject or not reject the null hypothesis, we calculated the distribution of K-S statistic values of EDFs drawn from the expected exponential distribution as follows. 
\begin{enumerate}
    \item First, we randomly draw 167 values uniformly distributed between 0 and 1 (sim$_{arr}$). Here, 167 refers to the total number of LSP continuum values between 1.5 d and 100 d excluding bins near 8.5 d.
    \item Then we evaluate the expression -Log$_{10}$(1-sim$_{arr}$) to give a simulated set of values that follow the expected 1-$e^{-z}$ distribution. Combined with the above step this procedure is sometimes referred to as the inverse sampling technique.
    \item We then compute the EDF of this simulated set of values drawn from 1-$e^{-z}$ distribution.
    \item Finally, we estimate the K-S statistic of this simulated set of values using its EDF.
\end{enumerate}
The above steps are repeated 100,000 times to get a distribution of the K-S test statistic values for a given sample size of 167. This is shown as a orange histogram in Fig.~\ref{fig:whitetests}c. \target's observed K-S test statistic (dashed vertical red line), which is a measure of maximum deviation between the observed EDF and the theoretical Cumulative Distribution Function (CDF), is within 1$\sigma$ deviation of the distribution. This indicates that the null hypothesis cannot be rejected even at the 90\% confidence level and suggests that LSP powers in the continuum are consistent with the expected exponential distribution, i.e., the LSP is consistent with being white between 1.5 d and 100 d.

To ensure the above conclusion is not dependent on the choice of the statistic used we also computed the Anderson-Darling statistic. Similar to above, we computed its distribution using bootstrap simulations (Fig.~\ref{fig:whitetests}d). Again, it is evident that the statistic computed from \target's observed LSP (vertical dashed red line) is consistent with the expected exponential  distribution. 

% To ensure that the CDFs computed above are not heavily biased by the high frequencies, i.e., $<$ 5 days, we repeated the above tests by only considering the LSP values that  corresponds to frequencies immediately surrounding the peak near 8.5 d, i.e., LSP powers between 5 and 20 days. This did not change the results. 

Based on the above tests we concluded that the ODR LSP values are consistent with white noise and proceeded to measure the global statistical significance based on this noise model.

\subsection{Monte Carlo Simulations to Estimate Global Statistical Significance}\label{supsec:whitemcsims}
The LSP power levels corresponding to the global 3 and 4$\sigma$ values can be estimated using Eq. 18 of \cite{scargle}. These correspond to 11.1 and 14.8, respectively. The highest bin near 8.5 d is above the 4$\sigma$ value with an adjacent/second highest bin crossing the 3$\sigma$ value. This suggests that the quasi-periodicity is statistically significant at greater than at least the 4$\sigma$ level.

However, because the signal we are trying to test is broad, i.e., over at least two frequency bins near 8.5 d, the standard approach of estimating significance based on just the highest bin will be inadequate. Because such an estimate will not include the contribution from multiple  frequency bins it will fail to capture the true significance estimate. By true significance we mean an estimate that accounts for the fact that the signal is distributed in multiple frequency bins. Therefore, we devise a methodology that can account for multiple frequency bins. This approach is similar to \cite{pashcow} with the additional complexity of irregular sampling. The mains steps are as follows:

\begin{enumerate}
    \item After establishing that the ODR curve's variability is white, i.e., frequency-independent noise, we simulate a uniformly sampled white noise light curve using the algorithm of \cite{timmerko}. The time resolution and temporal baseline of this light curve is 10-second and 150 d, respectively. 
    \item Next, we sample this light curve exactly as the window function of the real data. 
    \item We then extract an LSP of this data and identify the frequency bin with the highest power value.
    \item The LSP is normalized by the mean of all power values excluding those near the period corresponding to the maximum value in the LSP. 
    \item An array of sum of two neighboring LSP powers is generated from the normalized LSP from the step above. The maximum value of this array is saved.
\end{enumerate}

The above steps were repeated 500,000 times to get an array of 500,000 maximum LSP sums. From these measurements we computed the probability to exceed a certain LSP sum value, i.e., 1-CDF. This is shown in panel (c) of Fig.~\ref{fig:fig2}. The 3 and 4$\sigma$ confidence levels are indicated. The peak in the LSP near 8.5 d found in \nicer data of \target is statistically significant at $\approx$2$\times$10$^{-5}$ level which translates to $>$4.2$\sigma$ equivalent for a normal distribution.

For completeness, we also extracted the energy resolved light curves in the 0.3-0.55 keV and the 0.75-1.0 keV band. These are shown in Fig.~\ref{fig:lcs}.

Finally, we also tested the robustness of the peak in the LSP by changing the bandpass boundaries used in the definition of ODR by $\sim$20\%. A significant peak near 8.5 days was present in all the tested cases.

%++++++++++++++++++++++++++++++++++++++++++++++++++++++++++++++++++++++++++++
% ---------------------- XSTAR DESCRIPTION ------------
%++++++++++++++++++++++++++++++++++++++++++++++++++++++++++++++++++++++++++++
\section{XSTAR energy table models}\label{sisec:xstar}
We calculated physically-motivated XSPEC photoionization table models using the XSTAR code v. 2.39 \cite{2001ApJS..133..221K}. We produced a grid of photoionization models varying the column density and ionization parameter in a wide range of values of $N_H = 10^{19}$--$10^{23}$ cm$^{-2}$ and log$\xi ($erg s$^{-1}$ cm) $=$ 0--4, respectively. We considered an input spectral energy distribution consistent with the data, that is a blackbody continuum with a temperature T = 10$^6$ K (E $=$0.09 keV) and a mean unabsorbed ionizing luminosity in the 1--1000 Ryd band (1 Ryd $=$ 13.6 eV) of $1.5\times 10^{44}$ erg~s$^{-1}$, which is consistent with the observed narrow range between $1.2-2 \times 10^{44}$ erg~s$^{-1}$. We considered a constant density shell of $10^{10}$~cm$^{-3}$, although the actual value of the density is not strictly important for a geometrically-thin shell because the code would simply scale the distance in order to obtain the same ionization parameter (e.g., \cite{2011ApJ...742...44T}). All abundances were fixed to solar values. When modeled with an inverted Gaussian at E $\simeq$ 800 eV, the width of the absorption feature is very large, $\sigma_E \simeq70$ eV ($\sigma_v \simeq 25,000$ \kms), so we tested XSTAR grids with increasing velocity broadening from 100 \kms upwards, finding that the maximum velocity broadening of 30,000 \kms provides the best fit to the data. Such a high velocity broadening is not physically interpreted as due to turbulence but it is most likely indicating a rotation of the outflow launched close to the black hole (e.g., \cite{2019ApJ...885L..38F}). Indeed, a lower limit on the velocity broadening of $>$5,000 \kms is derived considering the rotational velocity at a distance of $<$6,000 $r_g$, given by the light-crossing-time of the 8 days modulation of the outflow.

A search for best-fit solutions was performed considering a wide range of redshifts for the XSTAR table, ranging from $z = -0.4$ to $z = 0.1$, in order to investigate the existence of rest-frame to high-velocity outflows. Importantly, the XSTAR tables self-consistently take into account all resonant lines and edges for a wide range of ionic species, from H up to Ni \cite{2001ApJS..133..221K}. The smoothness and lack of sharp edges clearly point to an interpretation of the absorption as due to a broadened and blue-shifted OVIII Ly$\alpha$ transition, with a rest-frame energy of 0.654 keV. We note that our photoionization modeling is consistent to the one adopted by \cite{14liufo} for the broad absorption feature in the \xmm spectrum of the tidal disruption event ASASSN-14li.

In order to model possible low outflow velocity X-ray warm absorber components in the high-energy resolution RGS data, we also calculated a separate XSTAR absorption table with a typical velocity broadening for warm absorbers of 100 \kms \cite{wa}.

The assumption of a single temperature black body is well justified as the difference in black body temperature is found to be within 15\% with respect to the average among the different spectra (Table \ref{tab: nicerxraydata}). We quantitatively tested for a possible dependence of the estimated parameters of the outflow on the black body temperature performing a fit of the XMM\#1 and \nicer's time-resolved spectra with a $kT = 0.12$ keV XSTAR table, corresponding to a extremely high temperature increase of 30 per-cent. We find that the outflow is always required and the best-fit values of the parameters are always consistent within the $\simeq 2 \sigma$ level independently of the considered black body temperature. 

\section{\target's optical/UV evolution}
After subtracting the host flux and correcting for foreground Galactic extinction we fit the \swift UVOT photometry as a blackbody using MCMC and forward-modeling methods to estimate the bolometric luminosity, temperature, and effective radius evolution of \target. We obtained the \swift UVOT filter response functions from the Spanish Virtual Observatory Filter Profile Service. This approach is similar to the methods of several previous studies on TDEs and ANTs (e.g., \cite{2022ApJ...933..196H, jason, hinkle22}). After obtaining the blackbody luminosity evolution, we estimated a bolometric light curve by scaling the ASAS-SN $g$-band light curve to match the bolometric luminosity evolution from the blackbody fits. Where there was no Swift data we assumed a constant scaling with time (i.e. a flat temperature evolution; see Fig.~\ref{fig:uvotevol}).

\clearpage
\setcounter{page}{1}
\renewcommand{\thetable}{S\arabic{table}}
\renewcommand{\thefigure}{S\arabic{figure}}
\setcounter{section}{0}
\setcounter{figure}{0}
\setcounter{table}{0}

\section*{{\Huge Supplementary Material.}}
Summary of Alternate models \\
Perturber-induced outflow scenario \\
GRMHD simulations of the perturbed accretion flow \\
On the nature of the orbiting perturber \\
Figures S1-S16 \\
Tables S1-S7\\
Supplementary Movie 1 \\
\clearpage
\section{Summary of alternative models}\label{sisec:othermodels}

We considered several models to interpret the above quasi-periodic variations. First, we consider a scenario consisting of a precessing accretion disk. This can, in principle, produce a quasi-periodicity for a certain black hole spin, outer radius of the precessing accretion disk, and a disk surface-density profile \cite{stone_loeb_2012}. However, a precessing disk should also result in a strong modulation of the continuum flux. In order to test this possibility, we calculated the Lomb-Scargle Periodogram of the continuum emission in the 0.3–0.55 keV energy band. As shown in Fig.~\ref{fig:lcs}b it is dominated by red noise and does not show statistically significant peaks, contrary to the clear peak at $\simeq$8.5 days in the outflow band (0.75-1.00 keV) shown in Fig.~\ref{fig:fig2}. We note that this result is not affected by a limited signal-to-noise, because the countrate in the continuum energy band is actually much higher than in the outflow band. The fact that there is no significant modulation in the continuum band strongly disfavor an interpretation as due to accretion disk precession. 

There are also other lines of argument that disfavor a precessing disk with a persistent outflow. As can be seen from Fig.~ \ref{fig:specplots}, going from the maxima to the minima, the outflow column density and the ionization parameter increase by about an order of magnitude, while the outflow velocity remains rather stable. This observed empirical pattern is inconsistent with a change of line of sight through any disk outflow model, which disfavors a disk outflow precession interpretation. Indeed, this statement is supported by three fundamental physical reasons: (i) if the outflow is radial, as expected in purely radiation-driven scenarios (e.g., \cite{kingpoundsufos}), we would expect no variability in the outflow parameters following a change in the line of sight; (ii) considering a conical geometry, as expected for line-driven outflows \cite{nomura16, elvis00}, following an increase in column density, we would expect a decrease in ionization parameter and an increase in outflow velocity, because the line of sight would intercept the main outflow streamline; (iii) in the MHD-driven case (e.g., \cite{Keigo10,keigo14}), the disk outflow has a specific stratified structure depending on the radial distance and polar angle, which predicts that an increase in column density, when the line of sight is closer to the equatorial region, will be followed by a significant decrease of the ionization parameter and outflow velocity. 

Another scenario one can envision is that of clumpy outflow with the individual clumps intervening our line of sight roughly once every 8.5 days. However, the chance probability of encountering such a uniformly clumpy configuration is $<$1 in a 50,000 as shown by Monte Carlo simulations above (see Methods section \ref{sisec:singleclump} and Fig.~\ref{fig:fig2}c) and hence unlikely to be the origin of the observed quasi-periodicity.

\target's QPOut properties appear distinct from the phenomenon of quasi-periodic eruptions (QPEs) found in a small sample of four systems \cite{Miniutti+2019:GSN069, Giustini+2020:RXQPEs, Arcodia+2021:eROSQPEs}. Firstly, QPEs are intense soft X-ray bursts with amplitudes in the range of 10-200 \cite{Arcodia+2021:eROSQPEs, Miniutti+2019:GSN069}. Such large amplitude changes in the soft X-ray flux are not seen here (see Fig.~\ref{fig:lcs}). Secondly, QPE spectra are thermal with no reported outflows similar to those seen here. A recent study found a warm absorber in the QPE source GSN~069 \cite{gsn069:QPEdisapp}. However, they concluded that this warm absorber was stable with no discernible changes over an extended period between 2010 and 2021 \cite{gsn069:QPEdisapp} (see Methods  section \ref{supsec:qpemodel}). 

The UFO is also detected in an \xmm observation taken around MJD 59554, roughly a year after the start of \nicer monitoring (see Fig.~\ref{fig:xmmlateufo} and Table \ref{tab:xmmxraydata}). The fact that the outflow is present even after the X-ray flux decreased by more than a factor of 200 at $\sim$0.01\% Eddington limit, suggests that it is unlikely to be driven by radiation pressure. There are two possibilities for the origin of the outflow in the low-luminosity state: (i) if the perturber-induced scenario described below is valid, the object is still expected to move around the SMBH after the decline of the accretion rate due to depletion of the inner TDE accretion disc. Thus, the outflows are still being launched from the outer ADAF flow and their presence would be revealed via absorption of the X-rays from inner parts of the disk. Because we have only one observation after the outburst has ended, we cannot confirm the quasi-periodicity of the outflow in the low accretion rate regime. ii) The alternative is that this could be the persistent magnetized outflow seen in our simulations of ADAF accretion flows oriented face-on. 

If you consider that the quasi-periodicity of the outflow during the outburst is due to the radiation-driven instability, this quasi-periodic outflow would have a different physical launching mechanism than the outflow seen after the outburst in the low-luminosity state. During the outburst, the outflow would originate from a small dense accretion disk with a relatively high accretion rate, while after the outburst, the outflow would originate from a magnetized, dilute, and much hotter ADAF accretion flow. Therefore, we would expect different parameters of the outflow in the two regimes, including the column density and the velocity. Because the measured properties of the outflow after the outburst are within the error bars more or less similar to the outflow during the outburst (taking into account that lower flux in the low-luminosity state means higher uncertainties and that we do not know, in which phase (ODR maxima or minima) the low-state measurement was taken), this is an argument against the radiation-driven instability origin for QPOuts unless an instability can operate over several orders of magnitude change in luminosity. Moreover, to achieve such a short period with the radiation pressure instability, the properties of the disc, including its size, would have to be quite fine-tuned, with a very small radial extent of the disk\cite{2023A&A...672A..19S}, which is supposed to radiate in the soft X-rays. In that case, however, the changes of the disk in the instability cycle would manifest as quasi-periodic variability of the inflow band, which is not seen for \target. The radiation-driven origin of the periodicity is thus unlikely.

High-resolution 3D GRMHD simulations of magnetically arrested disks (MADs) have suggested the possibility of variable outflows of heated plasma due to magnetic reconnection near the event horizon \cite{bart}. However, the long-term evolution of such MAD outflows, i.e., beyond a few cycles, is unclear at present due to computationally-limited integration time. Nevertheless, episodic outflows in MAD cycles based on longer time integration (of the order of tens of cycles) at lower resolutions, show a stochastic distribution of eruptions rather than a periodicity (see the blue curve in Figure 1 of \cite{2020MNRAS.497.4999D}). Furthermore, MADs  are estimated to produce pronounced variability in the continuum which is not seen in \target (Fig.~\ref{fig:lcs}).

\begin{sloppypar}

Concerning the Changing-look classification, typical traces of the changing-look phenomenon are not found during the observational campaign, i.e. the disappearance/reappearance of broad optical lines associated with the AGN type shift (e.g., from type 1 to type 2, and vice versa). The UV/optical continuum variability is rather typical of TDE power-law decay. However, we cannot exclude that the source indeed transitioned from a certain AGN optical type to a type 1 AGN before and after the outburst. During the observational campaign, after the outburst, the optical spectra are rather constant and they are not changing on timescales of a week, as instead clearly seen for the X-ray absorption, indicating that the AGN is not undergoing recurrent optical changing-look events. In any case, the possibility that the source went through an optical changing-look event before and after the outburst does not change the conclusions derived from its X-ray spectrum. On the other hand, the X-ray monitoring shows a quasi-periodic variability in its absorbed spectrum which could be classified as a changing-look behavior, in the sense that the source oscillates between highly- and lowly-absorbed states with a timescale of 8.5 days. The physical origin of this behavior would still be consistent with our theoretical interpretation of an orbiting object repeatedly perturbing the SMBH accretion disk.

\end{sloppypar}

\section{Perturber-induced outflow scenario}\label{msec:modeldiscussion}
We investigate whether \target’s observed outflow quasi-periodicity of P$_{\rm obs}$ $\approx$ 8.5 days can be induced by repetitive perturber—SMBH disk interactions \cite{Sukova+2021:QPOutSims}. The gravitationally bound perturber can either be a stellar object with an outflow (mass-losing star or a pulsar) or a black hole. This is motivated by the fact that for both early- and late-type galaxies with stellar mass approximately equal to that of \target’s host (log$(M_{\text{host}}/M_{\odot})$ = 10.13$^{+0.02}_{-0.01}$, see Section \ref{sec:host}), the occurrence rate of nuclear star clusters (NSCs) is high, between 60-80\% \cite{2020A&ARv..28....4N}. NSCs can be associated with stellar power-law density cusps around the SMBH \cite{2020A&A...641A.102S,2022NatAs...6.1008Z} which can result in a fraction of NSC objects (stars/compact objects) on tightly bound orbits around SMBHs in a manner similar to Sgr A*'s S stars \cite{2020ApJ...899...50P,2022ApJ...933...49P}.

Apart from in-situ star-formation, another way to build-up NSCs is the infall of massive star-forming or globular clusters which can also host intermediate-mass black holes (IMBHs; mass range of 10$^{2-5}$ M$_{\odot}$). These black holes can inspiral towards the central SMBH after their cluster gets disrupted by tides \cite{2002ApJ...576..899P,2004cbhg.symp..138R}, and it is predicted to result in SMBH-IMBH comoving merger rates of $\sim 10^{-5}$ - $3\times 10^{-4}$ Gpc$^{-3}$ yr$^{-1}$ in the local Universe \cite{smbh_imbhrates}. IMBHs with masses $\sim$10$^{3-4}$ M$_{\odot}$ can also be retained in NSCs as a result of repeated stellar mass black hole--stellar-mass black hole or stellar-mass black hole--star mergers \cite{2022ApJ...929L..22R,2022ApJ...927..231F}. 

A perturber (star or a compact object) is associated with its influence radius. This size is typically larger than the effective stellar radius in the case of a star or a compact remnant with a hard surface (neutron star or a white dwarf), and is larger than the event horizon in the case of a black hole. Theoretical work by \cite{Sukova+2021:QPOutSims} has shown that if the influence radius is large enough, the perturber can modulate both the inflow and the outflow rates with a characteristic periodicity. Under the ``unified model'' of AGN \cite{Netzer15}, the presence of broad optical emission lines suggests that we are viewing \target close to the SMBH’s rotational axis. In the repetitive perturber—disk interactions scenario two outflows are ejected per orbit \cite{Sukova+2021:QPOutSims}. However, because of the orientation only one outflow per orbit is expected to lead to an observable absorption event along our light of sight. The other outflow will not cause an absorption event, will be Doppler de-boosted, and will likely be obscured by the accretion disk. Therefore, in the following analysis, we set the rest-frame perturber orbital period equal to the QPOut period, i.e., P$_{\rm orb}$ = P$_{\rm obs}$/(1+z) $\approx$ 8.5/(1+0.056) $\simeq$ 8.05 days. Consistent with the estimates in section \ref{subsec:bhmass}, we consider a SMBH mass range of 10$^{7-8}$ M$_{\odot}$. 

Given this SMBH mass range and $P_{\rm orb}$, the characteristic radius of the perturber orbit is $r_{\rm per}=[P_{\rm orb}c^3/(2\pi G M_{\bullet})]^{2/3} r_{\rm g} \sim 37\,(10^8\,M_{\odot})-172\,(10^7\,M_{\odot})\,r_{\rm g}$ (in gravitational radii or $M$ assuming a zero SMBH spin; we use $r_{\rm g}$ or $M$ units interchangeably in derivations or plots). For the intermediate value of the supermassive black hole mass of $M_{\bullet}=10^{7.4}\,M_{\odot}$, which we adopt for most GRMHD simulations, the orbital distance of the perturber changes only slightly for the theoretically possible maximum range of SMBH spins -- from $r_{\rm per}=92.90\,r_{\rm g}$ for the maximum prograde spin, through $r_{\rm per}=92.97\,r_{\rm g}$ for a zero spin, up to $r_{\rm per}=93.04\,r_{\rm g}$ for the maximum retrograde spin. For the same SMBH mass range, this characteristic orbital radius does not exclude an orbiting, perturbing object, since the tidal disruption radius for a Solar-type star is,
\begin{equation}
    r_{\rm t}\simeq R_{\star} \left(\frac{M_{\bullet}}{m_{\rm per}} \right)^{1/3}\sim 4.37 \left(\frac{R_{\star}}{1\,R_{\odot}}\right)\left(\frac{M_{\bullet}}{10^{7.4}\,M_{\odot}}\right)^{\frac{1}{3}}\left(\frac{m_{\rm per}}{1\,M_{\odot}}\right)^{-\frac{1}{3}}\,r_{\rm g}\,,
    \label{eq_tidal_radius}
\end{equation}
where $R_{\star}$ is the stellar radius and $m_{\rm per}$ is the stellar (perturber) mass.
The tidal disruption radius given by Eq.~\eqref{eq_tidal_radius} shows that essentially only red giants with stellar radii of a few$\times$10$\,R_{\odot}$ would be partially disrupted. If one calculates the ratio of the tidal disruption radius to the stellar orbital distance, it is of the order of unity and less for the stellar radius of
\begin{equation}
    R_{\star}\lesssim 16.9\left(\frac{P_{\rm orb}}{8.05\,\text{d}} \right)^{2/3}\left(\frac{m_{\rm per}}{1\,M_{\odot}} \right)^{1/3}\,R_{\odot}\,,
    \label{eq_star_radius}
\end{equation}
regardless of the SMBH mass. In other words, stars that satisfy Eq. \eqref{eq_star_radius} are always beyond the tidal radius and are not disrupted. On the other hand, the GRMHD study of \cite{Sukova+2021:QPOutSims} indicates that the perturbers with influence radii
\begin{equation}
\mathcal{R}\gtrsim 1\,r_{\rm g}=53.2\,\left(\frac{M_{\bullet}}{10^{7.4}\,M_{\odot}} \right)R_{\odot}
\end{equation}
are necessary to induce significant perturbations (see SI sections \ref{sisec:grmhdsims} and \ref{sisec:naturepert}) This is only possible for large red giants and asymptotic giant-branch stars, whose envelopes would be tidally disrupted, and thus the remnant cores would become too small on the orbital timescale to induce significant perturbations. The other possibility are stars with powerful stellar outflows, which are rather unlikely as we show in subsequent sections. The restrictions for stars are less severe for smaller SMBH masses, in particular for $M_{\bullet}\lesssim 10^{6.9}\,M_{\odot}$, when $\mathcal{R}\lesssim 16.8\,R_{\odot}$, and hence the stars around such a SMBH would have large enough relative cross-sections of $\sim 1\,r_{\rm g}$ and would not be tidally disrupted.  

{\bf The necessity for having a perturber with the influence radius $\mathcal{R}\sim 1\,r_{\rm g}$ can also be inferred from the column density of the enhanced absorption due to QPOuts. If we consider the column density of the absorbing material at the ODR minima, $N_{\rm h}\approx 1.5\times 10^{22}\,{\rm cm^{-2}}$ (see Fig.~\ref{fig:specplots}), one can estimate the typical line-of-sight length-scale of absorbing gas clumps as $h\approx f_{\rm g} N_{\rm h}/n_{\rm flow}$, where $n_{\rm flow}$ is the number density of the accretion flow that is perturbed and from which the blob is ejected and $f_{\rm g}$ is the geometrical factor related to the perturber orbital inclination. If we consider the SMBH mass of $M_{\bullet}=10^{7.4}\,M_{\odot}$ and the distance $r\approx 93\,r_{\rm g}$ associated with the source-frame period $P_{\rm orb}\approx 8.05$ days, then the mean ADAF electron number density is $n_{\rm flow}\approx 6.3\times 10^{19}\alpha^{-1}(M_{\bullet}/M_{\odot})^{-1}\dot{m}\,[r_{\rm per}/(2r_{\rm g})]^{-3/2}\,{\rm cm^{-3}}\approx 3.95\times 10^9\,{\rm cm^{-3}}$ for the Eddington ratio of $\dot{m}=0.05$ \cite{2014ARA&A..52..529Y}. This gives the length-scale of the ejected blobs in terms of the gravitational radius
\begin{equation}
   \frac{h}{r_{\rm g}}\approx 1.02 f_{\rm g}\,\left(\frac{N_{\rm h}}{1.5 \times 10^{22}\,{\rm cm^{-2}}} \right) \left(\frac{\alpha}{0.1} \right) \left(\frac{\dot{m}}{0.05} \right)^{-1} \left(\frac{r_{\rm per}}{93\,r_{\rm g}} \right)^{3/2} \,.
\end{equation}

%The factor $f_{\rm exp}$ is an adiabatic expansion factor that takes into account the amount of the blob expansion along its trajectory to larger scale-heights where it provides an obscuration. After reaching the vertical position where the gas clump subtends the angle of $41^{\circ}$ with respect to the rotation axis, the expansion factor is $f_{\rm exp}\sim 4.2$, which results in $\mathcal{R}\approx 3\,r_{\rm g}$.
Under the assumption that the radius of influence of the perturbing body should be of the order of $h$ in order to eject the absorbing material of a comparable size, this estimate indicates the necessity for $\mathcal{R}$ to be at least of the order of a gravitational radius.}

{\bf Given the difficulties to have such large stellar perturbers (see the previous estimates with stellar radii), it is more likely that the perturber is of a compact nature. The mass range can be estimated from the Hill and Bondi radii as follows, }
\begin{align}
    m_{\rm per, Hill}&=\frac{12\pi^2 G^2}{c^6}\frac{M_{\bullet}^3}{P_{\rm orb}^2}\,\notag\\
    &\simeq 94 \left(\frac{M_{\bullet}}{10^{7.4}\,M_{\odot}} \right)^3 \left(\frac{P_{\rm orb}}{8.05\,{\rm days}} \right)^{-2}\,M_{\odot}\,,
\end{align}
\begin{align}
    m_{\rm per, Bondi}&=\frac{G^{2/3}(4\pi^2)^{1/3}}{c^2}M_{\bullet}^{5/3}P_{\rm orb}^{-2/3}\,\notag\\
    &\simeq 2.7\times 10^5\,\left(\frac{M_{\bullet}}{10^{7.4}\,M_{\odot}} \right)^{5/3} \left(\frac{P_{\rm orb}}{8.05\,{\rm days}} \right)^{-2/3}\,M_{\odot}\,,
\end{align}
{\bf hence the expected mass range of the compact body is  $m_{\rm per}\sim 10^2-10^5\,M_{\odot}$ or an intermediate-mass black hole (IMBH). The perturber can launch ultrafast outflows with the required column density, while at the same time, the perturbations of the X-ray continuum due to perturber-disk interactions and shocks are rather weak.

When we adopt the model set-up of \cite{2023ApJ...957...34L} with the perturber passing through a standard disk formed following the TDE, the flares due to ejected shocked gas blobs are typically at least an order of magnitude below the observed X-ray flux (0.3-1.1 keV) of ASASSN-20qc. Adopting the X-ray luminosity during the enhanced accretion period, $L_{X}\sim 5 \times 10^{43}\,{\rm erg\,s^{-1}}$, the Eddington ratio is $\dot{m}\sim \kappa_{\rm bol}L_{X}/L_{\rm Edd}\sim 0.1$, hence similar to the one assumed by \cite{2023ApJ...957...34L} for their standard-disk model. When we consider an interacting IMBH of $m_{\rm per}\sim 10^4\,M_{\odot}$, whose characteristic radius is given by the Bondi radius, the X-ray flare luminosity due to the expanding shocked ejecta can be estimated as (relation A3 in \cite{2023ApJ...957...34L}), 
\begin{equation}
   L_{\rm flare, IMBH}\lesssim 1.4 \times 10^{42} \left(\frac{\dot{m}}{0.1} \right)^{1/3}  \left(\frac{M_{\bullet}}{10^{7.4}\,M_{\odot}} \right)^{5/9} \left(\frac{m_{\rm per}}{10^{4}\,M_{\odot}} \right)^{2/3} \left(\frac{P_{\rm QPOut}}{8.05\,{\rm days}} \right)^{-2/9}\,{\rm erg\,s^{-1}}\,, 
   \label{eq_LM_IMBH}
\end{equation}
where we assumed that the IMBH interacts with the standard disk twice per orbit with the period set by quasiperiodic ultrafast outflows. Therefore, Eq.~\eqref{eq_LM_IMBH} can be treated as an upper limit and the flare luminosity will be weaker for a more diluted ADAF flow expected for ASASSN-20qc at the distance where the IMBH orbits the SMBH. For the same perturber mass of $10^4\,M_{\odot}$ as well as the relative accretion rate, the plasmoid model by \cite{2023A&A...675A.100F} predicts the flare luminosity of $L_{\rm flare,0.3-1.1keV}\lesssim 5.3\times 10^{42}\,{\rm erg\,s^{-1}}$ in the $0.3-1.1$ keV band, assuming the blackbody emission at $\sim 10^6$ K. Hence, in both models, the shock emission is at least an order of magnitude below the quiescent level. Since $L_{\rm flare,IMBH}<L_{X}$ by an order of magnitude, the modulation of the X-ray continuum flux is negligible. A stellar perturber of $m_{\rm per}\sim 1\,M_{\odot}$ interacting with the standard disk provides even smaller shock-generated X-ray flux of (relation 17 in \cite{2023ApJ...957...34L})
\begin{equation}
    L_{\rm flare, star}\lesssim 6.7 \times 10^{41} \left(\frac{R_{\star}}{1\,R_{\odot}} \right)^{2/3} 
    \left(\frac{M_{\bullet}}{10^{7.4}\,M_{\odot}} \right) \left(\frac{\dot{m}}{0.1} \right)^{1/3} \left(\frac{P_{\rm QPOut}}{8.05\,{\rm days}} \right)^{-2/3}   \,{\rm erg\,s^{-1}}\,.   
\end{equation}
Hence, the perturber-accretion disk shocks in ASASSN-20qc can generate only weak X-ray flares, generally below the quiescent X-ray level, mainly due to a relatively wide orbit of the perturber with the recurrence period of $\sim 8$ days. During interactions with the more diluted ADAF, the shock-induced X-ray flux will be even smaller. Hence, a massive compact perturber, such as an IMBH, orbiting at $\sim 100\,r_{\rm g}$ can naturally account for the generation of periodic ultrafast outflows with a high enough column density to cause absorption, while the continuum X-ray emission is not affected significantly by the passages.

}

\section{GRMHD simulations of the perturbed accretion flow}\label{sisec:grmhdsims}
As in \cite{Sukova+2021:QPOutSims}, we model the perturber-accretion disk interactions using the global general relativistic magneto-hydrodynamical (GRMHD) 2D and 3D simulations using the code \texttt{HARMPI} code \cite{2007MNRAS.379..469T,2015MNRAS.454.1848R} that is based on the original \texttt{HARM} code \cite{2003ApJ...589..444G,2006ApJ...641..626N}. The code with some of our applied modifications solves the equations of ideal magnetohydrodynamics on the curved Kerr spacetime background. The details on the numerical scheme can be found in \cite{Sukova+2021:QPOutSims}. The gas is polytropic with an index of $13/9$, i.e., a value between the relativistic and the non-relativistic case to capture both relativistic electrons and non-relativistic protons. 

This code proves to be efficient for simulating our scenario, in which the perturber comes very close to the central SMBH, hence the relativistic regime is essential, while the amount of accreted matter is negligible compared to the mass of the SMBH, therefore the assumption of fixed background described by Kerr metric is well justified. Because studying the repetitive transits of the star requires long integration times of the simulations, computational efficiency is desirable. Our version of the code uses non-uniform spacing of the grid, which allows us to use a single grid without mesh refinements and still resolve well the closest neighborhood of the black hole while having sufficiently large grid to capture the whole accreting torus. 

At the moment, we do not consider radiative transfer and associated feedback, hence the accretion flow does not cool radiatively. Therefore, the set-up is the most suitable for ADAF-type (Advection Dominated Accretion Flow) hot flows, i.e. SMBHs that accrete significantly below the Eddington limit. This seems to be the case of ASASSN-20qc, for which the integrated intrinsic luminosity in the range 1~eV--10~keV near the peak is $L_{\rm peak}=10^{44.2}\,{\rm erg\,s^{-1}}$. For a black hole mass in the range $M_{\bullet}=10^{7}-10^8\,M_{\odot}$, we obtain the Eddington ratio of $\lambda_{\rm Edd}=\kappa_{\rm bol} L_{\rm peak}/L_{\rm Edd}\sim 0.01-0.1$, where we consider the bolometric correction factor $\kappa_{\rm bol}\sim 1$. The Eddington ratio was clearly lower, $\lambda_{\rm Edd}\sim 10^{-5}$, before the detection of the optical/X-ray outburst, which implies a low-luminosity ADAF regime. On the other hand, the detected broad lines indicate the existence of an outer optically thick, standard disk \cite{2004A&A...428...39C}. Indeed, a hybrid accretion disk solution with an inner ADAF-type flow and an outer standard thin disk is often suggested as a phenomenological description of the emission for low-luminosity AGN (see, e.g., \cite{2014MNRAS.438.2804N}). 

In that case, we expect--that prior to the outburst--the inner part of the accretion disk should be geometrically thick and hot, i.e. ADAF-like optically thin solution, with the transition given by the ADAF principle, i.e. whenever the ADAF solution is permitted, the accretion transitions into the hot flow, \cite{1995ApJ...438L..37A,1996PASJ...48...77H,1998PASJ...50..559K},
\begin{equation}
    R_{\rm ADAF}=1600\, \alpha_{0.1}^4 \dot{m}_{0.05}^{-2} r_{\rm g}\,,
    \label{eq_ADAF_radius}
\end{equation}
where $\dot{m}_{0.05}$ is a dimensionless accretion rate expressed in Eddington units, $\dot{m}\equiv \dot{M}_{\bullet}/{\dot{M}_{\rm Edd}}$ and is scaled to 0.05 intermediate between $\lambda_{\rm Edd}\sim 0.01-0.1$ estimated above. The parameter $\alpha$ is a viscous parameter scaled to $0.1$. Another possibility of the ADAF formation is via the evaporation due to the electron conduction between the cold disk and the hot two-temperature corona. In that case, the transition radius would be smaller but still above the distance scale where we assume the star-disk interactions \cite{2004A&A...428...39C},
\begin{equation}
    R_{\rm evap}=191\,\alpha_{0.1}^{0.8}\beta^{-1.08}\dot{m}_{0.05}^{-0.53}\,r_{\rm g}\,,
    \label{eq_evap_radius}
\end{equation}
where $\beta=P_{\rm g}/(P_{\rm g}+P_{\rm m})$ is the magnetization parameter defined as the ratio of the gas pressure to the total pressure. For a negligible magnetic field $\beta=1$ while for large magnetic field, $\beta<1$, which leads to a larger scale of the ADAF.

The initial condition for our simulations is given by the solution of a thick torus \cite{Witzany_Jefremov-tori} yielding a large torus stretching between $r=20\,r_{\rm g}$ to $r=500\,r_{\rm g}$, which serves as a reservoir of matter for accretion. After the initial transient time $t_{\rm in}$, the perturber is added into the evolved state of the torus. 

The pertuber moves along a geodesic trajectory that is calculated using the 4th-order Runge-Kutta scheme simultaneously with the evolution of the gas using the time-step found via the adaptive GRMHD solver. This ensures a sufficient precision of the stellar position and the velocity within the evolving gaseous environment around the SMBH. The interaction of the perturber with the polytropic gas is modelled using a sphere of influence around the object with the radius $\mathcal{R}$ that mimics the stagnation radius around a wind-blowing or a magnetized star and/or the Hill radius or the synchronization radius around a dormant compact perturber (stellar or intermediate-mass black hole). As the perturber moves along its orbit, the grid cells that lie inside the sphere of influence adopt a velocity field comoving with the object, while other MHD parameters of the gas are kept intact. This effectively captures the perturber-gas interaction around the SMBH, in particular the development of a bow shock due to a supersonic motion and the propagation of density waves, while the internal properties of the perturbing object, in particular stellar evolution, are neglected.

%%%%%%%%%%%%%%

We performed several different runs with parameters tailored specifically to ASASSN-20qc (see Table~\ref{tab:GRMHD_runs}). The parameters differ in terms of the initial magnetic field configuration (one loop -- the disk is prone to develop the Magnetically Arrested disk or MAD in short, more loops yield a stable accretion and outflow rate, so-called Standard and Normal Evolution -- SANE), the distance of the perturber from the SMBH and its influence radius $\mathcal{R}$, and the inclination of its orbit. 
%The resolution is 384 $\times$ 256 cells in radial and angular direction, respectively. 
For all the runs, the stellar orbit is inclined with respect to the equatorial plane of the flow so that it effectively reaches the torus/funnel boundary, which leads to the blob ejection and the subsequent acceleration along this boundary. Since we do not have an estimate of the SMBH spin for ASASSN-20qc, we adopted a  fiducial value of $a \simeq 0.4$, i.e., the black hole rotates only mildly. The SMBH mass of $10^{7.4}\,M_{\odot}$ was adopted for most runs in accordance with the observationally inferred mean value, while the larger value of $10^{7.95}\,M_{\odot}$ reflects the uncertainty and was chosen so that the orbital period of $8.05$ days corresponds to the orbital distance of $40\,r_{\rm g}$.

We calculate the inflow rate $\dot{M}$ as well as the outflow rate $\dot{M}_{\rm out}$, which is split into the ``upper'' and the ``lower'' funnels. In particular, we integrate the outflow through the sector of the sphere along the symmetry axis with the opening angle of $45^{\circ}$ pointing ``upwards'' and ``downwards'' with a radius of $r_{\rm diag}=300\,r_{\rm g}$. To exclude the motion of the dense but slow gas inside the accreting torus, we integrate only the material which moves away from the black hole faster than a chosen threshold -- its Lorentz factor satisfies the sequence of conditions -- $\Gamma > 1.005$ ($v>0.1c$), $\Gamma > 1.02$ ($v>0.2c$), $\Gamma > 1.05$ ($v>0.3c$), $\Gamma > 1.091$ ($v>0.4c$) or $\Gamma > 1.155$ ($v>0.5c$). In configurations with different spin and magnetic field strength and geometry and orbital parameters, the outflow achieves different velocities; however, overall the simulation outcome is only weakly dependent on the SMBH spin value.

In Fig.~\ref{fig_run1}, we show a snapshot of Run 14 with the density distribution, the Lorentz factor of the outflow, and the outflow-rate distribution for $v>0.2c$. In the panel d), we also show the temporal evolution of the inflow and the outflow rates. The outflow periodicity is clearly visible, while the inflow rate is a combination of the stochastic red-noise variability and the periodic effect of the perturber.
We can notice one outflowing blob induced by the perturber interactions with the disk in panel c), which shows the spatial distribution of the mass-outflow rate and the bow shock formed by the motion of the perturber in the flow in panel a). The blobs are moving along the boundary between the dense accretion torus and the magnetized funnel, which half-opening angle is approximately $30^{\circ}$ depending on the parameters of the gas and the spacetime.

The trajectory of the perturber, in this case, is chosen to be mildly eccentric (with $e=0.5$). This causes the shift in the position of the individual peaks in the outflowing rate. Due to the launching mechanism in our scenario, the timing of the individual expelled blobs is given by the times, when the perturber flies out from the disk and pushes the gas into the funnel. This roughly corresponds to the maxima/minima of angle $\theta$. The frequency of oscillations in $r$ and $\theta$ directions differs and, as the orbit undergoes the precession, there is the phase shift of those maxima/minima. 
%This effect is more pronounced for eccentric orbits. 
Moreover, for such eccentric orbits, the radial distance of the perturber when achieving the turning point in $\theta$ angle differs, hence the blob is launched at different distances from the black hole. Therefore, the shape and amplitude of the individual peaks differ, depending on the gas condition at the place, where the perturber flies out from the disk. 

Runs 1-8 and 12-13 yield similar results as the one presented here, with different timing of the peaks due to the eccentricity.  The strength of the outflow varies with the size of the perturber as we discuss below.
 
Since the spin of the black hole is rather mild in our simulations, we can conclude that the fast rotation of the central black hole is not necessary for our scenario. The effect is viable for both the case of non-rotating, Schwarzshild SMBHs as well as for fast rotating Kerr SMBHs. 

All the simulations presented so-far were computed in 2D, i.e., under the assumption of the axisymmetry of the system. However, the perturber surely breaks this symmetry while transiting through the flow within a small azimuthal angle. In \target, the perturber is located quite far from the SMBH and its orbital period is long ($\sim 5630r_{\rm g}/c\simeq 8.05$ days in the source frame), therefore it would be computationally very demanding to perform full 3D simulations with these parameters. However, to overcome this caveat, we perform simulations with the perturber located close to the SMBH, $r=10\,r_{\rm g}, \mathcal{R}=1\,r_{\rm g}$ with the same parameters both in 2D and 3D and compare the strength of the measured outflow rate. We normalize the strength of the outflow from the 2D simulation by a relation expressing the ratio of the azimuthal width of the perturbing body to the full angle.

\begin{equation}
    \dot{M}_{\rm out} = \frac{2\mathcal{R}}{2 \pi r_{\rm min}} F_n \dot{M}_{\rm out}^{\rm 2D}.
    \label{eq_2D-factor}
\end{equation}

The factor $F_n$ is introduced due to the geometrical shape of the perturber and we fix its value to $F_n = 1/4$ by comparing with the outflow strength measured in the 3D simulation (Fig.~~\ref{fig_influence_radius}). 
The case with $\mathcal{R}=0.1\,r_{\rm g}$ is also shown and it yields the outflow rate smaller by more than three orders of magnitude. Hence, this demonstrates the strong dependence on the size of the perturber, which allows us to discuss its likely nature. 

In case of larger eccentricity (Run 15 with $e=0.67$), the perturber comes closer to the SMBH and has a larger impact on the flow with visible peaks in the accretion rate. However, we note that the influence on the accretion rate appears artificially enhanced in the 2D simulations in the manner as expressed by relation \ref{eq_2D-factor}. We are aiming to study these quantitative effects systematically with 3D simulations (follow-up work in progress).

\textbf{For our last simulation, Run 16, we shifted the perturber closer to the SMBH, so that we can follow the evolution for a similar number of orbits ($\sim 15$) as was observed in the case of ASSASN-20qc in 3D within a reasonable amount of computation time. The perturber with the influence radius $\mathcal{R}=2M$ moves on a  mildly elliptic ($e=0.18$), inclined ($\iota=67.7\degree$) orbit with pericenter $r_{\rm min}=10M$ and apocenter $r_{\rm max}=14.7M$. The characteristic periods, i.e. the inverse functions of the fundamental frequencies of geodesic motion in Kerr spacetime \cite{2002CQGra..19.2743S}, in $r$, $\theta$, and $\phi$ directions are $P_r = 370M, P_\theta = 273.3M, P_\phi = 269.7M$, respectively, while the total duration of the simulation was $4300M$. %so that we cover cca 16 revolutions of the body. 
}

\textbf{In Fig.~\ref{fig_run16} we show the time dependence of the accretion rate and outflowing rate in two velocity bins $0.3c<v<0.4c$ and $0.4c<v<0.5c$. Even though the perturber is much closer to the SMBH with similar $\mathcal{R}$ as in our previous 2D simulations, there is no clear sign of periodicity in the periodogram of the inflow corresponding to either $P_r, P_\theta$ or $P_\phi$, while the two peaks are visible in the periodogram of the outflow. 
It is possible, that with longer duration of the simulation, some signs of periodicity can emerge in the inflow rate, however, we are limited by computational resources at this point. However, the number of the covered cycles in this simulation is comparable to the observed one, hence constructing Lomb-Scargle periodograms from the longer runs would not correspond to the observational baseline.}

\textbf{It is reasonable to expect, that the effect of the perturber with the parameters derived for ASSASN-20qc
%, i.e. with pericenter $r_{\rm min}\sim45M$, apocenter $r_{\rm max}\sim 135M$ and influence radius $\mathcal{R}\sim3M$ 
on the accretion rate will be smaller than in this 3D case. This is because the flow is perturbed further away from the SMBH, there is more time for the density inhomogeneities to dissipate in the disc during their infall to the center. Therefore, we do not expect that the accretion rate in case of ASSASN-20qc would be significantly affected by the perturber.}

\section{On the nature of the orbiting perturber}\label{sisec:naturepert}

The order of magnitude estimate of the influence radius of the perturber $\mathcal{R}\sim 1\,r_{\rm g}$ is supported by our GRMHD simulations. In Fig.~\ref{fig_influence_radius}, we show the GRMHD-based ratio between the mass
outflow rate and the mass accretion rate in comparison with the measured one during the \nicer Min phases (see Table~\ref{tab:energetics}). We plot the same quantity for four runs, which differ only by the influence radius $\mathcal{R}$, ranging from $\mathcal{R}=2\,r_{\rm g}$ to $\mathcal{R}=8\,r_{\rm g}$. 
While the largest perturber yields the outflow rate ten times greater than the inflow rate, the smallest perturber barely reaches the minimal outflow/inflow ratio during outflow peaks seen in observations $(\dot{M}_{out}/\dot{M}_{acc})_{\rm min} = 0.03$.
The best agreement is achieved for $\mathcal{R}=3\,r_{\rm g}$. Due to the uncertainties in other parameters of the system, such as the spin of SMBH or the properties of the magnetic field, this still serves only as an order of magnitude estimate.

Since we can constrain the perturber influence radius to $\mathcal{R}\approx 3\,r_{\rm g}$, we can infer the likely nature of the perturber from mechanisms that can create such a large influence radius. For a compact object without any outflow, the largest scale of gravitational influence of the perturber within the two-body problem is given by the tidal or the Hill radius. Because the whole gas attracted by the perturber within the Hills sphere does not comove with the object, the Hills' relation gives us a lower limit on the mass of the perturber,

\begin{align}
    m_{\rm per, Hill}&=\frac{12\pi^2 G^2M_{\bullet}^3}{P_{\rm orb}^2c^6}\left(\frac{\mathcal{R}}{r_{\rm g}}\right)^3\,\notag\\
    &=2532\,\left(\frac{P_{\rm orb}}{8.05\,\text{d}}\right)^{-2} \left(\frac{\mathcal{R}}{3\,r_{\rm g}} \right)^3 \left(\frac{M_{\bullet}}{10^{7.4}\,M_{\odot}} \right)^3 M_{\odot}\,,
    \label{eq_hill_radius}
\end{align}
which is consistent with an intermediate-mass black hole of $\sim 10^2-10^5\,M_{\odot}$ given the uncertainty of $\mathcal{R}$ as well as that of the primary black hole mass in the range of $10^7-10^8\,M_{\odot}$.

The synchronization radius $R_{\rm sync}$ introduced by \cite{Sukova+2021:QPOutSims}, and based on the gas drag force resulting from the Bondi-Hoyle-Lyttleton accretion \cite{1939PCPS...35..592H,1944MNRAS.104..273B,1999ApJ...513..252O}, expresses the gas sphere that receives the full momentum from the star and starts comoving, which may not be entirely necessary for the ejection of larger blobs into the funnel region. In this regard, $R_{\rm sync}$ can be used to derive an upper limit for the perturber mass. Using the relation for the drag force acting on the IMBH for the interaction time $\Delta t$ derived by \cite{1999ApJ...513..252O}, we can derive the synchronization radius from the conservation of momentum, $|F_{\rm df}|\Delta t\approx M_{\rm gas} v_{\rm {\rm rel}}\sim (4/3)\pi R_{\rm syn}^3\rho_{\rm gas}v_{\rm rel}$, where $v_{\rm rel}\approx (2GM_{\bullet}/r_{\rm per})^{1/2}$ is the relative velocity of the IMBH with respect to the accretion disk for the perturber distance $r_{\rm per}$ from the SMBH and considering high inclinations with respect to the disk equatorial plane. By setting $\Delta t\approx P_{\rm orb}/4$, which corresponds to the duration of the IMBH passages through a thick flow with the scale-height to radius ratio of the order of unity, we obtain,
\begin{align}
    \frac{R_{\rm syn}}{r_{\rm g}}&\approx \left(\frac{3I}{8 \pi} \right)^{1/3} \frac{c^2}{\sqrt{2}} \frac{(Gm_{\rm per})^{2/3}}{(GM_{\bullet})^{4/3}}P_{\rm orb}^{2/3}\,,\notag\\
    m_{\rm per, syn}&\approx 35\,671 \left(\frac{P_{\rm orb}}{8.05\,{\rm d}} \right)^{-1}\left(\frac{\mathcal{R}}{3\,r_{\rm g}} \right)^{3/2}\left(\frac{M_{\bullet}}{10^{7.4}\,M_{\odot}} \right)^{2}\,M_{\odot}\,,
    \label{eq_synchronization_radius}
\end{align}
where the factor $I={\rm ln}{(r_{\rm max}/r_{\rm min})}$ and $r_{\rm max}$ and $r_{\rm min}$ correspond to the sizes of the surrounding gaseous medium and the perturbing object, respectively. For the supersonic motion of the perturber through the accretion disk on a highly inclined orbit, the factor $I$ can be approximated as $I\approx {\rm ln}{(M_{\bullet}/m_{\rm per})}\sim 7-9$ \cite{2000ApJ...536..663N}, which we approximate by setting $I\sim 10$ for further estimates.

The lower and upper mass limits given by the Hill and the synchronization radii, Eqs.~\eqref{eq_hill_radius} and \eqref{eq_synchronization_radius} respectively, give a broad range of masses $\sim 10^3-10^5\,M_{\odot}$ consistent with the intermediate-mass as well as a low-mass supermassive black hole, hence a massive, non-stellar perturber. The uncertainty is mainly given by the primary black hole mass as well as the size of the influence radius. We plot the influence radii $\mathcal{R}$ as a function of the primary SMBH in Fig.~\ref{fig_Hill_syn_radius} considering both the Hill radii as well as synchronization radii for limiting values of massive black-hole perturbers that still yield the influence radius of $\mathcal{R}=3r_{\rm g}$ for the lower and the upper limit of the SMBH mass ($10^7$-$10^8\,M_{\odot}$). Considering the Hill-radius mass estimate, we obtain the range of $m_{\rm per, Hill}=160-1.6\times 10^5\,M_{\odot}$, while the synchronization-radius relation gives a range of $m_{\rm per, syn}=5.65\times 10^3-5.65\times 10^5\,M_{\odot}$. As it can be inferred from Fig.~\ref{fig_Hill_syn_radius}, there is an overlap in the mass range between $\sim 10^{3}$ and $\sim 10^5\,M_{\odot}$. 

% FIG+++FIG+++FIG+++FIG+++FIG+++FIG+++FIG+++FIG+++FIG+++FIG+++FIG+++FIG+++
%                   Extended Data Figure 5: XMM RGS spectrum
% FIG+++FIG+++FIG+++FIG+++FIG+++FIG+++FIG+++FIG+++FIG+++FIG+++FIG+++FIG+++
However, the presence of the second supermassive black hole is less likely due to the short merger timescale for such a system. If we consider the initial perturber distance of $r_0$ that corresponds to its orbital period of $P_{\rm orb}\sim 8.05$ days, the orbit is well in the weak field of the primary and the gravitational radiation-reaction is weak. The merger timescale can then be estimated from the leading post-Newtonian formula as follows \cite{1964PhRv..136.1224P}
\begin{align}
    \tau_{\rm merge}&=\frac{5c^5}{256G^3}\frac{r_0^4}{M_{\bullet}m_{\rm per}(M_{\bullet}+m_{\rm per})}\,,\notag\\
    &=\frac{5c^5}{(2^{32}G^5\pi^8)^{1/3}}\frac{(M_{\bullet}+m_{\rm per})^{1/3}}{M_{\bullet}m_{\rm per}}P_{\rm orb}^{8/3}\,.
    \label{eq_timescale_merge}
\end{align}
 For the lower SMBH-IMBH pair mass limit, we obtain $\tau_{\rm merge}(10^{7.4}\,M_{\odot},10^3\,M_{\odot})\sim 143\,426$ years, while for the upper SMBH-IMBH pair mass limit, we get $\tau_{\rm merge}(10^{7.4}\,M_{\odot},10^5\,M_{\odot})\sim 1436$ years. Hence, the perturbation by a more massive IMBH is less likely, as it would necessarily result in the short merger timescale of $\lesssim 1000$ years. In bottom panel of Fig.~\ref{fig_influence_radius}, we plot the merger timescale in years as a function of the perturber mass. Since it is more plausible to have a concurrent tidal disruption event, which occurs with the rate of $\sim 10^{-4}\,{\rm yr^{-1}}$ per galaxy, to take place in the binary system with the merger timescale at least $\sim 10^4$ years, this statistical argument favours the perturbers with the mass of $\lesssim 10^{4}\,M_{\odot}$. Specifically, for the primary SMBH mass of $M_{\bullet}=10^7\,M_{\odot}$, we obtain $m_{\rm per}\lesssim 26500\,M_{\odot}$, for the intermediate value of $M_{\bullet}=10^{7.4}\,M_{\odot}$ we get $m_{\rm per}\lesssim 14\,300\,M_{\odot}$, while for $M_{\bullet}=10^8\,M_{\odot}$, we obtain $m_{\rm per}\lesssim 5700\,M_{\odot}$.
 
 We also checked the signal-to-noise ratios for a possible detection by LISA in the 2030s \cite{robson2019construction}. Since the gravitational radiation is weak for the SMBH-IMBH pair, the period of the binary will not evolve significantly in the next decade. Consequently, the source will be outside of the frequency range of LISA and the signal-to-noise ratios will be below $10^{-2}$ in all admissible scenarios for the primary and the secondary masses.
% FIG+++FIG+++FIG+++FIG+++FIG+++FIG+++FIG+++FIG+++FIG+++FIG+++FIG+++FIG+++
%                   Extended Data Figure 5: XMM RGS spectrum
% FIG+++FIG+++FIG+++FIG+++FIG+++FIG+++FIG+++FIG+++FIG+++FIG+++FIG+++FIG+++

A star with an outflow -- either a wind-blowing star or a pulsar whose stagnation radius can be analytically estimated \cite{Sukova+2021:QPOutSims} -- appears to be a much less suitable candidate for the perturber due to a small size of the associated stagnation radius across a wide range of parameters. Considering the SMBH mass of $M_{\bullet}=10^{7.4}\,M_{\odot}$ and assuming the hot-flow density profile for $\dot{m}\sim 0.05$ \cite{2014ARA&A..52..529Y}, we obtain the $R_{\rm wb}^{\rm ADAF}\sim 0.76 r_{\rm g}$ as the stagnation radius of the wind-driven shock for the mass-loss rate of $\dot{m}_{\rm w}=10^{-3}\,{\rm M_{\odot}\,yr^{-1}}$ and the terminal wind velocity of $v_{\rm w}=10^{3}\,{\rm km\,s^{-1}}$, which are rather large, unlikely values. The same stellar-wind parameters yield four orders of magnitude smaller stagnation radius $R_{\rm wb}^{\rm thin}=5.36 \times 10^{-5}\,r_{\rm g}$ for the case of a standard thin disk with the same accretion rate. For the larger black-hole mass of $M_{\bullet}=10^8\,M_{\odot}$, we obtain $R_{\rm wb}^{\rm ADAF}\sim 0.12 r_{\rm g}$ and $R_{\rm wb}^{\rm thin}=5.93\times 10^{-6}\,r_{\rm g}$ for the same stellar-wind parameters. 
% In the Extended Data Figure ~\ref{fig_stag_radius_wb}, we plot the stagnation radius of a stellar bow shock as a function of the mass-loss rate and a terminal wind velocity. 
The stagnation radii of the order of $0.1\,r_{\rm g}$ are plausible for either young massive stars (Wolf-Rayet stars) with fast winds or massive late-type stars (red supergiants) with large mass-loss rates. Hence, for a typical main-sequence Sun-like star, the stagnation radius is orders of magnitude below the gravitational radius.
For the case of an orbiting young pulsar with the large spin-down energy of $\dot{E}=10^{38}\,{\rm erg\,s^{-1}}$ comparable to the energetic Crab nebula pulsar, the stagnation radius of the pulsar wind bubble is $R_{\rm psr}^{\rm ADAF}=1.7\times 10^{-2}\,r_{\rm g}$ and $R_{\rm psr}^{\rm thin}=1.2\times 10^{-6}\,r_{\rm g}$ for $M_{\bullet}=10^{7.4}\,M_{\odot}$ for the hot ADAF and the thin-disk solutions with $\dot{m}=0.05$, respectively. For $M_{\bullet}=10^8\,M_{\odot}$, we obtain $R_{\rm psr}^{\rm ADAF}=2.8\times 10^{-3}\,r_{\rm g}$ and $R_{\rm psr}^{\rm thin}=1.4 \times 10^{-7}\,r_{\rm g}$ for the same spin-down energy.

An interesting possibility for the perturber would be a binary system with the semi-major axis of $a_{\rm bin}\sim 2\mathcal{R}$, i.e. the binary separation would correspond to the twice of the influence radius of the perturber. The semi-major axis of such a system is limited by the tidal field close to the SMBH. The binary needs to be located at the distance of $d_{\rm bin}\gtrsim r_{\rm T}\sim a_{\rm bin}(M_{\bullet}/m_{\rm bin})^{1/3}$, which puts an upper limit on the component separation at the distance $d_{\rm bin}=r_{\rm per}$ that corresponds to the orbital period of $P_{\rm orb}$,
\begin{align}
    a_{\rm bin} &\lesssim r_{\rm per}\left(\frac{m_{\rm bin}}{M_{\bullet}} \right)^{1/3}\,\notag\,\\
    \mathcal{R}_{\rm bin}&=\frac{1}{2}\frac{a_{\rm bin}}{r_{\rm g}}\lesssim \frac{c^2}{2G^{2/3}(4\pi^2)^{1/3}}P_{\rm orb}^{2/3}\frac{m_{\rm bin}^{1/3}}{M_{\bullet}}\,\notag\\
    &\simeq 0.34 \left(\frac{P_{\rm orb}}{8.05\,\text{d}}\right)^{2/3} \left(\frac{m_{\rm bin}}{10\,M_{\odot}}\right)^{1/3}\left(\frac{M_{\bullet}}{10^{7.4\,M_{\odot}}} \right)^{-1}\,.
    \label{eq_binary_separation}
\end{align}
Hence, a binary consisting of two approximately equally massive main-sequence stars of $5\,M_{\odot}$ each would have to be compact with $\mathcal{R}_{\rm bin}$ an order of magnitude below the limit of $\mathcal{R}\sim 3r_{\rm g}$. For a larger $\mathcal{R}_{\rm bin}$, the binary would disrupt (Hills mechanism, \cite{Hills88}). The only possibility to have $\mathcal{R}_{\rm bin}\sim 3\,r_{\rm g}$ is for $m_{\rm bin}\gtrsim 6\,800\,M_{\odot}$, i.e. for two IMBHs (e.g. of nearly equal mass of $\sim 3400\,M_{\odot}$) that we already proposed. However, such a system of two IMBHs orbiting each other with the period of $\sim 8.05$ days (in 1:1 resonance with the binary orbital period around the SMBH) is less likely and it is not necessary to explain QPOuts. Moreover, the required high inclination of the SMBH-IMBH system is consistent with the IMBH receiving the recoiling velocity kick following the merger with another IMBH, see e.g. \cite{2020PhRvL.124y1102G}. The IMBH binary as such would be relatively long-lived with the merger timescale of $\sim 20.4$ Myr.

Another possible set-up is that an IMBH of $\simeq 10^{4}\,M_{\odot}$ would be orbited by a star of mass $m_{\star}$ within $a_{\rm bin} \lesssim 6.84r_{\rm g}$ (here we consider the basic condition for the tidal stability), which corresponds to $\sim 1.69\,{\rm AU}$ or $\sim 364\,R_{\odot}$. Such a star would not be tidally disrupted by the IMBH up to the stellar radius of $\sim 16.9\,R_{\odot}$ (for the stellar mass of $1\,M_{\odot}$). In case the IMBH would be orbited by a larger star of $\gtrsim 10\,R_{\odot}$ at the distance of $\sim 1.69\,{\rm AU}$, such a binary system would disrupt as the IMBH descends further towards the SMBH due to gravitational-wave emission. Subsequently, the star is tidally disrupted by the SMBH. At the time of the IMBH-star separation due to the Hills mechanism, the maximum radius of the tidally stable star around the IMBH is equal to the tidal radius of a star around the SMBH,
\begin{align}
R_{\star}&\sim P_{\rm orb}^{2/3}(Gm_{\star}/4\pi^2)^{1/3}\,,\notag\\
&\sim 16.9\,\left(\frac{P_{\rm orb}}{8.05\,{\rm d}} \right)^{2/3}\left(\frac{m_{\star}}{1\,M_{\odot}} \right)^{1/3}R_{\odot}\,,
\label{eq_stellar_radius_condition}
\end{align}
assuming that the IMBH-star binary and the separated components share approximately the same orbit with the orbital period of $P_{\rm orb}$. It is clear that conditions given by Eqs.~\eqref{eq_stellar_radius_condition} and \eqref{eq_star_radius} are the same. In this regard, the TDE and the IMBH-induced recurrent outflow could be causally connected. Although this seems to be just a hypothetical, fine-tuned scenario, considering the scenario of an infalling stellar cluster that hosts the IMBH at the center, it appears to be quite plausible as the necessary final outcome of the cluster dissolution when only a single IMBH-star binary remains.  \cite{smbh_imbhrates}. 

Given the uncertainty in \target's SMBH mass, one can constrain the SMBH mass ranges where perturbers are more likely stars or black holes, using the condition that the influence radius needs to be relatively large to produce escaping blobs causing periodic absorbing events like those detected from \target, i.e. $\mathcal{R}\simeq 3 GM_{\bullet}/c^2$. In the left panel of Fig.~\ref{fig_perturber_regime}, we plot the influence radius $\mathcal{R}$ expressed in Solar radii as a function of the SMBH mass. Furthermore, we include the upper limits on the stellar radius given by the tidal stability condition as well as the estimates of bow-shock radii for the ADAF and the standard disks (for $\dot{m}=0.05$ and the stellar-wind parameters specified in the legend). Given the perturber distance around the SMBH corresponding to $P_{\rm orb}=8.05$ days, there is a limiting SMBH mass of $M_{\bullet}\simeq 10^{6.27}\,M_{\odot}$, below which perturbers with $\mathcal{R}\sim 3\,r_{\rm g}$ can likely be stars, either due to their physical cross-section or wind bow shock. For heavier SMBHs, stellar-mass and intermediate-mass black holes will have a large enough influence radius due to their gravitational influence, see the estimates given by Hill and synchronization radii in Fig.~\ref{fig_perturber_regime} (based on Eqs.~\eqref{eq_hill_radius} and \eqref{eq_synchronization_radius}, respectively). In the same SMBH mass range, stars would need to have large physical radii of $\gtrsim 11.7\,R_{\odot}$, which decreases the likelihood considering the fact that the star is found at a special evolutionary stage (a late-type red giant), and more importantly, these stars would not be tidally stable at the required distance given by the QPOut (orbital) period. Hence, in case the SMBH mass for ASASSN-20qc is $\gtrsim 10^7\,M_{\odot}$, an IMBH is the only possibility for a stable perturber with a large-enough cross-section and, at the same time, a long enough merger timescale so that a TDE is likely to occur concurrently.

\subsection{Orbital stability of the SMBH-IMBH system}
\label{subsec_directional_stability}

The SMBH-IMBH system is relatively stable in terms of the secular orbital changes due to gravitational-wave emission. According to Eq.~\eqref{eq_timescale_merge}, the gravitational-wave inspiral timescale is $\tau_{\rm merge}\sim 10\,246$ years for the circularized orbit of the IMBH of $14\,000\,M_{\odot}$ around the SMBH of $10^{7.4}\,M_{\odot}$ with the initial distance of $r_0\simeq 93\,r_{\rm g}$. The influence radius $\mathcal{R}$ of the IMBH for this configuration is $5.3\,r_{\rm g}$ and $1.6\,r_{\rm g}$ as given by the Hill and the synchronization radii, respectively. In other words, the timescale for a  semi-major axis decrease by one gravitational radius is $\sim 434$ years. During the inspiral time of the IMBH, e.g. from 500 to 93$\,r_{\rm g}$, which takes of the order of $8.56\times 10^6$ years, the orbit gets effectively circularized, e.g. starting with $e_0=0.9$ at $r_0=100\,r_{\rm g}$, it takes $\sim 156\,000$ years to reach $e=0.01$ \cite{1964PhRv..136.1224P}. However, a mildly eccentric orbit cannot be excluded at this point. A non-zero eccentricity can actually address a quasiperiodic nature of the outflow as indicated by the temporal evolution of the ODR, see Fig.~\ref{fig:fig2}, since the eccentric orbit undergoes a prograde relativistic (Schwarzschild) precession. This is supported by the GRMHD simulation runs with eccentric perturber orbits (see Fig.~\ref{fig_run1}).

The IMBH orbiting the SMBH on a mildly eccentric orbit with the semi-major axis $a_{\rm per}$ and the eccentricity $e_{\rm per}$ is subject to the prograde, relativistic Schwarzschild precession of the argument of the pericenter. The orbital plane of the IMBH and hence the position angle of the outflow is not changed, but the properties of the outflow could be affected, i.e. the outflow launch radius and hence the outflow velocity could differ depending on whether the perturber-disk interaction takes place close to the pericenter or the apocenter of the eccentric orbit. The Schwarzschild precession timescale for the change of the argument of the pericenter by $180^{\circ}$, i.e. when the absorbing blob launch radius would effectively be changed from the pericenter to the apocenter of the orbit, is,
\begin{align}
    \tau_{\rm S}(180^{\circ})&=\frac{c^2 a_{\rm per}(1-e_{\rm per}^2)P_{\rm orb}}{6GM_{\bullet}}\,\notag\\
    &=\frac{c^2(1-e_{\rm per}^2)P_{\rm orb}^{5/3}}{6(4\pi^2)^{1/3}(GM_{\bullet})^{2/3}}\,\notag\\
    &\simeq 109.5 \left(\frac{P_{\rm orb}}{8.05\,\text{d}} \right)^{5/3} \left( \frac{1-e_{\rm per}^2}{0.88}\right) \left(\frac{M_{\bullet}}{10^{7.4}\,M_{\odot}} \right)^{-2/3}\,\text{days},
\end{align}
where we scaled the eccentricity to $e_{\rm per}=0.35$ corresponding to Run~3 in Table~\ref{tab:GRMHD_runs}, hence $1-e_{\rm per}^2\simeq 0.88$. The Schwarzschild precession timescale shows that the argument of pericenter can in principle rotate by $180^{\circ}$ during the observational coverage of 12 QPOuts ($\sim 100$ days in the rest frame).

The orbiting massive perturber with the orbital distance of 100 $r_{\rm g}$ also provides a sufficient directional stability for the duration of QPOuts (12 cycles of total duration $\sim 100$ days in the rest frame). This is a basic requirement since the ejected outflows need to cross the line of sight once per orbital period. In fact, for the IMBH perturber orbiting $10^{7.4}\,M_{\odot}$ SMBH at $93\,r_{\rm g}$, the Lense-Thirring precession timescale, which corresponds to the rotation of the longitude of line of nodes by $90^{\circ}$, can be expressed as
\begin{align}
    t_{\rm LT}(90^{\circ})&=\frac{c^3 P_{\rm orb}^2}{16 \pi G M_{\bullet} a}\,\notag\\
    &=2255\,\left(\frac{P_{\rm orb}}{8.05\,{\rm d}} \right)^2 \left(\frac{M_{\bullet}}{10^{7.4}\,M_{\odot}} \right)^{-1} \left(\frac{a}{0.4} \right)^{-1}\,\text{days}\,,
    \label{eq_LT_timescale}
\end{align}
where the spin parameter $a$ is scaled to $0.4$ and the orbital eccentricity is set to zero. For the favored spin of $a=0.9$ given the inferred disk temperature of 0.085 keV (from analysis in Methods section \ref{sisec:temp}), the $90^{\circ}$ Lense-Thirring timescale gives an upper limit on the SMBH mass of $10^{8.4}\,M_{\odot}$, for which $t_{\rm LT}\sim 100$ days (i.e. 12 detected QPOuts). We show the distance dependencies of the Schwarzschild (for $e_{\rm per}=0.35$) and Lense-Thirring precession (for $e_{\rm per}=0$ and $a=0.4$) timescales in Fig.~\ref{fig_Hill_syn_radius} alongside other relevant dynamical timescales. The directional stability tends to disfavor the accretion-disk instability mechanisms, which are directionally stochastic, unless the instability would periodically launch a relativistic absorbing gas clump across a broad azimuthal range, i.e. an expanding ring-like blob, which is, however, unlikely.

With more X-ray data in the future, in particular of high-cadence where persistent low-outflow epochs can clearly be distinguished from enhanced-outflow epochs, the perturber model can further be tested for different orbital elements, in particular eccentricity, which results in different rates of Schwarzschild precession in the orbital plane. Therefore, it would modulate the exact timing of enhanced absorption events as well as the outflow velocity. In case the enhanced absorption events disappear for a certain period of time, it would be an indication of the precession of the line of nodes due to the Lense-Thirring effect, i.e. the perturber-induced outflow footpoint would essentially precess as well depending mostly on the SMBH spin.

\subsection{Formation channels for IMBHs and IMBH-SMBH pair statistics}
\label{subsec_formation_channels}
IMBHs can be formed via different channels, namely two basic formation mechanisms are (i) cosmological/primordial related to the direct collapse of gaseous clouds or remnants of population III stars \cite{2001ApJ...551L..27M,2014MNRAS.443.2410F} and (ii) repeated stellar and black-hole collisions and the subsequent growth by accretion and/or mergers inside massive stellar clusters (runaway scenario); see \cite{2020ARA&A..58..257G} for a review. Channel (ii) can lead to an increased occurrence of IMBHs in NSCs with respect to the rest of host galaxies due to
\begin{itemize}
    \item gradual build-up of NSCs via the infall of massive stellar clusters hosting IMBHs \cite{1969ApJ...158L.139S,2002ApJ...576..899P,2004cbhg.symp..138R,smbh_imbhrates},
    \item a series of mergers of stellar black holes \cite{2022ApJ...927..231F} or stellar black holes with other stars in the NSC \cite{2022ApJ...929L..22R}.
\end{itemize}
Thanks to the deeper gravitational potential of NSCs in comparison with other stellar environments, IMBHs as merger products can be retained and accumulated within galactic nuclei since the received post-merger recoil kicks are typically less than the required escape velocity from the NSC. Subsequently, they can form tight pairs with the SMBH as they descend within the SMBH sphere of influence on the dynamical friction timescale. The dynamical friction timescale is especially short for perturbers significantly more massive than field stars. In case the IMBH is moving at a comparable speed with respect to field stars at large distances from the SMBH, i.e. $v_{\rm per}\sim \sigma_{\star}$, then the dynamical friction time can be estimated as
\begin{align}
    T_{\rm df}&=\frac{3}{8}\sqrt{\frac{2}{\pi}} \frac{\sigma_{\star}^3}{G^2 \rho_{\star} m_{\rm per} {\rm ln}\Lambda}\,\notag\\
   & \approx 4\,\left(\frac{\sigma_{\star}}{135\,{\rm km\,s^{-1}}}\right)^3\left(\frac{\rho_{\star}}{5.8\times 10^4\,M_{\odot}{\rm pc^{-3}}} \right)^{-1}\left(\frac{m_{\rm per}}{10^4\,M_{\odot}} \right)^{-1}\left(\frac{{\rm ln}\Lambda}{17} \right)^{-1}{\rm Myr}\,,
   \label{eq_dynamical_friction_time}
\end{align}
where $\sigma_{\star}\sim 135\,{\rm km\,s^{-1}}$ is the stellar velocity dispersion estimated from $M_{\bullet}$-$\sigma_{\star}$ relation (i.e. \cite{2015ApJ...801...38W} for $M_{\bullet}=10^{7.4}\,M_{\odot}$), $\rho_{\star}$ is the stellar mass density inside the sphere of the SMBH gravitational influence (considering $M_{\star}\sim 2\times 10^{7.4}\,M_{\odot}$ inside the sphere of influence given by the velocity dispersion), and ${\rm ln}\Lambda$ is the Coulomb logarithm (${\rm ln}\Lambda\sim {\rm ln}(M_{\bullet}/M_{\odot})\sim 17$). 

%In case ASASSN-20qc is similar to the proposed set-up of GW190521 (merger of a black-hole binary in the accretion disk), the estimated event rate of such gravitational-wave events is $0.13^{+0.30}_{-0.11}\,{\rm Gpc^{-3}\,yr^{-1}}$ \cite{PhysRevLett.125.101102}, or $0.0088^{+0.0204}_{-0.0075}\,{\rm yr^{-1}}$ cumulative rates using the comoving volume within $z=0.06$, which gives $88^{+204}_{-75}$ SMBH-IMBH tight pairs if one considers the merger timescale of $\sim 10^4$ years.

There are several ways to estimate potential number of SMBH-IMBH sources within the redshift of \target, $z\lesssim 0.06$. Using the $N$-body model of the globular cluster disruptions within the NSC and subsequent interactions of the IMBHs with the SMBH, the SMBH-IMBH comoving merger rate estimate is $\Gamma_{\rm AC}=0.03\,{\rm Gpc^{-3}\,yr^{-1}}$ \cite{2019MNRAS.483..152A} within the local Universe. This translates into $N_{\rm SMBH-IMBH}=\Gamma_{\rm AC} V_{\rm com}(<z)\tau_{\rm merge}\sim 20.4$ pairs for the comoving volume within $z=0.06$ and their merger timescale of $\tau_{\rm merge}\sim 10^4$ years, during which a TDE is likely to occur. Within the semi-analytical framework of the disruptions of globular clusters hosting an IMBH, the SMBH-IMBH comoving merger rate is smaller than the previous estimate, $\Gamma_{\rm F}\sim 10^{-5}-3\times 10^{-4}\,{\rm Gpc^{-3}\,yr^{-1}}$ \cite{smbh_imbhrates}, where the range corresponds to the IMBH occupation fraction between 0.1 and 1.0 inside globular clusters, which gives $N_{\rm SMBH-IMBH}=\Gamma_{\rm F} V_{\rm com}(<z)\tau_{\rm merge}\sim 0.006-0.2$ pairs that are about to merge within $10^4$ years (this could be still an upper limit for this model since there is no evidence for globular clusters hosting IMBHs; \cite{2020ARA&A..58..257G}). There is a wide range of the number of SMBH-IMBH pairs in the local Universe, however, they are all consistent within the uncertainties with at least one tight SMBH-IMBH pair with $\tau_{\rm merge}\sim 10^4$ years within the redshift of ASASSN0-20qc.   

In addition, one can consider an estimate of the number of suitable SMBH-IMBH tight pairs based on the assumption that each NSC contains at least one IMBH within the sphere of influence of the SMBH. In that case the number of potentially detectable SMBH-IMBH tight pairs can be estimated as,
\begin{equation}
   N_{\rm pair}\sim n_{\rm NSC}\nu_{\rm infall}\tau_{\rm merge}V_{\rm com}(<z)\,,
   \label{eq_pair_number}
\end{equation}
where $n_{\rm NSC}$ is an approximate number density of galaxies hosting an NSC, $\nu_{\rm infall}\equiv 1/T_{\rm df}$ is the infall rate of IMBHs within an NSC as given by the dynamical friction timescale (see Eq.~\eqref{eq_dynamical_friction_time}), $\tau_{\rm merge}$ is the SMBH-IMBH merger timescale (see Eq.~\eqref{eq_timescale_merge}), and $V_{\rm com}(<z)$ is the comoving volume within the redshift $z$. The number of galaxies per year that host a SMBH-IMBH pair and in which a TDE occurs can be expressed as follows
\begin{equation}
    N_{\rm pair,TDE} [{\rm yr^{-1}}]=N_{\rm pair}\dot{N}_{\rm TDE}\,,
\end{equation}
where $\dot{N}_{\rm TDE}$ is a mean TDE rate per galaxy.
These quantities can be estimated in the following way:
\begin{itemize}
    \item $n_{\rm NSC}\sim 0.037\,{\rm Mpc^{-3}}$ by integrating the Schechter function
    \begin{equation}
    \Phi(M_{\rm \star,gal})=\frac{\Phi_{\star}}{M_{\rm c}}\left(\frac{M_{\rm \star, gal}}{M_{\rm c}} \right)^{\alpha_{\rm c}}\exp \left(-\frac{M_{\rm \star, gal}}{M_{\rm c}} \right)\,,    
    \end{equation}
     over the range of galactic stellar masses $M_{\rm \star,gal}=10^8-10^{10}\,M_{\odot}$ (with $\Phi_{\star}=0.84\times 10^{-3}\,{\rm Mpc^{-3}}$, $M_{\rm c}=10^{11.14}\,M_{\odot}$, and $\alpha_{\rm c}=-1.43$ fixed; see e.g. \cite{smbh_imbhrates}), i.e. galaxies that have an NSC occupation fraction in the range $60\%-80\%$ \cite{2020A&ARv..28....4N}, 
%    \item $f_{\rm AGN}\sim 0.1$, i.e. $\sim 10\%$ of AGN in the local Universe,
%    \item $f_{I}\sim 0.2$ according to the fraction among SDSS sources close to $z=0.06$, see \cite{2017ApJ...850...74K},
    \item $\nu_{\rm infall}=1/T_{\rm df}$, where $T_{\rm df}\propto \sigma_{\star}^3/(\rho_{\star}m_{\rm per}) \sim M_{\bullet}^2/(\sigma_{\star}^3 m_{\rm per})$. Considering the range of SMBH masses, and related differences in the stellar velocity dispersion as well as in the SMBH influence radii, $T_{\rm df}(10^7\,M_{\odot})\sim 5\times 10^5$ years, $T_{\rm df}(10^{7.4}\,M_{\odot})\sim 3\times 10^6$ years, and $T_{\rm df}(10^8\,M_{\odot})\sim 40\times 10^6$ years for SMBHs of $10^7$, $10^{7.4}$, and $10^8\,M_{\odot}$, respectively, 
    \item $\tau_{\rm merge}\sim 10^4$ years, i.e. selection of SMBH-IMBH pairs that have a long enough merger timescale for a TDE to take place, i.e for $M_{\bullet}=10^7-10^8\,M_{\odot}$, the perturber mass is in the range $m_{\rm per}=26\,500-5700\,M_{\odot}$,
    \item $V_{\rm com}(<0.06)\sim 0.068\,{\rm Gpc^3}$, which is a comoving volume within $z=0.06$ for flat $\Lambda CDM$ with $\Omega_{\rm m}=0.3$. 
\end{itemize}
Inserting these estimates into Eq.~\eqref{eq_pair_number} results in $N_{\rm pair}\sim 53200-700$ SMBH-IMBH pairs considering $M_{\bullet}=10^7-10^8\,M_{\odot}$, with $N_{\rm pair}\sim 9200$ for $M_{\bullet}\sim 10^{7.4}\,M_{\odot}$ (for all the galaxies, considering both AGN and quiescent nuclei). Selecting those where a TDE takes place (considering the rate of $\dot{N}_{\rm TDE}\sim 10^{-4}\,{\rm yr^{-1}}$), we obtain $N_{\rm pair,TDE}=0.07-5.3$ sources per year ($M_{\bullet}=10^7-10^8\,M_{\odot}$) and $N_{\rm pair,TDE}=0.9$ sources per year for $M_{\bullet}\sim 10^{7.4}\,M_{\odot}$. Hence, out of the total number of galaxies in a given cosmological volume ($\sim 2.5$ million galaxies), we expect the TDE occurrence in a galaxy hosting the SMBH-IMBH pair in 1 out of $\sim 5\times 10^5-36\,\times 10^6$ galaxies.  

Using the TDE rate of $\dot{N}_{\rm TDE}\sim 10^{-4}\,{\rm yr^{-1}}$ per galaxy, we can estimate the timescale on which it is expected we detect a TDE flare in galaxies hosting a tight SMBH-IMBH pair every $\tau_{\rm TDE-IMBH}\sim (N_{\rm pair}\dot{N}_{\rm TDE})^{-1}\sim 0.2-14.3$ years for the whole range $M_{\bullet}=10^7-10^8\,M_{\odot}$, with $\tau_{\rm TDE-IMBH}\sim 1.1$ years for $M_{\bullet}=10^{7.4}\,M_{\odot}$. 

From an observational point of view, the number of sources with IMBH-induced QPOuts is $N_{\rm QPOut}=f_{\rm I}f_{\rm inc}N_{\rm pair}$, i.e. from the total number of galaxies with tight SMBH-IMBH pairs we are selecting those that we observe sufficiently close to the rotation axis (the fraction $f_{I}$, i.e. sources with the viewing angle less than $45^{\circ}$ from the rotation axis so that the accretion flow is viewed close to face-on) as well as at the same time, the IMBH is highly inclined so that the ejected blob obscures the underlying flow (the fraction $f_{\rm inc}$, i.e. sources with the IMBH inclined at more than $45^{\circ}$ from the equatorial plane). Assuming the uniform distribution of viewing-angle/inclination cosines, we get $f_{\rm I}\sim 0.71$ and $f_{\rm inc}\sim 0.29$, which yields $N_{\rm QPOut}\sim 0.21 N_{\rm pair}$ or 11\,000--150 sources hosting SMBH-IMBH pairs can be revealed via QPOuts, i.e. one in 250 up to 5000 galaxies can exhibit QPOuts triggered by an IMBH (massive perturber).

Although there is a large uncertainty of nearly two orders of magnitudes in terms of the TDE occurrence in NSCs hosting a SMBH-IMBH pair, crude estimates provided here show that such an event is not entirely unlikely given the long-term monitoring of nearby AGN. We note that according to Eq.~\eqref{eq_pair_number} the number of expected SMBH-IMBH pairs does not depend e.g. on the considered merger timescale or the perturber mass since both $\tau_{\rm merge}$ and $T_{\rm df}$ are inversely proportional to $m_{\rm per}$, though in reality there likely is a dependency considering the fact that e.g. black holes of different masses are produced via different formation channels. On the other hand, the estimate of $N_{\rm pair}$ can be considered as a lower limit since we considered only AGN, while the TDE phenomenon occurring in NSCs hosting an IMBH is also relevant for quiescent nuclei, such as the Galactic center. Considering sources at an even larger redshift would also significantly extend the sample of the sources similar to \target.         

\subsection{Inclination and Distance of the IMBH}
A higher inclination of the perturber with respect to the SMBH's accretion disk is required to perturb the region close to the funnel/disk boundary where the material can be pushed into the outflow region and further accelerated by the ordered magnetic field in the funnel. When the IMBH migrates to the innermost regions of a few$\times$100 to $\sim$1000 gravitational radii from the NSC, the inclination distribution of the IMBH perturbers can be broad with a probability of $\sim 0.7$ for the perturber to be inclined between $45$ and $135$ degrees, assuming the isotropic distribution of orbits within the NSC, i.e. the uniform distribution of inclination cosines. Hence, inclined orbits with respect to the accretion disk are generally more likely than aligned orbits for an IMBH from within the NSC or beyond due to the isotropic massive cluster infall. In case there is a population of aligned compact remnants within the disk plane, e.g. due to the migration trap \cite{2016ApJ...819L..17B}, then the inclination can be increased due to (a) recoil velocity kick due to black hole-black hole merger or (b) Kozai-Lidov eccentricity-inclination oscillations due to the presence of a massive body/disk at larger distances.

Case (a) is based on the fact that gravitational waves carry away linear momentum flux, hence during the binary black hole merger and the ring-down, the merger product receives a recoil velocity kick that can reach several hundred to thousand km/s \cite{2007ApJ...662L..63S}. Assuming initially a circular orbit for a black-hole binary as well as a circular orbit for the inclined orbit of the formed IMBH, the required velocity kick to change the orbital inclination by $\Delta \iota$ is $\Delta v_{\rm kick}\sim 2v_{\rm orb}\sin{(\Delta \iota/2)}$, where the orbits are also assumed to have a comparable semi-major axis and $\Delta v_{\rm kick}$ is perpendicular to the orbital velocity vector. For the inclination change of $\Delta \iota=60^{\circ}$ from the disk plane to the inclined orbit crossing the disk, the required velocity kick is approximately equal to the orbital velocity, $\Delta v_{\rm kick}\approx v_{\rm orb}$. For $\Delta v_{\rm kick}\approx 1000\,{\rm km\,s^{-1}}$ and general $\Delta \iota$, this implies the effective distance from the SMBH where gravitational-wave recoil kicks can lead to highly inclined orbits, 
\begin{align}
   r_{\rm inc} &= \left[2\frac{c}{\Delta v_{\rm kick}} \sin{\left(\frac{\Delta \iota}{2} \right)}\right]^2\notag\\
    &=5.3\times 10^4\left(\frac{\Delta v_{\rm kick}}{1000\,{\rm km\,s^{-1}}} \right)^{-2}\sin^2{\left(\frac{1}{2}\frac{\Delta \iota}{45^{\circ}}\right)}\,r_{\rm g}\,,\label{eq_rinc}
\end{align}
which is at least two orders of magnitude further than the inferred distance of the IMBH perturber. The merger timescale given by Eq.~\eqref{eq_timescale_merge} for the distance $r_{\rm inc}$ in Eq.~\eqref{eq_rinc} is much longer than other relevent dynamical timescales, $\tau_{\rm merge}\sim 2.35 \times 10^{16}\,{\rm yr}$. It is therefore more likely that the inclination increased due to repetitive mergers. For illustration, mergers taking place at $440\,r_{\rm g}$ with the average velocity kicks of $\sim 500\,{\rm km\,s^{-1}}$ can change the inclination step-wise by 2 degrees. Hence, between 10 to 100 consecutive mergers are needed for the initial stellar black hole to increase its mass to the intermediate mass range, while at the same time the inclination changes can add up to reach high values above the disk plane, depending on the mass ratios and spin distribution of merging black holes. A single high kick velocity of $\Delta v_{\rm kick}\sim 5000\,{\rm km\,s^{-1}}$ indicated by the recent gravitational-wave event analysis \cite{2022PhRvL.128s1102V}, which is perpendicular to the orbital plane, can increase the inclination by $17^{\circ}$ with respect to the disk plane at the orbital distance of $\sim 300\,r_{\rm g}$ where the migration trap with several accumulated stellar black holes can be located \cite{2016ApJ...819L..17B}.

\begin{sloppypar}
Case (b) process -- Kozai-Lidov mechanism -- is based on the preservation of the $z$-component of the specific angular momentum in the inner three-body problem. Specifically, $(1-e_{\rm per}^2)^{1/2}\cos\iota_{\rm per}$ is constant, which implies the periodic oscillations from highly-inclined circular orbits to disk-embedded eccentric orbits. However, for the case of ASASSN-20qc and the likely perturbation by the distant dusty torus with the mass of $m_{\rm t}=0.1M_{\bullet}\sim 10^{6.4}\,M_{\odot}$ and the torus distance of $r_{\rm t}\sim 10\,{\rm pc}$ \cite{2019ApJ...884..171H}, the oscillation timescale is longer than the Hubble time at the current distance of the IMBH \cite{2005A&A...433..405S}, 
\begin{align}
   T_{\rm KL}&\sim 4\pi^2 \frac{(GM_{\bullet})^{1/2}}{Gm_{\rm t}}\frac{r_{\rm t}^3}{r_{\rm per}^{3/2}}\,,\notag\\
   &=8\pi^3 \frac{r_{\rm t}^3}{Gm_{\rm t}P_{\rm orb}}\,\notag\\
   &\simeq 10^{15}\left(\frac{r_{\rm t}}{10\,{\rm pc}} \right)^3 \left(\frac{m_{\rm t}}{10^{6.4}\,M_{\odot}} \right)^{-1}\left(\frac{P_{\rm orb}}{8.05\,{\rm d}} \right)^{-1}\,{\rm yr}\,,
   \label{eq_KL_timescale}
\end{align}
which implies that an inclined IMBH will likely stayed inclined during the inspiral unless the IMBH experiences an orbital decay due to the disk drag. The Kozai-Lidov timescale can be shortened in case the IMBH would be perturbed by another massive body closer than the torus, which is, however, speculative.

\end{sloppypar}

The grinding mechanism was explored extensively in the context of orbiting bodies around the SMBH.
\cite{1991MNRAS.250..505S} were the first to recognize potential importance of hydrodynamical interaction between stars and accretion disk in the nuclei of active galaxies and quasars fed via accretion. Starting from order-of-magnitude arguments and semi-analytical estimates these authors suggested that a long-term effect of the interaction should lead to secular changes of the stellar trajectories around a supermassive black hole: circularization (decrease of the osculating eccentricity) of the orbits and their monotonic sinking towards the center (decay of the semimajor axis) accompanied by the gradual ``grinding'' (decrease of inclination). The process eventually brings stars into the accretion disk plane. Relevant time-scales depend on several factors, most importantly, the ratio between the surface density of the accretion disk to the projected surface density of the star. For example, in the case of standard-type, geometrically thin, planar accretion disk \cite{1973A&A....24..337S}, the orbital decay takes place over a vast range of $10^4$--$10^7$ Keplerian periods of the stellar orbiter at the corresponding distance from the SMBH \cite{1998MNRAS.293L...1V}. After that time the star should become fully embedded into the accretion disk. Let us note that the effective cross-sectional area for the hydrodynamical interaction of a $10^5M_{\odot}$ IMBH is comparable with that of a solar-type star usually considered in the quoted papers.

Evaluation of the grinding time requires to specify the accretion disk density profile as a function of radius; see eq. (24) in \cite{1999A&A...352..452S}. For example, for a solar-type orbiter at radius $10^4R_{\rm g}$ and the standard accretion disk with viscosity parameter $\alpha\approx10^{-3}$, accretion rate $\dot{M}_{\bullet}=1M_{\odot}$ per year one obtains the grinding time about $10^6$ revolutions, corresponding to $\simeq10^8$ years. Precise temporal dependencies can be determined by direct numerical evaluation. 

The above-mentioned simplistic scenario is based on a number of assumptions, which were further studied by various authors. First of all, the collisions should not expel excessive amount of gaseous material out of the accretion disk \cite{1994ApJ...434...46Z,2002A&A...387..804V}, thereby threatening the accretion system. On the other hand, if the accretion flow becomes self-gravitating and its total mass is not negligible with respect to the central black hole, the orbiters are affected by Kozai-Lidov mechanism that causes continued oscillations of the osculating elements (eccentricity and inclination) \cite{1998MNRAS.298...53V,2005A&A...433..405S,2020ApJ...901..170L}, see Eq.~\eqref{eq_KL_timescale} for the Kozai--Lidov timescale. Hence, the perturber does not align  with the disk plane. Similarly, a secondary massive black hole could induce a non-spherical perturbation of the central gravitational field, so that the orbital decay of the stellar satellites proceeds via continued exchange between eccentricity and inclination. 

%The monotonic decrease of eccentricity and inclination is overlaid with faster oscillations if the perturbation to the central field is non-negligible. Furthermore, the satellites inclination converges to a different distribution (instead of a strictly flattened disk) when the two-body gravitational relaxation is taken into account \cite{2002A&A...387..804V}.

Several authors explored the process of orbital decay due to emission of gravitational waves by a gradually sinking stellar-mass orbiter; see, e.g., \cite{2000ApJ...536..663N,2001A&A...376..686K,2010MNRAS.402.1614D}, and further references cited therein. They find that the direct hydrodynamical interaction with an accretion disk plays a dominant role in the orbital evolution of such satellites. However, geometrically thick ADAF structure is less efficient in grinding the inclination as the projected surface density is {\em orders of magnitude lower} than that of a standard accretion disk; the corresponding time-scales of orbital evolution are thus proportionally longer; at larger distances of $\gtrsim10^2R_{\rm g}$ it exceeds the Hubble time (even if it was argued \cite{1996ApJ...470..652Z} that turbulence within the medium may enhance the drag effects). IMBH has about the same effective cross-sectional area for the hydrodynamical interaction with gas as a solar-mass star but the mass of IMBH is five orders of magnitude higher than a normal star. Hence, the effect of collisions is weaker by that factor.

Here we remind the reader that in the case of \target during the outburst, the outflow rate at maxima reaches the value of $\dot{m}_{\rm out} \simeq 0.002 M_{\odot}/yr$ (see Sect.\ \ref{sec:energetics}), while the accretion rate during the outburst was about thousand times higher than in the previous quiescent time. Assuming that the expelled amount of gas scales with the density of the accretion flow, the pre-outburst outflow rate can be estimated $\sim 10^{-6} M_{\odot}/yr$. Hence, the ratio of the mass expelled by the IMBH during one year to its own mass is $\Delta m_{\rm out}/m_{\rm per}\sim 10^{-10}$ and thus the dynamical effect on the IMBH orbit is negligible. 

A quantitative comparison between the outer standard accretion disk and the inner ADAF can be done based on the analytical estimate of the grinding timescale as derived by \cite{1991MNRAS.250..505S},
\begin{equation}
    t_{\rm grind}\sim \chi \frac{m_{\rm per}}{\Delta m} \frac{1}{(\Omega_{\rm K}P_{\rm orb})^{2/3}}P_{\rm orb}\,,
    \label{eq_grind_timescale}
\end{equation}
where $\Delta m\sim \rho_{\rm flow}h_{\rm flow} \mathcal{R}^2$ is the mass pushed from the accretion flow with the mass density of $\rho_{\rm flow}$ and the scale-height of $h_{\rm flow}$ by the perturber with the influence radius $\mathcal{R}$. In Eq.~\eqref{eq_grind_timescale}, the Keplerian angular velocity at the perturber distance $r_{\rm per}$ is denoted as $\Omega_{\rm K}$ and the corresponding orbital period by $P_{\rm orb}$. The numerical factor $\chi$ is set to 5 according to \cite{1991MNRAS.250..505S}. Considering the thick ADAF with the scale-height to radius ratio close to unity, we can set $h_{\rm flow}\sim r_{\rm per}$. Furthermore, we set the viscosity parameter to $\alpha=0.1$, the accretion rate normalized with respect to the Eddington rate to $\dot{m}=0.05$, and the perturber influence radius to $\mathcal{R}=3\,r_{\rm g}$ (its mass is set to $m_{\rm per}=10^4\,M_{\odot}$, while the SMBH mass is kept at $M_{\bullet}=10^{7.4}\,M_{\odot}$). At the distance of $r_{\rm per}\sim 93\,r_{\rm g}$, the grinding timescale for the ADAF flow is $t_{\rm grind}\sim 4.6\times 10^9$ years, while for the standard thin disk, the grinding timescale is $t_{\rm grind}\sim 10^4$ years. Hence, for the ADAF, the timescale for the perturber alignment is $4.5 \times 10^5$ times longer than for the standard thin disk, independent of the perturber distance from the SMBH. The radial profiles of the grinding timescales for both the ADAF and the standard disk are shown in Fig.~\ref{fig_Hill_syn_radius}. It is clear that the grinding timescale for the IMBH-ADAF interaction is always larger than the merger timescale for the distance range of interest. This suggests that an initially inclined orbit of the IMBH with respect to the ADAF will not become aligned with the equatorial plane of the flow during the inspiral. The situation is more complex for the mixed flow with the thermal component since then the grinding timescale is shorter than the merger timescale outside $100\,r_{\rm g}$ (however, it is longer inside $\sim 100\,r_{\rm g}$, see Fig.~\ref{fig_Hill_syn_radius}). However, the detailed inclination evolution depends on the initial conditions and the ADAF extent. It would also be necessary to perform numerical calculations following orbital evolution during multiple passages through the standard disk and then the inner ADAF, which is beyond the scope of the current study.

There could be physical reasons why the SMBH-IMBH pair is detected at the separation of $100\,r_{\rm g}$ that we briefly discuss here. One is related to the variable density of the accretion flow along the radial direction, specifically the accretion disk transitions from the standard disk to ADAF at the radius given by Eq.~\eqref{eq_ADAF_radius} or \eqref{eq_evap_radius}. In addition, this radius is variable due to changing accretion rate with time. The density of the flow directly affects the hydrodynamic drag timescale, i.e. the e-folding timescale of the IMBH specific angular momentum \cite{2000ApJ...536..663N}, 
\begin{equation}
    t_{\rm hd}=\frac{v_{\rm per}v_{\rm rel}^2}{4\pi I G^2 m_{\rm per}\rho_{\rm flow}}\,,
    \label{eq_hydro_timescale}
\end{equation}
where $v_{\rm per}$ is the Keplerian velocity of the perturber, $v_{\rm rel}\approx(2GM_{\bullet}/r)^{1/2}$ is the relative velocity with respect to the gas, $\rho_{\rm flow}$ is the mass density of the accretion flow, which can either correspond to a standard thin disk at larger distances or an ADAF closer in, and the factor $I$ was already defined below Eq.~\eqref{eq_synchronization_radius}. To obtain order-of-magnitude quantitative comparison, we adopt $\dot{m}=0.05$ for the accretion rate, the radiative efficiency is set to 0.1 and the viscosity parameter to 0.1 for both types of accretion flows. For the IMBH mass of $m_{\rm per}=10^{4}\,M_{\odot}$, the ratio of timescales is $\sim 2\times 10^8$ at $93\,r_{\rm g}$, and it does not evolve much with distance. Hence, when the disk transitions to ADAF, the hydrodynamic drag is eight orders of magnitude weaker and the e-folding timescale is prolonged from $\sim 300$ years to $\sim 5\times 10^{10}$ years, which results in the IMBH stalling within the ADAF flow. The further orbital decay is then dominated by the gravitational emission. We depict the radial dependency of hydrodynamic timescales in Fig.~\ref{fig_Hill_syn_radius}, where we also show the grinding timescales corresponding to the inclined IMBH, its merger timescale, the Kozai-Lidov timescale due to the AGN torus perturbation, and general relativistic Schwarzschild and Lense-Thirring timescales.

Another mechanism that can lead to the IMBH orbiting at $\sim 100\,r_{\rm g}$ is the potential formation of migration traps in AGN disks, which are regions of equilibrium orbits with zero torque from the differentially rotating disk, or in other words, these are the distance ranges where the inward migration of embedded objects meets the outward migration. Bellovary et al. \cite{2016ApJ...819L..17B} used two steady-state, analytical disk solutions and derived the location of the migration trap between 40--600 $r_{\rm g}$, i.e. the region where the putative IMBH perturber in the \target source is located. Within the migration trap, stellar black holes tend to accummulate, merge and scatter, which naturally leads to the efficient IMBH formation within 10 Myr. Mergers and dynamical scattering could eventually naturally create inclined orbits with respect to the underlying disk.

\subsection{Caveats of the perturber--accretion disk interaction model}\label{sisec:caveats}

The basic caveat of 2D simulations in this study is the progressive weakening of the magneto-rotationally instability due to the lack of the toroidal flow. This is noticeable for the times later than $\sim 30\,000GM_{\bullet}/c^3$. 3D simulations do not exhibit such a problem, however, they are computationally much more demanding for studying perturbations at $\sim 100\,r_{\rm g}$ due to much longer orbital period; this was the reason for performing the 3D runs with the perturber orbiting at 10\,$r_{\rm g}$, for which a few cycles could have been studied. We tested the effect of the disappearing MRI in 2D runs by placing the perturber earlier to the state resulting from the unperturbed 2D flow -- at 20\,000$r_{\rm g}/c$ (Run 1) and later at $50\,000\,r_{\rm g}/c$. In both cases, periodic outbursts in the outflow rate, i.e. QPOuts, are always present regardless of the MRI that is progressively weakening. This indicates that the QPOuts are induced by the perturber and the background accretion flow is less relevant, i.e. even at late stages when MRI is suppressed, the perturber still causes QPOuts that appear comparable as to when the MRI is still active. However, here we note that after $50\,000\,r_{\rm g}/c$ or 71.5 days the mass-accretion rate drops consistently by two orders of magnitude with respect to the initial value, which is indicative of the magnetorotational instability weakening and hence the comparison of the simulation data with realistic accretion flows is limited at this stage. In contrast to the weak dependence of the fast outflow formation on the background accretion flow, the ordered and stable poloidal magnetic field in the funnel region appears necessary for accelerating the ejected blobs to relativistic velocities, therefore ordered poloidal magnetic field is a crucial element of the model.

2D and 3D GRMHD simulations presented here and performed using the HARM code neglect radiative feedback and radiative cooling. In case the inner part of the accretion flow transitions into the hot radiatively inefficient accretion flow (of ADAF type) {\bf at the radius comparable to or larger than the IMBH distance}, this does not pose a problem. However, in case the inner disk does cool radiatively and still contains a thermal component, the accretion-disk evolution, perturbed by the IMBH, would deviate from the one presented here. 
Indeed, in case of a geometrically thin accretion disk, the dynamics of outflows is (almost) symmetrical with respect to the upper and lower hemispheres. This situation was originally modelled within the hydrodynamic framework \cite{Ivanov1998}. 
Considering both intersections of the trajectory with the disk plane would increase the effective rate of star-disk collisions and thus it would influence the model parameters, although the overall qualitative picture of the scenario remains unchanged. Moreover, it appears that strong jets powered by the MAD state and the associated semiregular magnetic eruptions  cannot be sustained by a geometrically thin disk \cite{Liska2022,Curd2023}. On the other hand, the backflow is expected to be relatively weaker for geometrically thick, magnetized tori, which we expect to persist from the prior low-luminous phase. Even if some material is expelled from the disk in the backward direction with respect to the orbiter's velocity vector, the forward push dominates. Furthermore, the flares due to shocks as the star impacts the thin standard disk should be revealed in the continuum emission (X-ray, UV, optical), which is not seen in this source (see also estimates below). Instead, we see quasiperiodic enhancements in the absorbing material that is ultrarelativistic. The perturber-thin disk interaction assumes a standard, optically thick disk, where the shock is more prominent due to a denser, cooler gas. In our model, we argue that the interaction takes place in the warmer and much more diluted ADAF component that is present due to a previous low-luminosity state of ASASSN-20qc, or more precisely, the standard outer disk transitions into ADAF outside the orbit of the perturber because of the lower Eddington ratio of $\sim 0.05$. 

{\bf In addition, if we adopt the thin-disk setup, a star on a circular orbit would cross the disk twice, hence $P_{\rm orb}\sim 16.6$ days. This implies the semi-major axis of $a_{\star}/r_{\rm g}\sim 150.62\,r_{\rm g}$ and the Keplerian velocity of $v_{\rm K}\sim c/\sqrt{a_{\star}/r_{\rm g}}\sim 0.08c$. The shocked gas should expand above and below the disk plane at the velocity of $v_{\rm sh}\sim \sqrt{2}v_{\rm K}\sim 0.1\,{\rm c}$, which is below the detected QPOut velocity of $0.3c$. Hence, the model would need some fine-tuning in terms of a significantly higher eccentricity, implying that QPOuts take place only at the pericenter. In that case, the orbital period is $P_{\rm orb}\sim 8.3$ days as originally assumed and the semi-major axis is $(a_{\star}/r_{\rm g})\sim 95\,r_{\rm g}$. The eccentricity of the orbit would have to be $e\sim 0.7$ to reach shock speeds of $v_{\rm sh}\sim \sqrt{2}v_{\rm K}\sim 0.35c$. However, in that case, the pericenter of the orbit would be $28.5\,r_{\rm g}$, which would result in more profound quasiperiodic variability of the inflowing matter, which should be revealed in the X-ray domain. However, no significant quasiperiodicity of the X-ray continuum emission was found. In general, we do not expect significant periodic X-ray flux enhancements due to either IMBH or star interactions with the accretion flow, given the orbital radius of $\sim 100\,r_{\rm g}$ of the perturber corresponding to the QPOut periodicity of $8.05$ days. When we use the model set-up involving the standard disk as in \cite{2023ApJ...957...34L}, the X-ray flares due to the launched shocked ejecta are below the quiescent X-ray level of ASASSN-20qc (see also the estimates above in Section~\ref{msec:modeldiscussion}). X-ray flares due to shocks are expected to be even fainter for the case of a more diluted ADAF.}  Moreover,
 the constraint for the perturber influence radius is expected to differ {\bf for the standard-disk scenario}, though the relations for the influence radius--perturber mass correspondence, see Eqs.~\eqref{eq_hill_radius} and \eqref{eq_synchronization_radius}, would still hold since they do not depend explicitly on the accretion disk density. Also, stars and pulsars interacting with the denser disk would produce even smaller stagnation radii, hence their potential to reproduce the observed ultrafast outflow would be even smaller.

Observed X-ray spectra indicate the ASASSN-20qc accretion disk likely consists of both a colder, thermal component further away and a warmer, diffused ADAF-like comptonizing medium in the inner region. This is implied by the best-fit black-body temperature of the thermal continuum, $kT\sim 0.086\,{\rm keV}$, which corresponds to $T\sim 998\,000\,{\rm K}$. This temperature is higher than the characteristic temperature corresponding to the standard thin disk around the SMBH of $10^7-10^8\,M_{\odot}$, $T(10^7\,M_{\odot})\sim 274\,000\,{\rm K}$ ($0.024\,{\rm keV}$) and $T(10^8\,M_{\odot})\sim 87\,000\,{\rm K}$ ($0.007\,{\rm keV}$), which is accreting with the Eddington ratio of $0.1-0.01$, respectively, with $10\%$ radiative efficiency and the inner radius at $6\,r_{\rm g}$. For the range of the SMBH masses and Eddington ratios of $\lambda_{\rm Edd}\lesssim 0.1$, the standard thin disk is expected to make a transition to the hot ADAF, e.g. due to thermal conduction, at the radius of $R_{\rm evap}\sim 132-448\,r_{\rm g}$, see Eq.~\eqref{eq_evap_radius}. The ADAF is characterized by the nearly virialized temperature profile for ions, $T_{\rm i}\sim 3.6\times 10^{12}(r/r_{\rm g})^{-1}\,{\rm K}$ ($313\,{\rm MeV}(r/r_{\rm g})^{-1}$), while the electrons cool down via bremsstrahlung, synchrotron, and Compton processes to $T_{\rm e}\sim 10^9\,{\rm K}$ ($86\,{\rm keV}$). This hot diluted medium can then serve as the comptonizing environment for the underlying thermal disk emitting softer photons. Given the Eddington rate of $\sim 0.05-0.5$ during the X-ray outburst, and the low Eddington rate of $2\times 10^{-5}$ before that, ASASSN-20qc likely transitioned from the hard to the soft state, and back again, with the ADAF region shrinking and extending again with the change in the accretion rate according to Eq.~\eqref{eq_ADAF_radius}. Its outburst Eddington ratio also appears to indicate that the source could be found in the intermediate state, i.e. when the ADAF recondenses back to the thin disk at the recondensation radius of $r_{\rm con}=25 (\alpha/0.2)^{-28/3} (\dot{m}/0.01)^{8/3} r_{\rm g}$ as the accretion rate decreases \cite{2007A&A...463....1M}. The inflection point of the 2-10 keV photon index-Eddington ratio correlation lies close to $\lambda_{\rm Edd} \sim 0.05$ \cite{2009MNRAS.399..349G}, which indicates the mixed, non-typical accretion state of \target sharing the properties of both low-luminosity and high-luminosity AGN. The peak of the Eddington-ratio distribution of changing-look AGN also lies close to $\lambda_{\rm Edd} \sim 0.01$ \cite{2022arXiv220610056P}. The thin disk solution including general relativistic effects, in particular the Kerr metric, can indeed address the higher temperature of $0.086$ keV for a highly accreting SMBH (see Methods section \ref{sisec:temp}). Such a thin disk emitting thermal X-rays is rather compact, extends within the tidal radius of the tidally disrupted star, and is embedded in the pre-existing optically thin ADAF. 

In summary, to fully capture the dynamics of the perturbed flow, we would need to address the increased accretion rate following the TDE, which presumably leads to the inner thin disk formation on the length-scale of the tidal radius that cools radiatively. The radiatively inefficient ADAF component, which is a remnant of the previous low-luminosity state, continues on the length-scales beyond the tidal radius. It is likely that the accretion flow transitions to the standard thin disk further out as implied by the detected broad lines indicative of the standard disk \cite{2004A&A...428...39C} because of the established radius-luminosity relations.

The assumption of fully synchronized motion of the gas with the perturber within the influence radius does not capture the detailed structure of the flow in the vicinity of the secondary body. In particular, we do not follow the radial outflow of the stellar wind in case of a star, nor the motion of the gas directly governed gravitationally by the secondary inside the bow shock and wake in case of a black hole. The velocity distribution may be quite complex and may change on a small length scale compared with the influence radius, which is,however, beyond the resolution of our simulations. However, since the bow shock is comoving with the perturber, its main dynamical effect on the surrounding medium is captured reasonably well by our simulations. 

In spite of several simplifications in our GRMHD simulations, the lead-order general relativistic magnetohydrodynamical effects of the pertuber-disk interaction are captured with a sufficient precision and imply the recurrent ultrafast outflow generation at the perturber-disk interaction site as long as the magnetic field is sufficiently ordered in the inner region. The ejected blobs are outflowing along the funnel-accretion disk boundary, where they cross the line of sight and cause periodic absorption of the underlying thermal disk.  

In addition, since in the inclined perturber model, the IMBH is expected to be misaligned with respect to the equatorial plane, its orbit would precess due to the frame-dragging, as we discussed in Subsection~\ref{subsec_directional_stability}. Though the $90^{\circ}$ Lense-Thirring timescale, during which the outflow is expected to change the direction away from the observer, is at least an order of magnitude longer than the observed QPOut cycle (2255 days for the SMBH spin of 0.4, see Eq.~\eqref{eq_LT_timescale}), it could in principle lead to the cessation of QPOuts within a few year timescale, especially for a fast rotating black hole (in 1002 days or 2.7 years, the outflow direction would change by $90^{\circ}$ for the spin of $a=0.9$). In case QPOuts would cease to be detected within several years, it would strengthen the case for the perturber-induced scenario, and it would provide a unique way to constrain the SMBH spin (given that the SMBH mass is well constrained). On the other hand, the ejected absorbing gas clump will also expand as it is accelerated downstream, making the perturber-induced outflow less sensitive to the orbital orientation. In addition, the blob trajectory is not necessarily linear as it is accelerated by the magnetic field, it can follow e.g. the helical magnetic-field configuration, which would lead to the absorbing events for any orientation of the perturber orbit.  However, high-cadence monitoring of \target is challenging because of the decrease in the X-ray flux density following the TDE. 

\clearpage
%%%%%%%%%%%%%%%%%%%%%%%%%%%%%%%%%%%%%%%%%%%%%%
%%%%%% radio/vast observations 
%%%%%%%%%%%%%%%%%%%%%%%%%%%%%%%%%%%%%%%%%%%%%%
\begin{table}[!htp]
    \centering
    \begin{tabular}{ccccc}
    Date & MJD & $\Delta t$ & $F_{\nu}$ & $\Delta F_{\nu}$ \\
dd/mm/yyyy &  & $\textrm{Days}$ & $\textrm{mJy}$ & $\textrm{mJy}$\\ [0.4ex] \hline
    04/05/2019 & $58607$ & $-596$ & $0.96$ & $0.34$ \\ [0.1ex]
27/08/2019 & $58723$ & $-480$ & $0.77$ & $0.25$ \\ [0.1ex]
29/10/2019 & $58786$ & $-417$ & $0.94$ & $0.24$ \\ [0.1ex]
30/10/2019 & $58786$ & $-416$ & $1.40$ & $0.25$ \\ [0.1ex]
10/01/2020 & $58859$ & $-344$ & $0.80$ & $0.47$ \\ [0.1ex]
24/01/2020 & $58873$ & $-331$ & $1.25$ & $0.22$ \\ [0.1ex]
25/01/2020 & $58874$ & $-330$ & $1.21$ & $0.21$ \\ [0.1ex]
20/06/2020 & $59020$ & $-183$ & $1.45$ & $0.25$ \\ [0.1ex]
28/08/2020 & $59090$ & $-113$ & $1.24$ & $0.24$ \\ [0.1ex]
24/07/2021 & $59419$ & $216$ & $1.35$ & $0.24$ \\ [0.1ex]
22/08/2021 & $59449$ & $246$ & $1.13$ & $0.22$ \\ [0.1ex]
        \end{tabular}
    \caption{\target's radio (VAST) observations. $\Delta t$ is the time since December 20, 2020. $F_{\nu}$ is the peak flux density in mJy, and $\Delta F_{\nu}$ is its uncertainty in mJy.}
    \label{tab:Radio_Observations}
\end{table}

\clearpage
% TABLE_TABLE_TABLE_TABLE_TABLE_TABLE_TABLE_TABLE_TABLE_TABLE_TABLE_TABLE
%                   Extended Data Table 1: Black Hole mass
% TABLE_TABLE_TABLE_TABLE_TABLE_TABLE_TABLE_TABLE_TABLE_TABLE_TABLE_TABLE
\begin{table}[htp!]
    \centering
    \resizebox{1\linewidth}{!}{%
    \begin{tabular}{l l l l l l l l} \hline
         Quantities   &   2021/01/11            &     2021/07/30       & 2021/08/19       & 2021/11/09       & 2022/01/25       & 2022/03/10 &  2022/08/26 \\\hline
L5100                      & 43.97 $\pm$ 0.01   &   43.84 $\pm$ 0.01   & 43.79 $\pm$ 0.01 & 43.66 $\pm$ 0.01 & 43.60 $\pm$ 0.01 & 43.60 $\pm$ 0.01 & 43.56$\pm$0.01\\
%$\alpha_{\lambda}$        & -1.76 $\pm$ 0.01   &   -0.5 $\pm$ 0.02    & 0.77 $\pm$ 0.02  & -0.56 $\pm$ 0.01 & -0.29 $\pm$ 0.01 & -0.29 $\pm$ 0.01 & -0.78 $\pm$ 0.02\\
FWHM(H$\beta_{bc}$)        & 2108  $\pm$ 183    &   2722 $\pm$ 204     & 2087 $\pm$ 396   & 2664 $\pm$ 75    & 2718 $\pm$ 56    & 2444 $\pm$ 53 & 3199 $\pm$ 96\\
L(H$\beta_{bc}$)           & 41.48 $\pm$ 0.03   &   42.14 $\pm$ 0.03   & 42.31 $\pm$ 0.02 & 42.14 $\pm$ 0.01 & 42.19 $\pm$ 0.01 & 42.21 $\pm$ 0.01 & 41.62 $\pm$ 0.06\\   
L(H$\beta_{nc}$)           & 40.60 $\pm$ 0.19   &   39.05 $\pm$ 1.63   & 40.54 $\pm$ 1.04 & 40.83 $\pm$ 0.04 & 41.04 $\pm$ 0.03 & 40.93 $\pm$ 0.04 & 40.83 $\pm$ 0.03\\
L(O[III]5007)              & 40.97 $\pm$ 0.07   &   41.18 $\pm$ 0.08   & 41.40 $\pm$ 0.04 & 41.17 $\pm$ 0.01 & 41.26 $\pm$ 0.01 & 41.32 $\pm$ 0.01 & 41.36 $\pm$ 0.08\\
L(HeI5876$_{bc}$)          & --                 &   41.76 $\pm$ 0.03   & 41.75 $\pm$ 0.01 & 41.43 $\pm$ 0.01 & 41.48 $\pm$ 0.01 & 41.48 $\pm$ 0.01 & 41.52 $\pm$ 0.03\\
FWHM(HeI5876$_{bc}$)       & --                 &   17724 $\pm$ 1914   & 9386 $\pm$ 113   & 5020 $\pm$ 56    & 6242 $\pm$ 134   & 6323  $\pm$ 278 & 22489 $\pm$ 2304\\
FWHM(H$\alpha_{bc}$)       & 2654 $\pm$ 441     &   3090 $\pm$ 151     & 2680 $\pm$ 211   & 3009 $\pm$ 77    & 2751 $\pm$ 553    & 2748 $\pm$ 103 & 2870 $\pm$ 8\\
L(H$\alpha_{bc}$)          & 41.81 $\pm$ 0.04   &   42.59 $\pm$ 0.04   & 42.95 $\pm$ 0.04 & 42.69 $\pm$ 0.01 & 42.84 $\pm$ 0.05 & 42.87 $\pm$ 0.04 & 42.65 $\pm$ 0.01\\
L(H$\alpha_{NA}$)          & 41.41 $\pm$ 0.02   &   41.55 $\pm$ 0.05   & 41.93 $\pm$ 0.06 & 41.61 $\pm$ 0.01 & 41.61 $\pm$ 0.27 & 41.67 $\pm$ 0.07 & 41.3 $\pm$ 0.01\\
L([NII]6585)               & 41.28 $\pm$ 0.03   &   41.45 $\pm$ 0.27   & 41.13 $\pm$ 1.47 & 41.21 $\pm$ 0.32 & -- & -- & 41.09 $\pm$ 0.01\\
$\log M_{\bullet}$       & 7.50 $\pm$ 0.08  &    7.66 $\pm$ 0.07   &  7.40 $\pm$ 0.16 & 7.54 $\pm$ 0.02  & 7.53 $\pm$ 0.02  & 7.43 $\pm$ 0.02 & 7.64 $\pm$ 0.03\\\hline
    \end{tabular}}
    \caption{Spectral properties at different epochs after flux re-scaling based on photometry. The luminosity are given in logarithm scale and in erg/s unit. The FWHM and $M_{\bullet}$ are in the units of \kms and M$_{\odot}$, respectively.}
    \label{tab:mbh}
\end{table}

\clearpage

\begin{table}[h]
    \centering
    \begin{tabular}{c|cc}
        Band & Observed & Model\\
        & (AB mag) & (AB mag) \\\hline
       GALEX FUV & 19.84 $\pm$ 0.13 & 19.96	$\pm$ 0.14 \\
       GALEX NUV & 19.27 $\pm$ 0.06 & 19.30	$\pm$ 0.06 \\
       DES g& 16.36 $\pm$ 0.01& 16.36 $\pm$	0.01 \\
       DES r& 15.81 $\pm$ 0.01& 15.83	$\pm$	0.01 \\
       DES i& 15.60 $\pm$ 0.01& 15.59	$\pm$	0.01 \\
       DES z& 15.41 $\pm$ 0.01& 15.43	$\pm$	0.01 \\
       DES Y& 15.30 $\pm$ 0.05& 15.36	$\pm$	0.01 \\
       2MASS J& 15.17 $\pm$ 0.09& 15.27	$\pm$	0.01 \\
       2MASS H& 15.08 $\pm$ 0.12& 15.17	$\pm$	0.01 \\
       2MASS Ks& 15.23 $\pm$ 0.13& 15.40	$\pm$	0.01 \\
       WISE W1 & 15.85 $\pm$ 0.25&  15.82	$\pm$	0.01 \\
       WISE W2 & 16.05 $\pm$ 0.25& 16.08	$\pm$	0.03  \\\hline
       UVOT W2 & ---& 19.45	$\pm$ 0.07 \\
       UVOT M2 & ---& 19.32	$\pm$ 0.06 \\
       UVOT W1 & ---& 18.94	$\pm$ 0.04 \\
       UVOT U & ---& 17.89 $\pm$	0.01 \\
       UVOT B & ---& 16.69 $\pm$	0.01  \\
       UVOT V & ---& 16.09 $\pm$	0.01  \\
    \end{tabular}
    \caption{Results from host galaxy SED model fitting. }
    \label{tab:hostgalphot}
\end{table}

\clearpage

\begin{sidewaystable}[!htp]
\ttabbox[\linewidth]{
\resizebox{\textwidth}{!}{\begin{tabular}{*{16}{l}}
\toprule
\toprule
\multicolumn{15}{c}{Best-fit parameters from \xmm energy spectral modeling.} \\
\bottomrule
{\bf ID} & {\bf Observation ID} & {\bf Start} & {\bf End} & {\bf Exposure} & {\bf nH$_{\rm neutral}$} & {\bf nH$_{\rm abs}$} & {\bf Log$\xi$} & {\bf $v_{\rm out}$} & {\bf kT} & {\bf Norm} & {\bf Log(Integ. Lum.)} & {\bf Log(Obs. Lum.)} & {\bf Log(unabs. Lum.)} & {\bf best-fit statistic/dof } & {\bf Fit statistic } \\
  & & (MJD) & (MJD) & (ks) & (10$^{21}$ cm$^{-2})$ & (10$^{21}$ cm$^{-2})$ & & & (keV) & & (1eV-10 keV) & (0.3-1.1 keV) & (0.3-1.1 keV) & used \\
\midrule
XMM1 & 0852600301 & 59287.359 & 59288.039 & 16.6 & 0.0 & 4.0$^{+3.7}_{-1.0}$ & 1.8$^{+0.2}_{-0.2}$ & -0.332$^{+0.013}_{-0.012}$ & 0.096$^{+0.002}_{-0.002}$ & 1.86$^{+0.18}_{-0.16}$ & 44.34$^{+0.011}_{-0.011}$ & 43.53$^{+0.01}_{-0.02}$ & 43.81$^{+0.026}_{-0.013}$ & 19.6/16 & $\chi^{2}$\\
XMM2 & 0891800101 & 59416.775 & 59417.057 & 6.6 & 0.0 & - & - & - & 0.091$^{+0.012}_{-0.012}$ & 0.02$^{+0.03}_{-0.01}$ & 42.30$^{+0.20}_{-0.20}$ & 41.52$^{+0.08}_{-0.05}$ & 41.73$^{+0.10}_{-0.10}$ & 33/31 & C-statistic \\
XMM3 & 0891803701 and 0891803801 & 59552.545 & 59556.749 & 36 & 0.0 & 8.3$^{+1.0}_{-2.0}$ & 1.0$^{+0.1}_{-0.1}$ & -0.22$^{+0.05}_{-0.05}$ & 0.117$^{+0.025}_{-0.011}$ & $<$0.2 & 43.18$^{+0.29}_{-0.55}$ & 41.59$^{+0.34}_{-0.55}$ & 42.79$^{+0.27}_{-0.48}$ & 99.1/88 & C-statistic\\
XMM4 & 0893810701 & 59615.36 & 59615.54 & 10 & 0.0 & - & - & - & 0.1$^{+0.03}_{-0.03}$ & 0.02$^{+0.02}_{-0.01}$ & 42.0$^{+0.20}_{-0.16}$ & 41.48$^{+0.06}_{-0.10}$ & 41.64$^{+0.08}_{-0.08}$ & 52/38 & C-statistic \\
% XMM5 & 0893811001 & 59677.97 & 59678.50 & 28 & 0.0 & 97.7$^{+1.2}_{-1.2}$ & 3.0$^{+0.2}_{-0.3}$ & -0.378$^{+0.041}_{-0.040}$ & 0.38$^{+0.13}_{-0.10}$ & $<$0.11 & 43.0$^{+0.3}_{-0.2}$ & 41.52$^{+0.08}_{-0.03}$ & 42.7$^{+0.3}_{-0.1}$ & 9.8/10 & $\chi^{2}$ \\
 \bottomrule
\end{tabular}}
 }
{\caption{{\bf Summary of \xmm X-ray energy spectral modeling of \target}. Spectra are fit with the same model used for time-resolved \nicer spectral analysis, i.e., {\it tbabs*ztbabs*xstar*zashift(diskbb)}. {\bf ID} is the identifier used to refer to this observation. {\bf Observation ID} is the ID assigned by \xmm. {\bf Start} and {\bf End} represent the start and end times (in units of MJD) of the exposures used. {\bf Exposure} is the total exposure time after removing intervals of background flaring. {\bf nH$_{\rm neutral}$} is the neutral column density (in units of 10$^{21}$ cm$^{-2}$). Cases where {\bf nH$_{\rm neutral}$} is pegged to zero by XSPEC are indicated by ``0.0''. {\bf nH$_{\rm abs}$} is the column density of the ionized gas (in units of 10$^{21}$ cm$^{-2}$). {\bf Log$\xi$} is the Logarithm of the ionization parameter. $v_{\rm out}$ is the outflow speed with respect to us in units of c, the speed of light. {\bf kT} is the temperature of the ionizing blackbody disk (in units of keV). {\bf Norm} is the normalization of the {\it diskbb} model in units of 10$^{4}$. {\bf Log(Integ. Lum.)} is the logarithm of the integrated blackbody luminosity in 1 eV-10 keV in units of erg s$^{-1}$. {\bf Log(Obs. Lum.)} and {\bf Log(unabs. Lum.)} are the logarithms of the observed and unabsorbed 0.3-1.1 keV luminosities in units of erg s$^{-1}$, respectively. {\bf best-fit statistic/dof} is the best-fit value of statistic ($\chi^{2}$ or C-statistic) over the degrees of freedom. All errorbars represent 1-$\sigma$ uncertainties. {\bf statistic} used for fitting ($\chi^{2}$ for high count rate spectra and C-stat when the count rate was low). $^{*}$Fixed to the best-fit value. }\label{tab:xmmxraydata}}
\end{sidewaystable}

%%%%%%%%%%%%%%%%%%%%%%%%%%%%%%%

\begin{sidewaystable}[!htp]
\ttabbox[\linewidth]{
\resizebox{\textwidth}{!}{\begin{tabular}{*{16}{l}}
\toprule
\toprule
\multicolumn{15}{c}{Best-fit parameters from fitting spectra from {\bf minima} in outflow deficit ratio} \\
\bottomrule
{\bf Start} & {\bf End} & {\bf Exposure} & {\bf FPMs} & {\bf Phase} & {\bf nH$_{\rm neutral}$} & {\bf nH$_{\rm abs}$} & {\bf Log$\xi$} & {\bf $v_{\rm out}$} & {\bf kT} & {\bf Norm} & {\bf Log(Integ. Lum.)} & {\bf Log(Obs. Lum.)} & {\bf Log(unabs. Lum.)} & {\bf Count rate } & {\bf $\chi^{2}$/dof } \\
 (MJD) & (MJD) & (ks) & & & (10$^{21}$ cm$^{-2})$ & (10$^{21}$ cm$^{-2})$ & & & (keV) & & (1eV-10 keV) & (0.3-1.1 keV) & (0.3-1.1 keV) & (0.3-1.1 keV) \\
\midrule
59266.93 & 59273.8 & 3.73 & 50 & Min1 & 0.0 & 16.0$^{+4.1}_{-8.0}$ & 2.0$^{+0.2}_{-0.2}$ & -0.35$^{+0.023}_{-0.019}$ & 0.082$^{+0.004}_{-0.004}$ & 11.46$^{+5.34}_{-6.5}$ &44.25$^{+0.11}_{-0.14}$ & 43.16$^{+0.13}_{-0.16}$ & 43.72$^{+0.08}_{-0.19}$ & 0.038$\pm$0.001 & 3.5/9 \\
59275.6 & 59285.32 & 6.95 & 49 & Min2 & 0.0 & 12.6$^{+2.2}_{-2.2}$ & 2.3$^{+0.4}_{-0.2}$ & -0.366$^{+0.011}_{-0.01}$ & 0.093$^{+0.002}_{-0.002}$ & 1.76$^{+0.22}_{-0.18}$ &44.19$^{+0.05}_{-0.05}$ & 43.38$^{+0.03}_{-0.07}$ & 43.73$^{+0.1}_{-0.06}$ & 0.064$\pm$0.0007 & 9/10 \\
59287.47 & 59290.79 & 3.02 & 48 & Min3 & 0.0 & 10.1$^{+2.7}_{-2.9}$ & 2.0$^{+0.2}_{-0.2}$ & -0.37$^{+0.012}_{-0.012}$ & 0.106$^{+0.003}_{-0.003}$ & 1.27$^{+0.15}_{-0.12}$ &44.29$^{+0.09}_{-0.08}$ & 43.51$^{+0.08}_{-0.06}$ & 43.91$^{+0.11}_{-0.09}$ & 0.0892$\pm$0.0014 & 10.5/9 \\
59292.3 & 59302.5 & 15.21 & 46 & Min4 & 0.0 & 9.9$^{+1.6}_{-1.6}$ & 2.2$^{+0.2}_{-0.1}$ & -0.359$^{+0.009}_{-0.009}$ & 0.096$^{+0.002}_{-0.002}$ & 1.56$^{+0.11}_{-0.1}$ &44.2$^{+0.05}_{-0.04}$ & 43.42$^{+0.04}_{-0.04}$ & 43.76$^{+0.06}_{-0.05}$ & 0.0702$\pm$0.0004 & 6.5/11 \\
59303.7 & 59317.38 & 5.85 & 50 & Min5 & 0.0 & 7.1$^{+2.4}_{-3.1}$ & 2.2$^{+0.7}_{-0.2}$ & -0.351$^{+0.02}_{-0.017}$ & 0.09$^{+0.002}_{-0.002}$ & 1.88$^{+0.16}_{-0.13}$ &44.15$^{+0.06}_{-0.05}$ & 43.37$^{+0.08}_{-0.1}$ & 43.66$^{+0.07}_{-0.06}$ & 0.0629$\pm$0.0007 &  12.1/10 \\
59318.94 & 59335.15 & 5.72 & 47 & Min6 & 0.0 & 12.6$^{+2.6}_{-2.8}$ & 2.1$^{+0.2}_{-0.1}$ & -0.372$^{+0.012}_{-0.011}$ & 0.095$^{+0.002}_{-0.002}$ & 2.09$^{+0.25}_{-0.12}$ &44.29$^{+0.08}_{-0.08}$ & 43.42$^{+0.05}_{-0.06}$ & 43.84$^{+0.09}_{-0.09}$ & 0.0695$\pm$0.0009 & 14.6/10 \\
59337.12 & 59343.15 & 7.01 & 48 & Min7 & 0.0 & 2.8$^{+0.8}_{-0.8}$ & 1.8$^{+0.2}_{-0.3}$ & -0.344$^{+0.016}_{-0.013}$ & 0.103$^{+0.002}_{-0.001}$ & 1.16$^{+0.08}_{-0.07}$ &44.2$^{+0.03}_{-0.01}$ & 43.55$^{+0.02}_{-0.05}$ & 43.79$^{+0.01}_{-0.01}$ & 0.0965$\pm$0.0007 & 6.5/10 \\
59345.74 & 59351.99 & 5.36 & 48 & Min8 & 0.0 & 21.4$^{+5.5}_{-4.9}$ & 2.9$^{+0.1}_{-0.1}$ & -0.306$^{+0.021}_{-0.017}$ & 0.109$^{+0.002}_{-0.002}$ & 0.94$^{+0.08}_{-0.06}$ &44.21$^{+0.06}_{-0.02}$ & 43.56$^{+0.03}_{-0.08}$ & 43.83$^{+0.06}_{-0.05}$ & 0.0986$\pm$0.0009 &  7.5/10\\
59352.96 & 59359.88 & 2.48 & 48 & Min9 & 0.0 & 21.5$^{+7.5}_{-2.9}$ & 3.2$^{+0.2}_{-0.1}$ & -0.228$^{+0.019}_{-0.022}$ & 0.101$^{+0.002}_{-0.002}$ & 1.08$^{+0.18}_{-0.1}$ &44.12$^{+0.07}_{-0.01}$ & 43.5$^{+0.01}_{-0.15}$ & 44.03$^{+0.47}_{-0.15}$ & 0.087$\pm$0.0014 &  9.8/9\\
59361.57 & 59366.53 & 15.0 & 48 & Min10 & 0.3$^{+0.05}_{-0.05}$ & 15.0$^{+-15.0}_{-15.0}$ & 2.4$^{+0.0}_{-0.2}$ & -0.324$^{+0.009}_{-0.009}$ & 0.075$^{+0.002}_{-0.002}$ & 6.66$^{+1.0}_{-1.5}$ &44.33$^{+0.05}_{-0.05}$ & 43.13$^{+-0.0}_{-0.04}$ & 43.71$^{+0.06}_{-0.04}$ & 0.0359$\pm$0.0003 & 18.8/10 \\
 \bottomrule
\\
\toprule
\multicolumn{15}{c}{Best-fit parameters from fitting spectra from {\bf maxima} in outflow deficit ratio} \\
\bottomrule
59273.8 & 59275.6 & 1.37 & 50 & Max1 & - & - & - & - & 0.084$^{+0.002}_{-0.002}$ & 2.2$^{+0.3}_{-0.3}$ &44.07$^{+0.02}_{-0.02}$ & 43.35$^{+0.01}_{-0.01}$ & 43.54$^{+0.01}_{-0.01}$ & 0.0611$\pm$0.001 & 15.7/11.0 \\
59285.32 & 59287.47 & 0.6 & 48 & Max2 & - & 2.5$^{+0.0}_{-2.5}$ & 1.8$^{+0.5}_{-0.3}$ & -0.341$^{+0.029}_{-0.026}$ & 0.116$^{+0.005}_{-0.005}$ & 0.6$^{+0.2}_{-0.1}$ &44.17$^{+0.04}_{-0.02}$ & 43.59$^{+0.0}_{-0.09}$ & 43.82$^{+0.05}_{-0.04}$ & 0.1096$\pm$0.002 & 6.9/8.0 \\
59290.79 & 59292.3 & 1.15 & 48 & Max3 & - & 1.0$^{+1.1}_{-0.4}$ & 0.8$^{+0.2}_{-0.3}$ & -0.331$^{+0.02}_{-0.021}$ & 0.116$^{+0.002}_{-0.003}$ & 0.6$^{+0.1}_{-0.0}$ &44.11$^{+0.03}_{-0.01}$ & 43.53$^{+0.03}_{-0.15}$ & 43.76$^{+0.03}_{-0.02}$ & 0.0976$\pm$0.0014 & 11.7/8.0 \\
59302.5 & 59303.7 & 2.78 & 48 & Max4 & - & 1.2$^{+1.5}_{-0.3}$ & 0.8$^{+0.5}_{-0.1}$ & -0.332$^{+0.015}_{-0.014}$ & 0.115$^{+0.002}_{-0.003}$ & 0.6$^{+0.1}_{-0.1}$ &44.09$^{+0.06}_{-0.05}$ & 43.47$^{+0.04}_{-0.12}$ & 43.79$^{+0.06}_{-0.06}$ & 0.0848$\pm$0.0008 & 10.9/9.0 \\
59317.38 & 59318.94 & 1.04 & 52 & Max5 & - & 2.3$^{+3.3}_{-0.9}$ & 2.0$^{+0.5}_{-0.5}$ & -0.304$^{+0.036}_{-0.032}$ & 0.11$^{+0.004}_{-0.004}$ & 0.8$^{+0.1}_{-0.1}$ &44.17$^{+0.02}_{-0.02}$ & 43.58$^{+0.02}_{-0.03}$ & 43.8$^{+0.02}_{-0.03}$ & 0.1079$\pm$0.0014 & 6.6/9.0 \\
59335.15 & 59337.12 & 1.85 & 48 & Max6 & - & 2.6$^{+3.2}_{-0.7}$ & 1.6$^{+0.2}_{-0.5}$ & -0.344$^{+0.014}_{-0.014}$ & 0.11$^{+0.003}_{-0.007}$ & 0.8$^{+0.7}_{-0.1}$ &44.19$^{+0.14}_{-0.02}$ & 43.56$^{+-0.0}_{-0.11}$ & 43.82$^{+0.11}_{-0.02}$ & 0.1007$\pm$0.0011 & 7.8/9.0 \\
59343.15 & 59345.74 & 2.83 & 47 & Max7 & - & 1.0$^{+1.2}_{-0.5}$ & 0.9$^{+0.4}_{-0.2}$ & -0.347$^{+0.03}_{-0.02}$ & 0.109$^{+0.001}_{-0.003}$ & 1.0$^{+0.2}_{-0.1}$ &44.23$^{+0.05}_{-0.02}$ & 43.63$^{+0.04}_{-0.11}$ & 43.86$^{+0.05}_{-0.03}$ & 0.1206$\pm$0.001 & 6.4/7.0 \\
59351.99 & 59352.96 & 0.6 & 49 & Max8 & - & - & - & - & 0.094$^{+0.002}_{-0.002}$ & 1.5$^{+0.2}_{-0.2}$ &44.14$^{+0.02}_{-0.02}$ & 43.5$^{+0.01}_{-0.01}$ & 43.68$^{+0.01}_{-0.01}$ & 0.0846$\pm$0.0018 & 9.4/9.0 \\
59359.88 & 59361.57 & 0.24 & 48 & Max9 & - & - & - & - & 0.095$^{+0.004}_{-0.004}$ & 1.4$^{+0.4}_{-0.3}$ &44.13$^{+0.04}_{-0.04}$ & 43.51$^{+0.01}_{-0.02}$ & 43.69$^{+0.02}_{-0.02}$ & 0.0896$\pm$0.003 & 6.6/7.0 \\

\bottomrule
\end{tabular}}
 }
{\caption{{\bf Summary of time-resolved X-ray energy spectral modeling of \target~}. Here \nicer spectra corresponding to the mimina and the maxima in outflow deficit ratio (ODR) are fitted with {\it tbabs*ztbabs*xstar*zashift(diskbb)} model (See SI). {\bf Start} and {\bf End} represent the start and end times (in units of MJD) of the interval used to extract a combined \nicer spectrum. {\bf Exposure} is the accumulated exposure time during this phase/time interval. {\bf FPMs:} The total number of active detectors minus the ``hot'' detectors. {\bf Phase} is the name used to identify the epoch. {\bf nH$_{\rm neutral}$} is the neutral column density (in units of 10$^{21}$ cm$^{-2}$). Cases where {\bf nH$_{\rm neutral}$} is pegged to zero by XSPEC are indicated by ``0.0". {\bf nH$_{\rm abs}$} is the column density of the ionized outflow (in units of 10$^{21}$ cm$^{-2}$). {\bf Log$\xi$} is the Logarithm of the outflow's ionization parameter. $v_{\rm out}$ is the outflow speed with respect to us in units of c, the speed of light. {\bf kT} is the temperature of the ionizing blackbody disk (in units of keV). {\bf Norm} is the normalization of the {\it diskbb} model in units of 10$^{4}$. {\bf Log(Integ. Lum.)} is the logarithm of the integrated blackbody luminosity in 1 eV-10 keV in units of erg s$^{-1}$. {\bf Log(Obs. Lum.)} and {\bf Log(unabs. Lum.)} are the logarithms of the observed and unabsorbed 0.3-1.1 keV luminosities in units of erg s$^{-1}$, respectively. {\bf Count Rate } is the background-subtracted \nicer count rate in 0.3-1.1 keV in units of counts/sec/FPM. All errorbars represent 1-$\sigma$ uncertainties. {\bf $\chi^{2}$/dof} represents the best-fit $\chi^{2}$ and the degrees of freedom. }\label{tab: nicerxraydata}}
\end{sidewaystable}

\clearpage

\begin{table}[H]
\centering
\begin{tabular}{l c c c c c}
\toprule
\toprule
\multicolumn{6}{c}{NICER Min phases} \\
\bottomrule
Phase & $r$ & $\dot{M}_{out}$ & $\dot{E}_{out}$ & $\frac{\dot{E}_{out}}{L}$ & $\frac{\dot{M}_{out}}{\dot{M}_{acc}}$\\
  & ($r_g$) & ($10^{-3} M_{\odot}/yr$) & ($10^{42}$ erg/s) & & \\
\midrule
Min1 & 16.4 & 2.3 & 8.2 & 0.16 & 0.26\\ 
Min2 & 15.0 & 1.9 & 7.2 & 0.13 & 0.20\\
Min3 & 14.6 & 1.4 & 5.6 & 0.07 & 0.10\\
Min4 & 15.4 & 1.4 & 5.3 & 0.09 & 0.14\\
Min5 & 16.2 & 1.1 & 3.7 & 0.08 & 0.13\\
Min6 & 14.4 & 1.7 & 6.9 & 0.10 & 0.14\\
Min7 & 16.8 & 0.4 & 1.4 & 0.02 & 0.03\\
Min8 & 21.4 & 3.6 & 9.7 & 0.14 & 0.31\\
Min9 & 38.4 & 4.9 & 7.2 & 0.07 & 0.27\\
Min10 & 19.0 & 2.4 & 7.1 & 0.14 & 0.26\\
\toprule
\multicolumn{6}{c}{NICER Max phases} \\
\midrule
Max1 & - & - & - & - & -\\
Max2 & 17.2 & 0.4 & 1.3 & 0.02 & 0.03\\
Max3 & 18.2 & 0.1 & 0.5 & 0.01 & 0.01\\
Max4 & 18.2 & 0.2 & 0.6 & 0.01 & 0.02\\
Max5 & 21.6 & 0.4 & 1.0 & 0.02 & 0.04\\
Max6 & 16.8 & 0.4 & 1.3 & 0.02 & 0.03\\
Max7 & 16.6 & 0.1 & 0.5 & 0.01 & 0.01\\
Max8 & - & - & - & - & -\\
Max9 & - & - & - & - & -\\
\toprule
\multicolumn{6}{c}{XMM phases} \\
\midrule
XMM1 & 18.1 & 0.6 & 2.0 & 0.03 & 0.05\\
XMM2 & - & - & - & - & -\\
XMM3 & 26.6 & 3.3 & 7.1 & 0.82 & 2.17\\
XMM4 & - & - & - & - & -\\
\bottomrule
\end{tabular}
\caption{Conservative estimates of the location and energetics of the outflow detected in the time-resolved NICER analysis and in the XMM-Newton spectra. Phase is the name used to identify the epoch. r is the launching radius in units of the gravitational radius ($r_g = GM_{\bullet}/c^2$). $\dot{M}_{out}$ is the mass outflow rate in units of $10^{-3} M_{\odot}/yr$. $\dot{E}_{out}$ is the outflow kinetic power in units of $10^{42}$ erg/s. $\dot{E}_{out}/L$ is the ratio between the outflow kinetic power and the unabsorbed luminosity in the 0.3-1.1 keV band. $\dot{M}_{out}/\dot{M}_{acc}$ is the ratio between the mass outflow rate and the mass accretion rate estimated as $\dot{M}_{acc} = L/\eta c^2$, with $\eta = 0.1$.}   
\label{tab:energetics}
\end{table}

\begin{table}[]
\centering
\begin{tabular}{l c c c c c c c c c}
\toprule
\toprule
\multicolumn{7}{c}{GRMHD simulations} \\
\bottomrule
Run & $a$ & B  &  $t_{\rm in}$ &  $t_{\rm f}$ & $r_{\rm per}$ & $\log{(M_{\bullet})}$ & $\mathcal{R}$ & $\iota$  & Resolution \\
  & & & ($r_g/c$) & ($r_g/c$) & ($r_g$) & [M$_{\odot}$] & ($r_g$) & (\degree) & \\
\midrule
Run 1 & 0.4 & a) & $20\,000$ & $100\,000$& 93 & 7.4  & 2 & 72& 384 x 256\\ %MIT10-R1

Run 2 & 0.4 & b) & $50\,000$ & $200\,000$ & 93 & 7.4  &2 & 72 & 384 x 256 \\ 
Run 3 & 0.4 & a) & $50\,000$ & $200\,000$ & 93 & 7.4  &2 & 72& 384 x 256\\ %V150-B4-R1
Run 4 & 0.4 & a) & $50\,000$ & $200\,000$ & 60 -- 125 & 7.4  & 2 & 66 & 384 x 256 \\ %V150-B4-R2
Run 5 & 0.4 & a) & $50\,000$ & $200\,000$ & 40 & 7.95 & 1 & 72 & 384 x 256\\ %V150-B4-R3
Run 6 & 0.4 & a) & $50\,000$ & $200\,000$ & 93 & 7.4 & 3 & 72 & 384 x 256\\ %V150-B4-R6
Run 7 & 0.4 & a) & $50\,000$ & $200\,000$ & 93 & 7.4 & 4 & 72 & 384 x 256\\ %V150-B4-R4 
Run 8 & 0.4 & a) & $50\,000$ & $200\,000$ & 93 & 7.4 & 8 & 72 & 384 x 256\\ %V150-B4-R5
\toprule
%\multicolumn{6}{c}{} \\
\midrule
Run 9 & 0 & b)  & $50\,000$ & $53\,000$ & 10 & -- & 1 & 81 & 384 x 256 x 128 \\ %V181-R1
Run 10 & 0 & b)  & $50\,000$ & $200\,000$ & 10 & -- & 1 & 81 & 384 x 256 \\ %V181-R2
Run 11 & 0 & b)  & $50\,000$ & $200\,000$ & 10 & -- & 0.1 & 81 & 384 x 256 \\ %V181-R3
\toprule
\midrule
Run 12 & 0.4 & a) & $20\,000$ & $100\,000$ & 45 -- 136 &7.4 & 2 & 67 & 512 x 320 \\ %MIT30-R12
Run 13 & 0.4 & a) & $20\,000$ & $100\,000$ & 45 -- 136 &7.4 & 1 & 67 & 640 x 512 \\ %MIT31-R11 
Run 14 & 0.4 & a) & $20\,000$ & $100\,000$ & 45 -- 136 &7.4 & 3 & 67 & 512 x 320 \\ %MIT30-R13 
Run 15  & 0.4 & a) & $20\,000$ & $100\,000$ & 30 -- 151 &7.4 & 2 & 66 & 512 x 320 \\ %MIT30-R2
\toprule
\midrule
\textbf{Run 16} & 0.5 & a) & $10\,000$ & $14\,300$ & 10 -- 14.7 &-- & 2 & 68 & 384 x 256 x 96 \\ %MV200-31
\bottomrule
\end{tabular}
\caption{\textbf{Parameters of GRMHD runs.} We list the dimensionless spin of SMBH $a$, initial magnetic field configuration: {\bf a) more loops, b) 1 loop },  transient time after which the perturber is added into the flow $t_{\rm in}$, {\bf final time $t_{\rm f}$} the distance of the perturber from SMBH $r_{\rm per}$, the logarithm of the SMBH mass considered to derive the distance, influence radius of the perturber~$\mathcal{R}$, inclination $\iota$ of the perturber orbit with respect to the equatorial plane, and the resolution of the run in terms of the number of radial logarithmic bins times the number of bins in $\theta$ direction ($r\times \theta$). For the 3D Runs 9 and 16, the resolution pertains to the number of bins in $r$, $\theta$, and $\phi$ directions.}
\label{tab:GRMHD_runs}
\end{table}

\newpage

FIG+++FIG+++FIG+++FIG+++FIG+++FIG+++FIG+++FIG+++FIG+++FIG+++FIG+++FIG+++
%                   Extended Data Figure 2: All optical spectra
% FIG+++FIG+++FIG+++FIG+++FIG+++FIG+++FIG+++FIG+++FIG+++FIG+++FIG+++FIG+++
\begin{figure}[!htbp]
\centering
\resizebox{12.0cm}{16cm}{\includegraphics{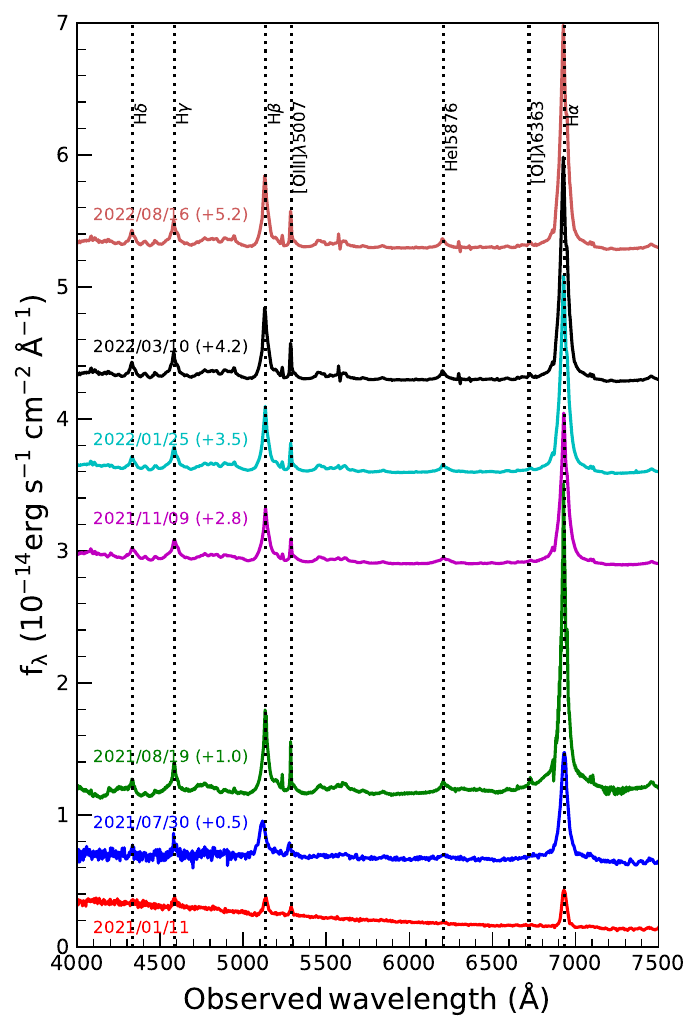}}
\caption{Optical spectra of different epochs after flux rescaling based on photometry. Strong broad Balmer emission lines, e.g., H$\alpha$, H$\beta$, H$\gamma$ are clearly visible (marked in the plot).}\label{Fig:optical spectra} 
\end{figure}
% \vfill\eject
% 

% FIG+++FIG+++FIG+++FIG+++FIG+++FIG+++FIG+++FIG+++FIG+++FIG+++FIG+++FIG+++
%                   Extended Data Figure 3: All optical spectra
% FIG+++FIG+++FIG+++FIG+++FIG+++FIG+++FIG+++FIG+++FIG+++FIG+++FIG+++FIG+++
% \clearpage
\begin{figure}[!htbp]
\centering
\resizebox{14.0cm}{18cm}{\includegraphics{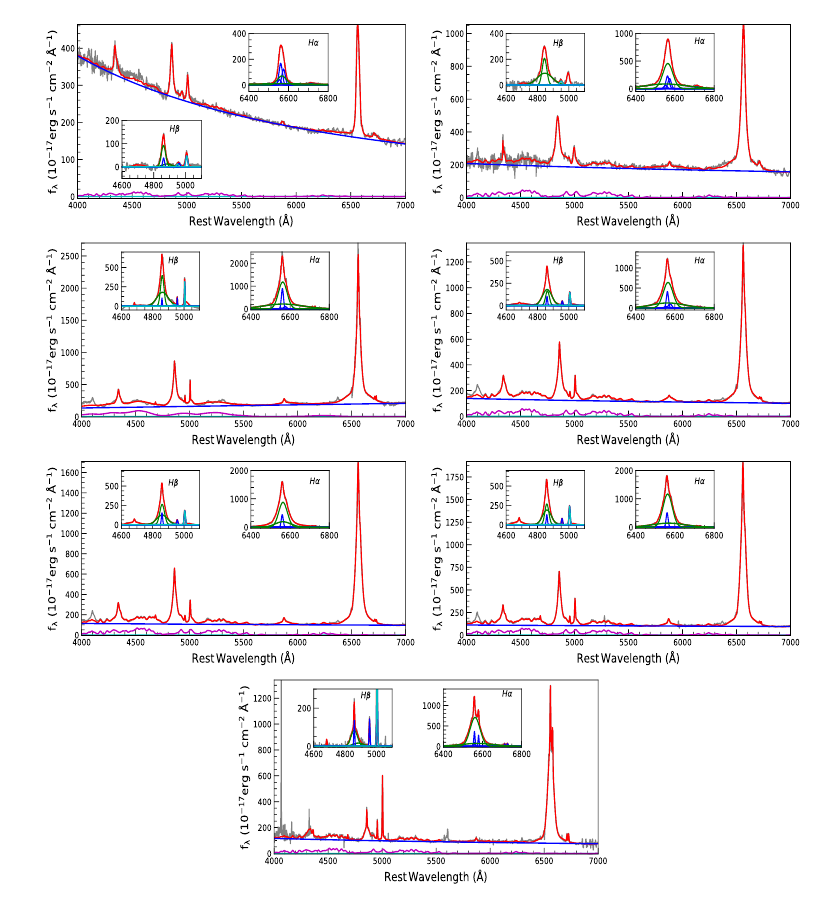}}
\caption{The optical spectral decomposition are shown for different epochs (from top to bottom 2021/01/11, 2021/07/30, 2021/08/19, 2021/11/09, 2022/01/25, 2022/03/10 and 2022/08/16). The data is shown in black while the best-fit entire model is shown in red. The decomposed continuum power-law and Fe II emission are shown by blue and magenta colors, respectively.  The inset plots shows the zoomed-in version of H$\beta$ and H$\alpha$ complex where broad components are shown in green and narrow components are in blue.}\label{Fig:optical_spectral_fit} 
\end{figure}
% FIG+++FIG+++FIG+++FIG+++FIG+++FIG+++FIG+++FIG+++FIG+++FIG+++FIG+++FIG+++
% FIG+++FIG+++FIG+++FIG+++FIG+++FIG+++FIG+++FIG+++FIG+++FIG+++FIG+++FIG+++
%                   Extended Data Figure 4: BPTWHAM
% FIG+++FIG+++FIG+++FIG+++FIG+++FIG+++FIG+++FIG+++FIG+++FIG+++FIG+++FIG+++
\begin{figure}[ht]
\centering
\includegraphics[width=\textwidth]{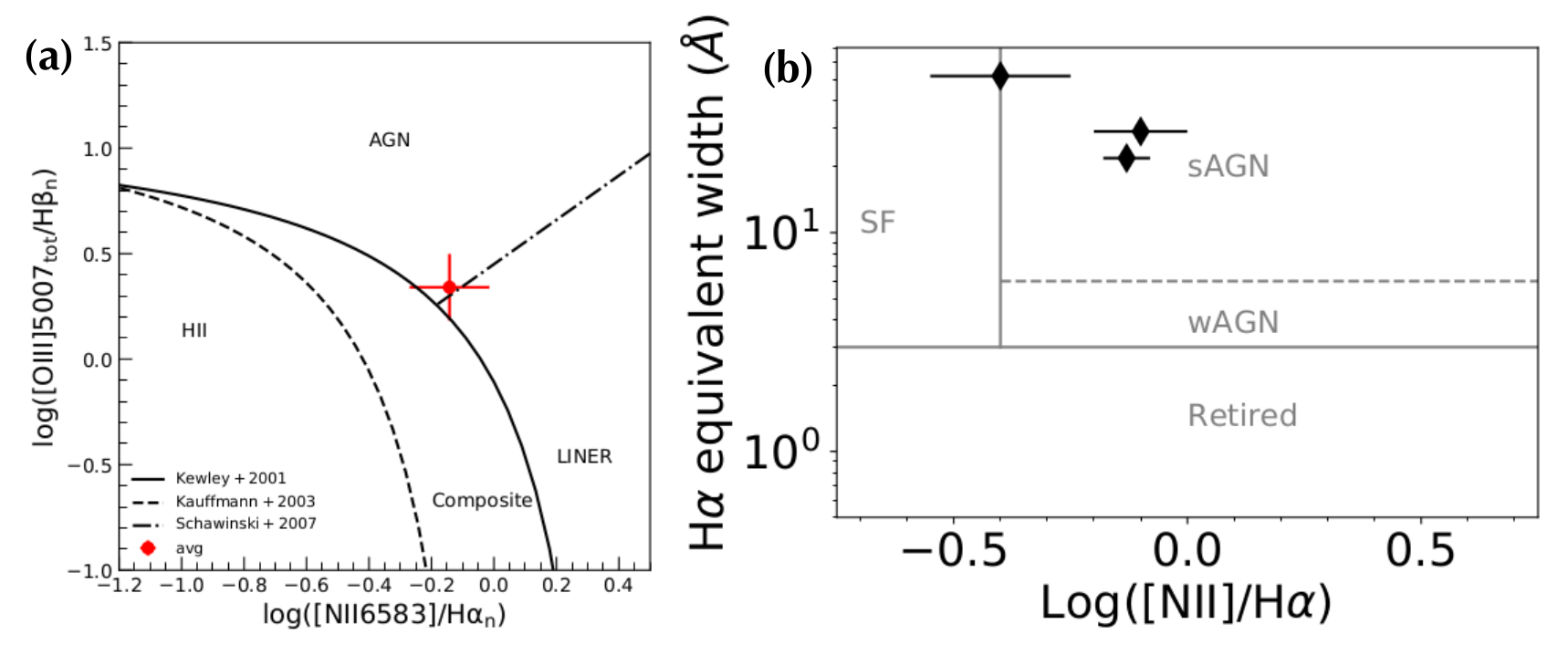}
\caption{{\bf \target's BPT and WHAN diagnostic diagrams.} {\bf (a)} \target's position in the BPT diagram is based on the weighted average of all the reliable measurements. The Kewley et al. (2001) \cite{Kewley2001}, Kauffmann et al. (2003) \cite{Kauffmann2003}, and Schawinski et al. (2007) \cite{Schawinski2007} separation line of LINER and AGNs are also shown. {\bf (b)} WHAN diagram showing the strong AGN (sAGN), weak AGN (wAGN), star formation (SF) and retired galaxies (see \cite{Cid2011} for definitions). The 3 data points correspond to the 3 optical spectra with well-constrained line measurements. Both the BPT and the WHAN diagrams indicate that  \target hosts an AGN.}\label{Fig:BPT} 
\end{figure}
% 
% FIG+++FIG+++FIG+++FIG+++FIG+++FIG+++FIG+++FIG+++FIG+++FIG+++FIG+++FIG+++
%                   Extended Data Figure 1: XMM residuals
% FIG+++FIG+++FIG+++FIG+++FIG+++FIG+++FIG+++FIG+++FIG+++FIG+++FIG+++FIG+++
% \clearpage
\begin{figure}[!htp]
\begin{center}
\includegraphics[width=0.65\textwidth, angle=0]{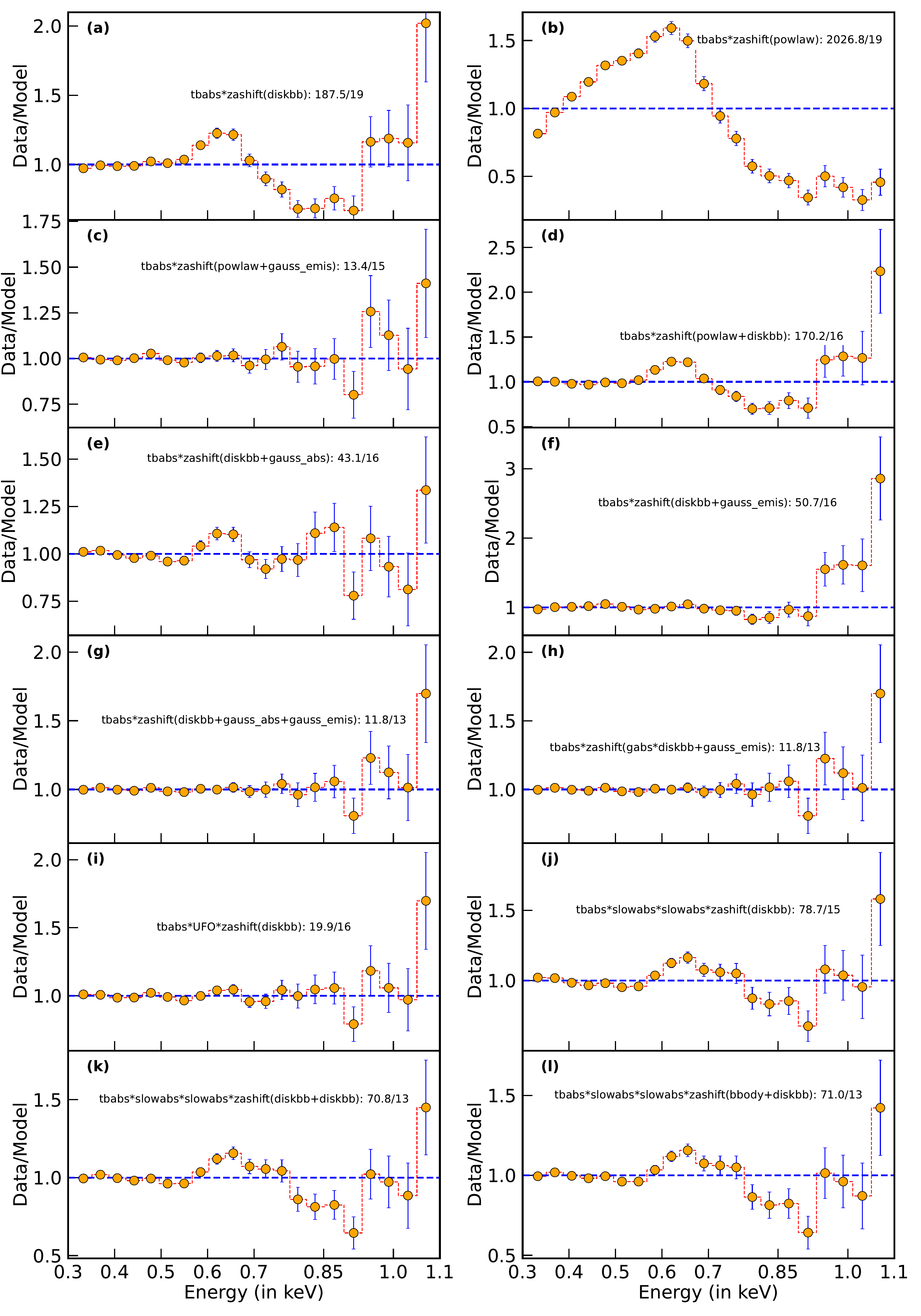}
\end{center}
\vspace{-.35cm} 
\caption{{\textbf{Ratio of the observed X-ray spectrum over the best-fit model.}} EPIC-pn's 0.3-1.1 keV energy spectrum was modeled with various phenomenological models. Each panel shows the ratio: Data/best-fitting model. The models are: (a) thermal accretion disk; (b) powerlaw; (c) powerlaw plus a Gaussian line; (d) powerlaw plus a thermal disk; (e) thermal disk plus inverted Gaussian line; (f) a disk plus a Gaussian emission line; (g) thermal disk plus an inverted Gaussian plus a Gaussian line; (h) Gaussian absorption line modifying the thermal disk plus a Gaussian line; (i) ionized absorber; (j) two warm absorbers from RGS and a disk; (k) two warm absorbers and two disks; (l) two warm absorbers from RGS, a disk and a blackbody component. The exact model used in {\it XSPEC} is shown on each panel. See Methods section \ref{sisec:spectral_model} for a detailed description of the spectral fits. The errorbars represent 1-$\sigma$ uncertainties.}
\label{fig:xmmresiduals}
\end{figure}
% FIG+++FIG+++FIG+++FIG+++FIG+++FIG+++FIG+++FIG+++FIG+++FIG+++FIG+++FIG+++
\clearpage
% FIG+++FIG+++FIG+++FIG+++FIG+++FIG+++FIG+++FIG+++FIG+++FIG+++FIG+++FIG+++
%                   Extended Data Figure 2: Continuum LSP --> no QPOs
% FIG+++FIG+++FIG+++FIG+++FIG+++FIG+++FIG+++FIG+++FIG+++FIG+++FIG+++FIG+++
\begin{figure}[htbp!]
\begin{center}
\includegraphics[width=\textwidth, angle=0]{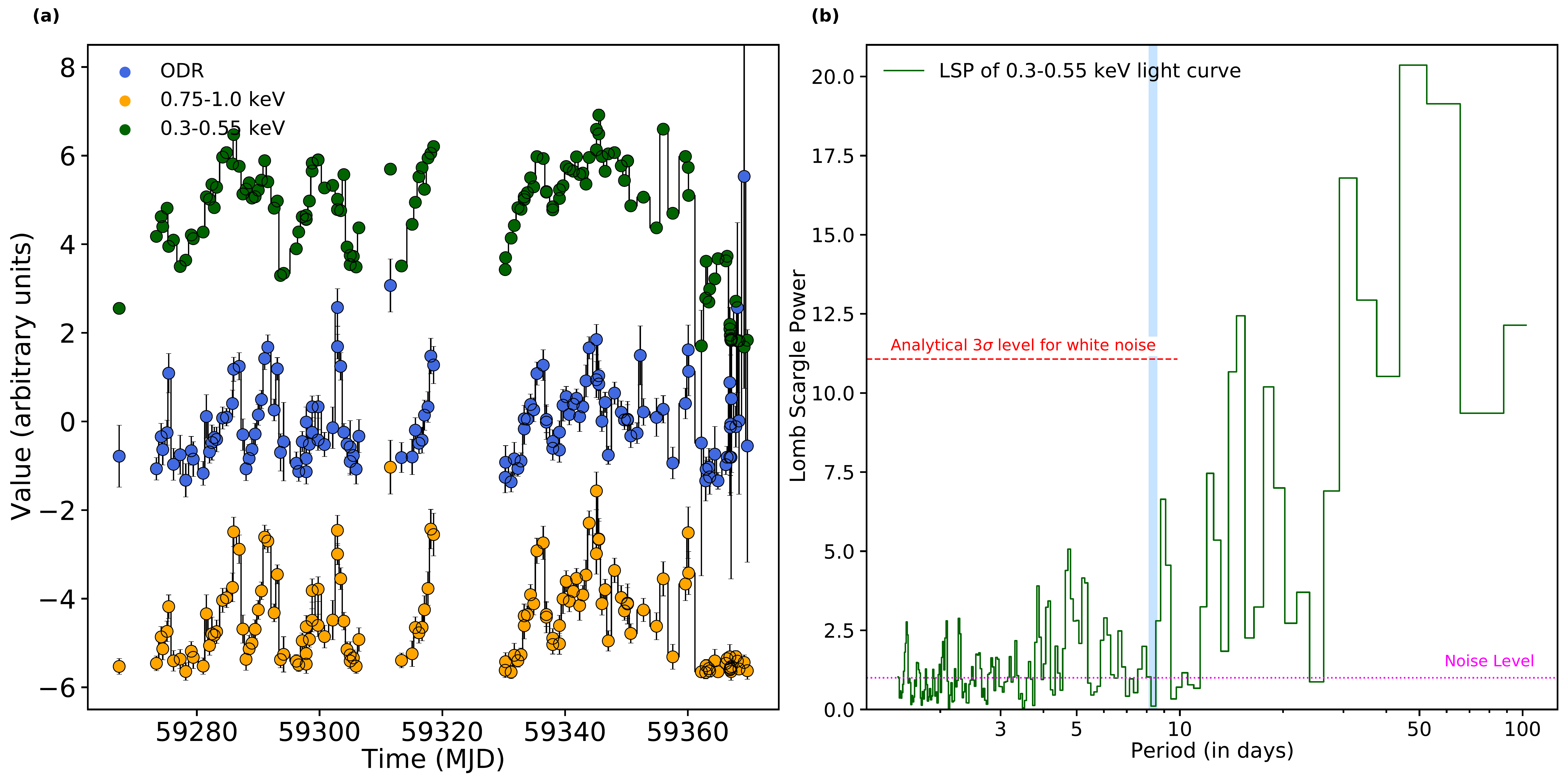}
\end{center} 
\caption{{\textbf{Background-subtracted soft (0.3--0.55 keV), hard (0.75--1 keV) and ODR curves and LSP of the 0.3-0.55 keV light curve}}. {\bf (a)} The 0.3-0.55 keV represents the continuum flux while the 0.75-1.0 keV tracks the outflow's strength. {\bf (b)} The LSP of the continuum is dominated by red noise and does not show a statistically significant peak near 8.5 days (blue shaded column). } 
\label{fig:lcs}
\end{figure}
% FIG+++FIG+++FIG+++FIG+++FIG+++FIG+++FIG+++FIG+++FIG+++FIG+++FIG+++FIG+++
\clearpage

% FIG+++FIG+++FIG+++FIG+++FIG+++FIG+++FIG+++FIG+++FIG+++FIG+++FIG+++FIG+++
\begin{figure}[htbp!]
\begin{center}
\includegraphics[width=\textwidth, angle=0]{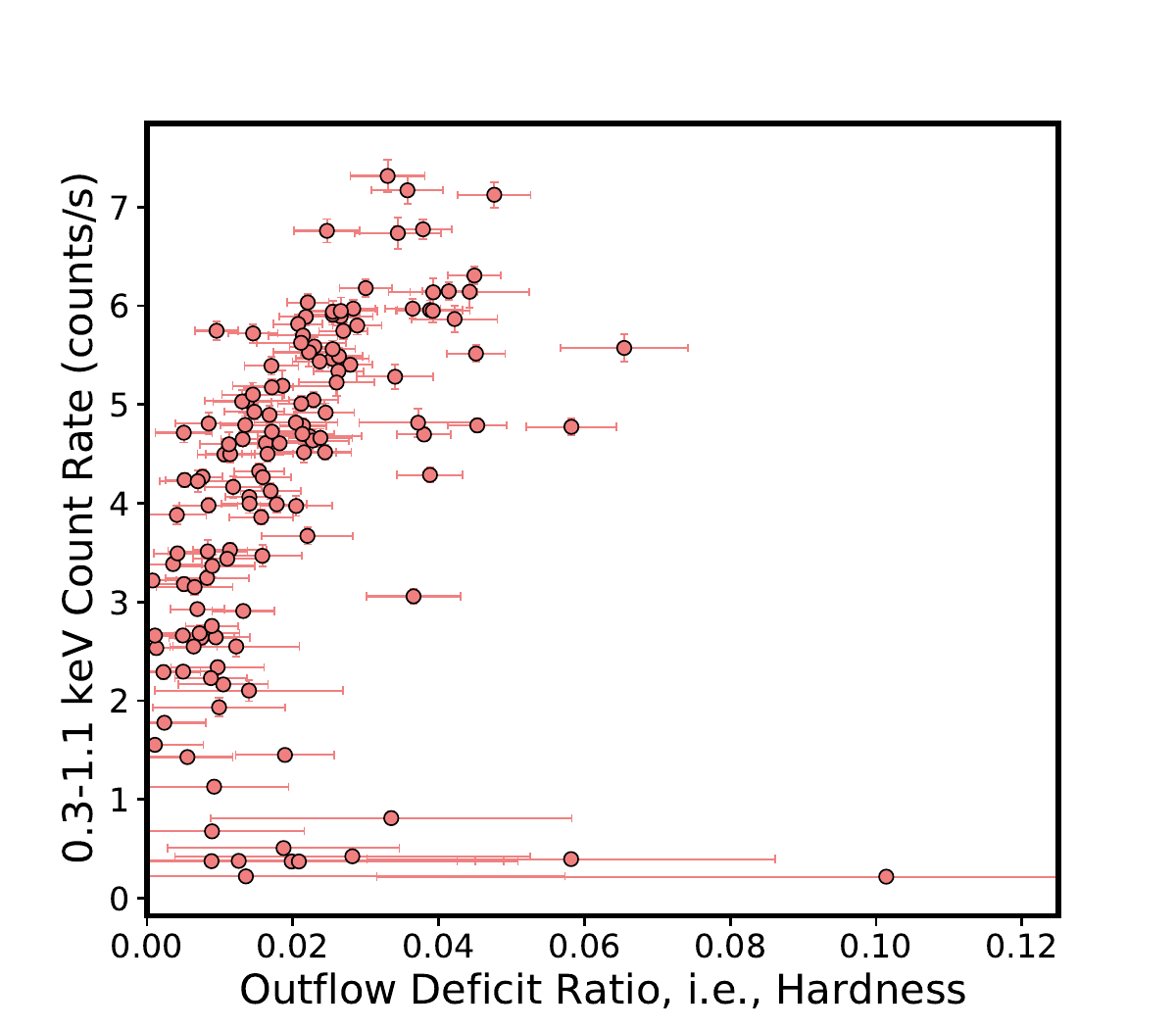}
\end{center} 
\caption{{\textbf{ODR vs 0.3-1.1 keV count rate}}. ODR represents the strength of the broad absorption line with respect to the continuum and the 0.3-1.1 keV count rate represents the overall X-ray intensity. A lower ODR value indicates stronger absorption dip and stronger outflow. } 
\label{fig:hid}
\end{figure}
% FIG+++FIG+++FIG+++FIG+++FIG+++FIG+++FIG+++FIG+++FIG+++FIG+++FIG+++FIG+++
\clearpage
\begin{figure}
    \centering
    \includegraphics[width=\textwidth]{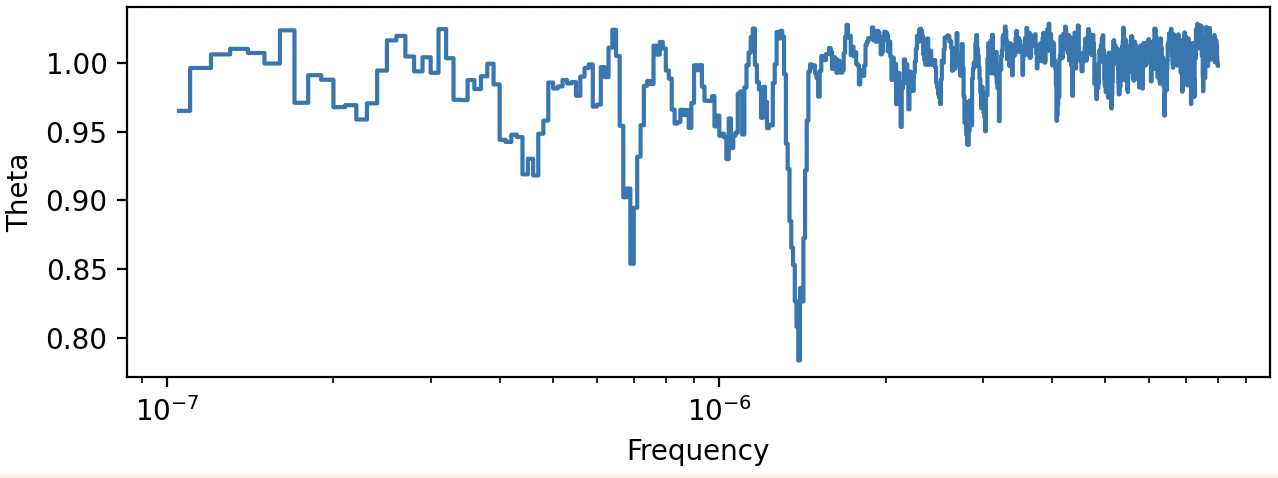}
    \caption{{\bf Results from application of the phase dispersion minimization algorithm on the ODR time series}. The strongest dip is at $1.4 \times 10^{-6}$ Hz (8.5 days), which is consistent with the LSP analysis.}
    \label{fig_pdm}
\end{figure}
% FIG+++FIG+++FIG+++FIG+++FIG+++FIG+++FIG+++FIG+++FIG+++FIG+++FIG+++FIG+++
\clearpage
\begin{figure}
    \centering
    \includegraphics[width=\textwidth]{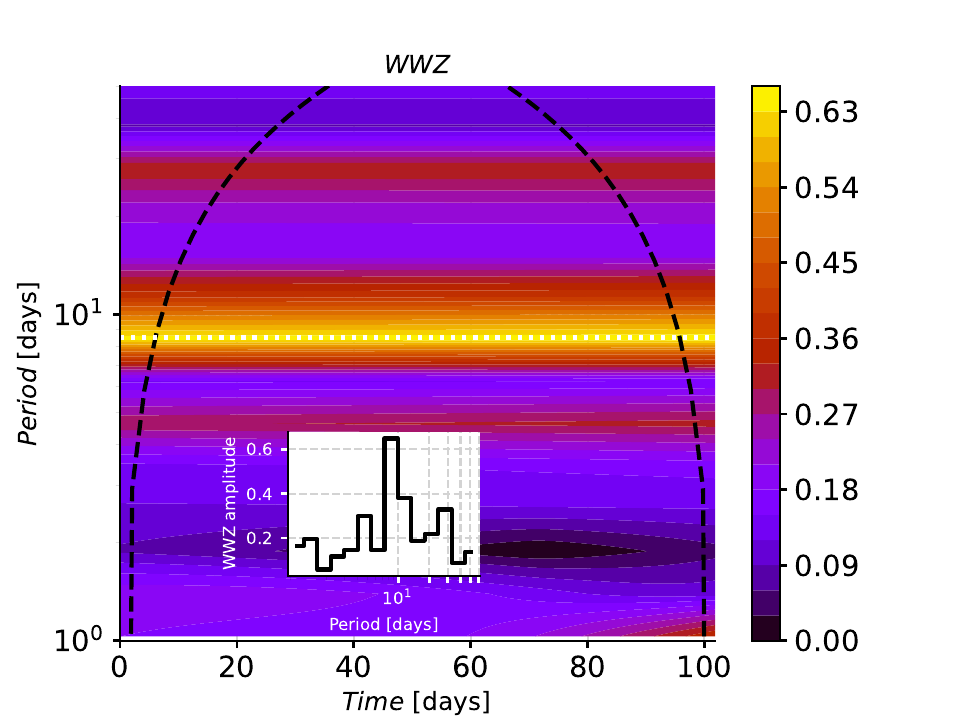}
    \caption{{\bf ODR periodicity analysis using Weighted Wavelet Z-transform (WWZ)}. The plot depicts the colour-coded WWZ amplitude in the time--period plane (both in days). The dotted white line is at 8.5 days, which is consistent with the LSP peak within the uncertainties for the whole time range. The figure inset depicts the WWZ amplitude vs. period (in days).}
    \label{fig_wwz}
\end{figure}
% FIG+++FIG+++FIG+++FIG+++FIG+++FIG+++FIG+++FIG+++FIG+++FIG+++FIG+++FIG+++
%                   Extended Data Figure 5: white noise tests
% FIG+++FIG+++FIG+++FIG+++FIG+++FIG+++FIG+++FIG+++FIG+++FIG+++FIG+++FIG+++
\clearpage
\begin{figure}[!htp]
\begin{center}
\includegraphics[width=\textwidth, angle=0]{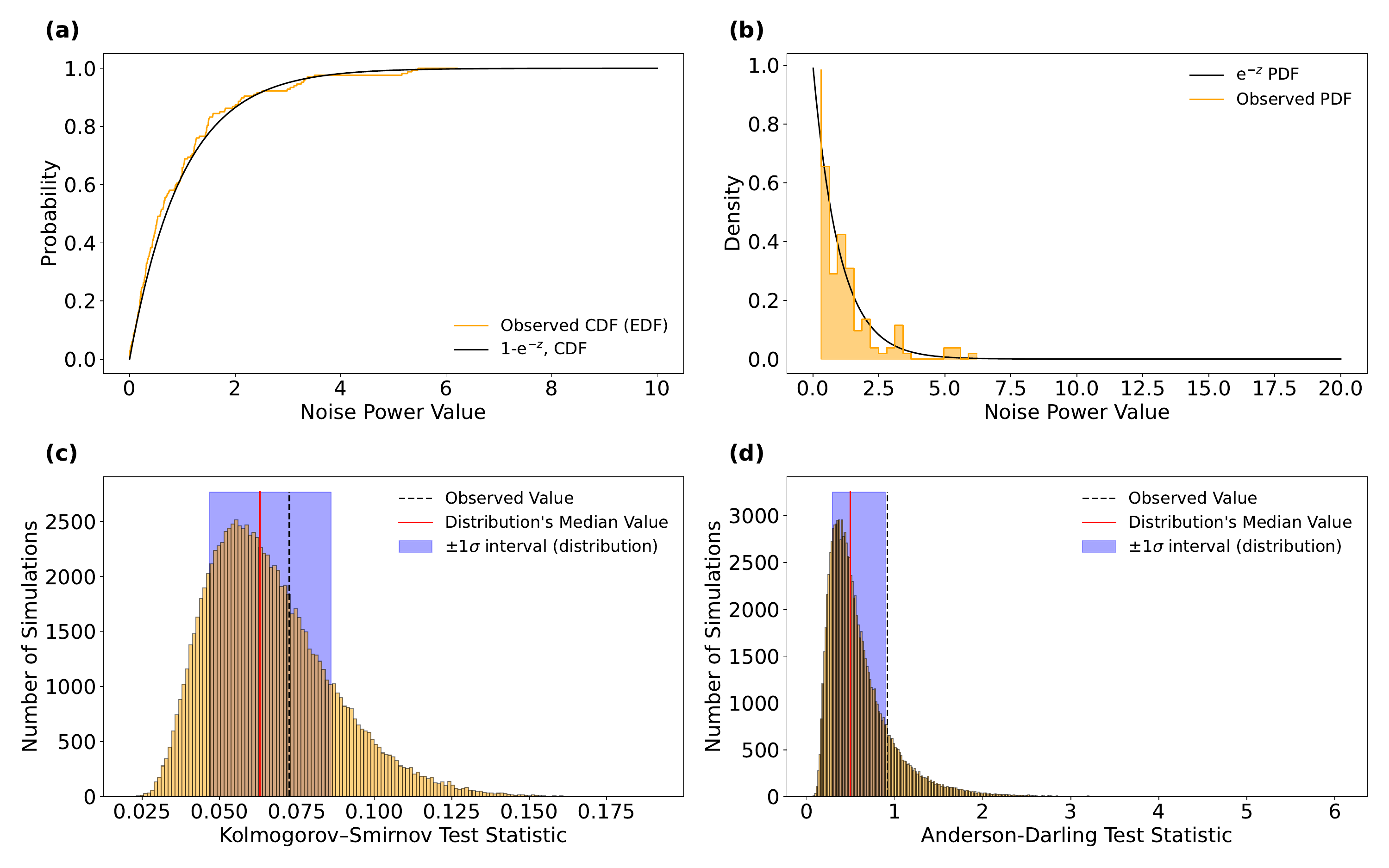}
\end{center}
\vspace{-.35cm} 
\caption{{\textbf{White noise tests for the distribution of noise powers in the observed Lomb Scargle periodogram.}} \textbf{(a) Comparison of the cumulative distribution functions (CDFs) of the observed noise powers and the expected exponential distribution.} The orange histogram is the CDF of the observed Lomb Scargle Periodogram of the ODR curve while the solid black line shows the CDF of white noise powers, i.e., the exponential distribution. \textbf{(b) Comparison of the probability density functions of the observed noise powers and the expected exponential distribution.} The shaded orange histogram represents the PDF of the noise powers in the observed LSP of ODR curve while the solid black line is the PDF of white noise distribution, i.e., the exponential distribution. \textbf{(c) Distribution of the Kolmogorov-Smirnov test statistic derived from simulations}. The solid red and the dashed black lines represent the median of the distribution and the observed value, respectively. \textbf{(d) Distribution of the Anderson-Darling test statistic using simulations}. The solid red and the dashed black lines represent the median of the distribution and the observed value, respectively. In both {\bf (c)} and {\bf (d)} the shaded blue regions indicate the $\pm$1$\sigma$ values of their respective distributions.} 
\label{fig:whitetests}
\end{figure}
% FIG+++FIG+++FIG+++FIG+++FIG+++FIG+++FIG+++FIG+++FIG+++FIG+++FIG+++FIG+++
\clearpage
% FIG+++FIG+++FIG+++FIG+++FIG+++FIG+++FIG+++FIG+++FIG+++FIG+++FIG+++FIG+++
%                   Extended Data Figure 4: Time resolved spectral evolution
% FIG+++FIG+++FIG+++FIG+++FIG+++FIG+++FIG+++FIG+++FIG+++FIG+++FIG+++FIG+++
\begin{figure}[ht]
\begin{center}
\includegraphics[width=\textwidth, angle=0]{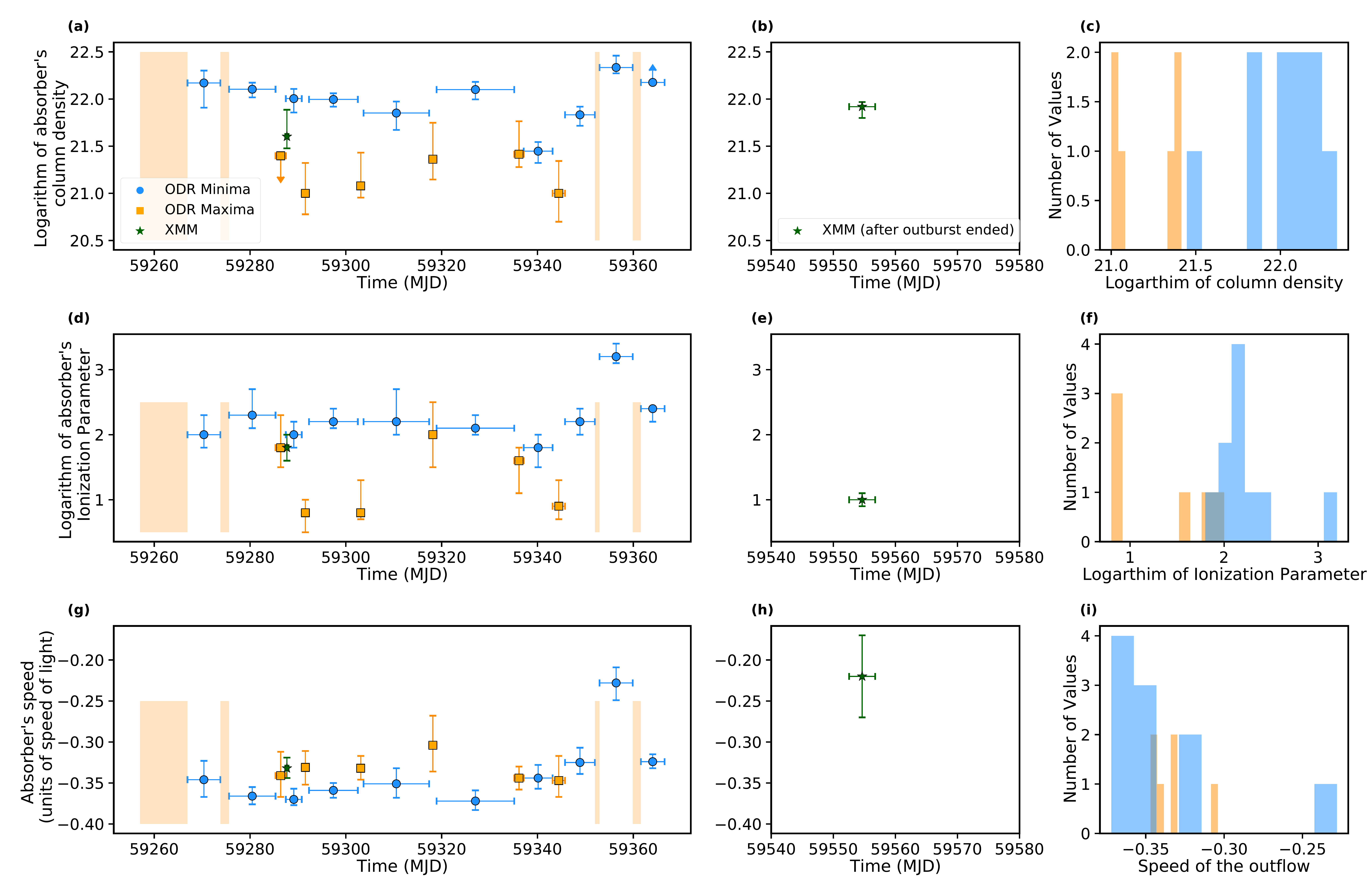}
\end{center}
\caption{{\bf Properties of the Quasi-Periodic Outflow (QPOuts) and their time evolution.} {\bf (a), (b) Logarithm of the absorbing column density of the outflow vs time.} The blue filled circles and the orange squares represent data from the ODR minima and maxima, respectively. The shaded orange regions represent the epochs of ODR maxima where an outflow was not detected in the energy spectra. Data from \xmm~ are shown as green stars.  {\bf (c) Histogram of the logarithm of outflow's column density during ODR maxima and minima.} Blue and orange histograms represent the data from ODR minima and maxima, respectively. The outflow has an order of magnitude higher column during the ODR minima compared to the maxima. {\bf (d), (e). Same as (a) and (b) but here the evolution of the logarithm of the ionization parameter of the outflow is shown.} {\bf (f) Histogram of the logarithm of the ionization parameter during the ODR maxima and minima.} Same color scheme as (c). The ionization parameter is roughly an order of magnitude larger during the ODR minima compared to ODR maxima. {\bf (g), (h) Evolution of the outflow's line of sight speed with time.} Same color scheme as in (a) and (b). {\bf (i) Histogram of the outflow speed during ODR maxima and minima.} The outflow speed is consistent between the ODR maxima and minima. }\label{fig:specplots}
\end{figure}
\clearpage
%%%%%%%%%%%%%%%%%%%%%%%%%%%%%%%%%%%%%%%%%%%%%%%%%%%%%%%%%%%%%%%%%%%%%%%%%%%%%%%%%%%%%%%%%%
%%%%%%%%%%%%%%%%%%%%%%%%%%%% XMM RESIDUALS ENERGY SPECTRA %%%%%%%%%%%%%%%%%%%%%%%%%%%%
%%%%%%%%%%%%%%%%%%%%%%%%%%%%%%%%%%%%%%%%%%%%%%%%%%%%%%%%%%%%%%%%%%%%%%%%%%%%%%%%%%%%%%%%%%
\clearpage
\begin{figure}[!ht]
\begin{center}
%\hspace{-1.35cm}
\includegraphics[width=0.5\textwidth, angle=0]{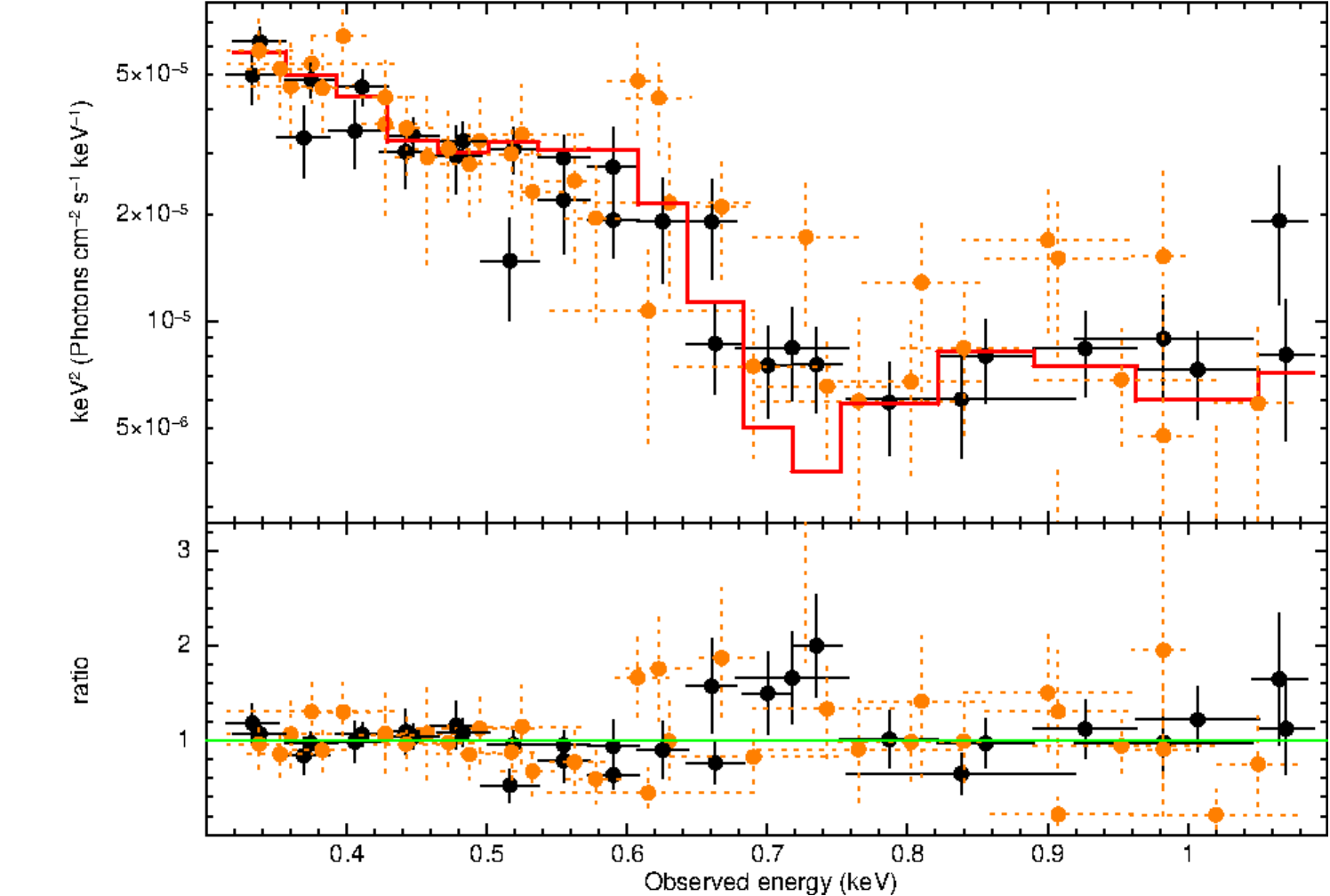}
\end{center}
\vspace{.35cm} 
\caption{\textbf{\xmm~ EPIC spectra at late times (after the initial outburst ended).} The black and the orange data points represent the pn and the combined MOS spectra, respectively. The spectra are combined and rebinned for visual purpose only. The red histogram is the best-fit model containing the outflow.} 
\label{fig:xmmlateufo}
\end{figure}

% FIG+++FIG+++FIG+++FIG+++FIG+++FIG+++FIG+++FIG+++FIG+++FIG+++FIG+++FIG+++
\clearpage
\begin{figure}[!htp]
\begin{center}
%\hspace{-1.35cm}
\includegraphics[width=0.9\textwidth, angle=0]{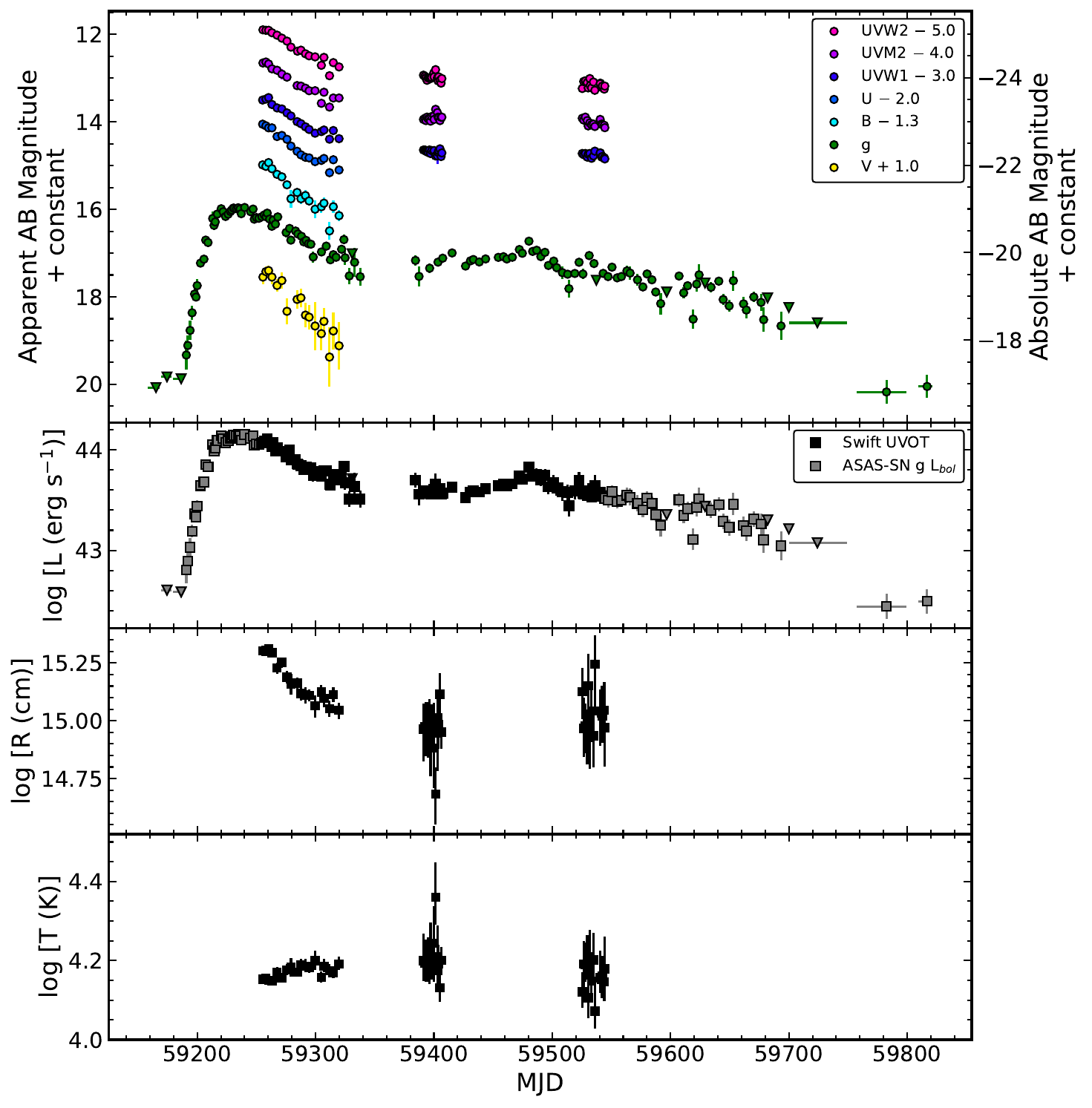}
\end{center}
%\vspace{-.35cm} 
\caption{\textbf{Host-subtracted and Galactic extinction-corrected UV/optical light curves of ASASSN-20qc (top panel) from the Swift UVOT (UV+$UBV$) and ASAS-SN ($g$)}. Evolution of the UV/optical blackbody luminosity (second panel), radius (third panel), and temperature (bottom panel) for ASASSN-20qc. The gray squares indicate where ASAS-SN $g$-band data outside the temporal range covered by Swift has been bolometrically corrected using nearby Swift data. The ASAS-SN data has been stacked in 10-day bins prior to the flare for deep limits, 1-day bins during the rise and peak of the flare, 3-day bins after the seasonal break, and 50-day bins for the final deeper points.}\label{fig:uvotevol}
\end{figure}

% FIG+++FIG+++FIG+++FIG+++FIG+++FIG+++FIG+++FIG+++FIG+++FIG+++FIG+++FIG+++
%                   Extended Data Figure 5: XMM RGS spectrum
% FIG+++FIG+++FIG+++FIG+++FIG+++FIG+++FIG+++FIG+++FIG+++FIG+++FIG+++FIG+++
\begin{figure}[!htp]
    \centering
    \includegraphics[width=0.7\textwidth]{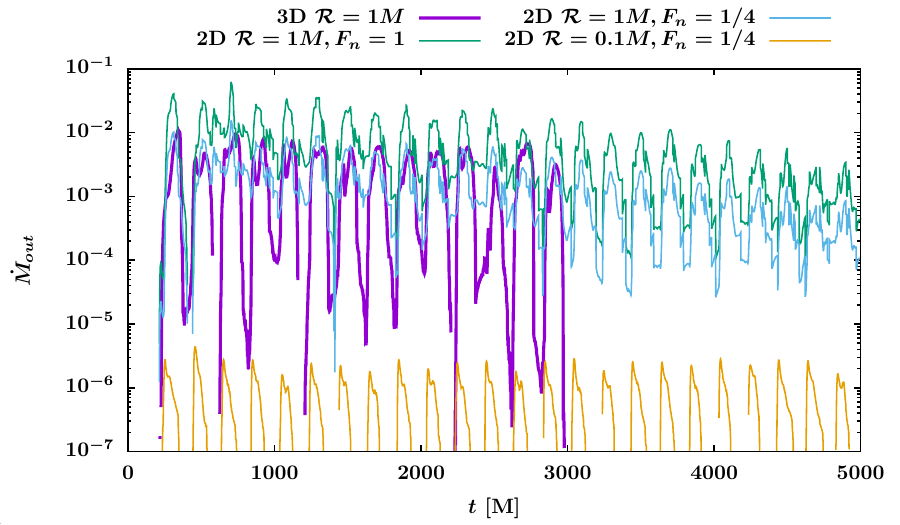}
    \includegraphics[width=0.7\textwidth]{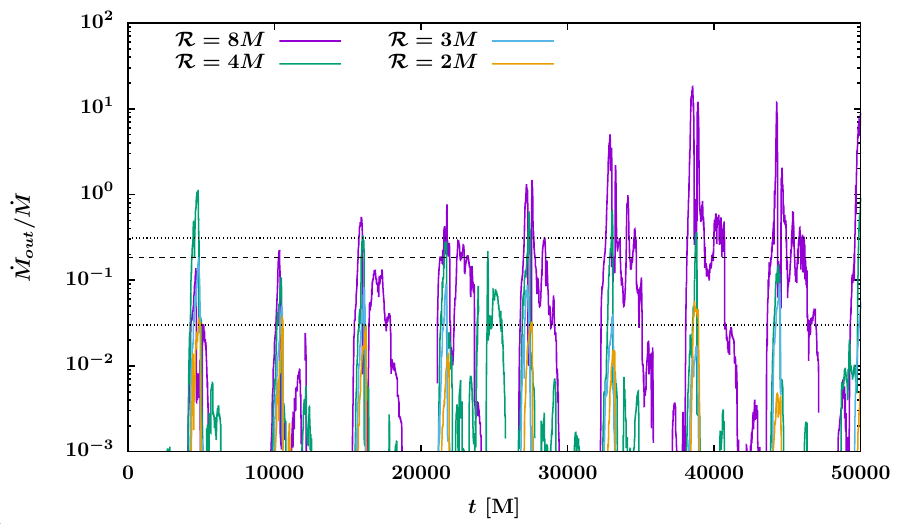}
    \caption{\textbf{Constraining the influence radius of the perturber using the ratio of the outflow rate to the inflow rate.} {\bf Top panel:} Outflow rate $\dot{M}_{\rm out}$ with $v>0.5c$ for the simulation corresponding to a perturber on a circular inclined orbit with $r=10M$. The purple line shows the result from the 3D GRMHD simulation with a perturber influence radius of $\mathcal{R}=1M$ (Run 9) while the green line shows the result from the 2D GRMHD simulation scaled by $\frac{ 2  \mathcal{R} } { (2\pi r)} F_{n} $ with $\mathcal{R}=1M, F_{n} = 1$ (Run 10). The blue line shows results from a 2D run with $\mathcal{R}=1M, F_{n} = 1/4$ (Run 10), and align better with the 3D results. The orange line depicts a 2D run with low perturber radius of $\mathcal{R}=0.1M$ and $F_{n} = 1/4$ (Run 11), and highlights the $>$3 orders of magnitude weaker outflow compared to $\mathcal{R}=1M$. {\bf Bottom panel:} The ratio of the outflow to the inflow rate in simulations of ASASSN-20qc with different $\mathcal{R}$ (Runs 2, 6, 7 and 8). The long-dashed line represents the average of observationally inferred ratio $\dot{M}_{out}/\dot{M}_{acc}$ during \nicer ODR Min phases (0.184), while the short-dashed lines represent the minimal (0.03) and maximal values (0.31) of the ratio, see Table~\ref{tab:energetics}. }
    \label{fig_influence_radius}
\end{figure}
% FIG+++FIG+++FIG+++FIG+++FIG+++FIG+++FIG+++FIG+++FIG+++FIG+++FIG+++FIG+++
\clearpage

% FIG+++FIG+++FIG+++FIG+++FIG+++FIG+++FIG+++FIG+++FIG+++FIG+++FIG+++FIG+++
%                   Extended Data Figure 5: XMM RGS spectrum
% FIG+++FIG+++FIG+++FIG+++FIG+++FIG+++FIG+++FIG+++FIG+++FIG+++FIG+++FIG+++
\begin{figure}[!htp]
    \centering
   \includegraphics[width=0.49\textwidth]{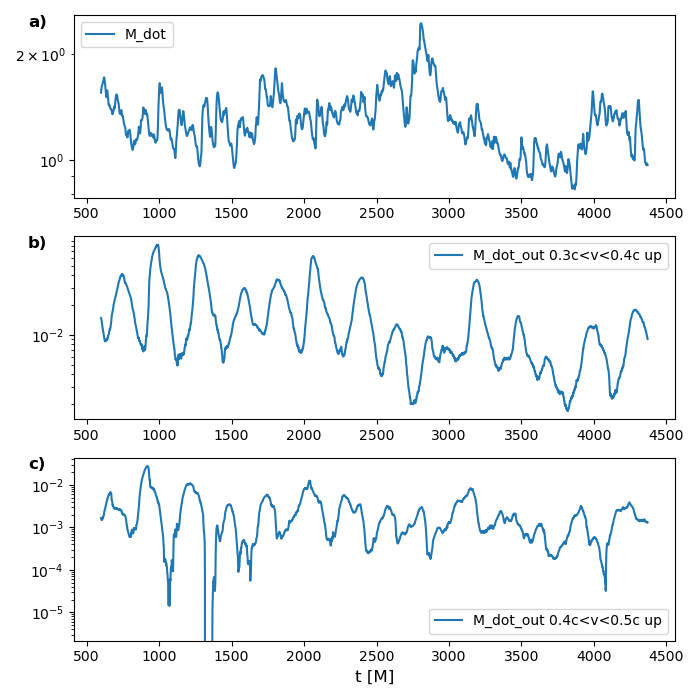}
   \includegraphics[width=0.49\textwidth]{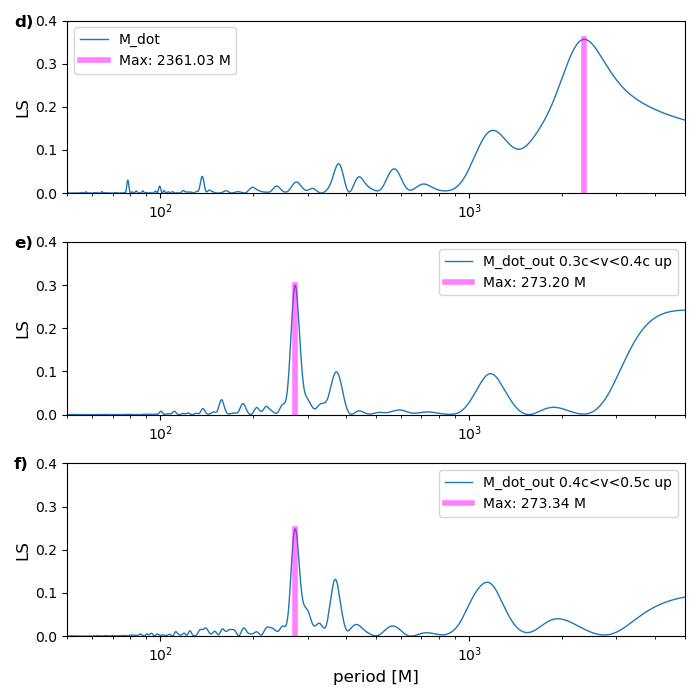}
  \includegraphics[width=0.49\textwidth]{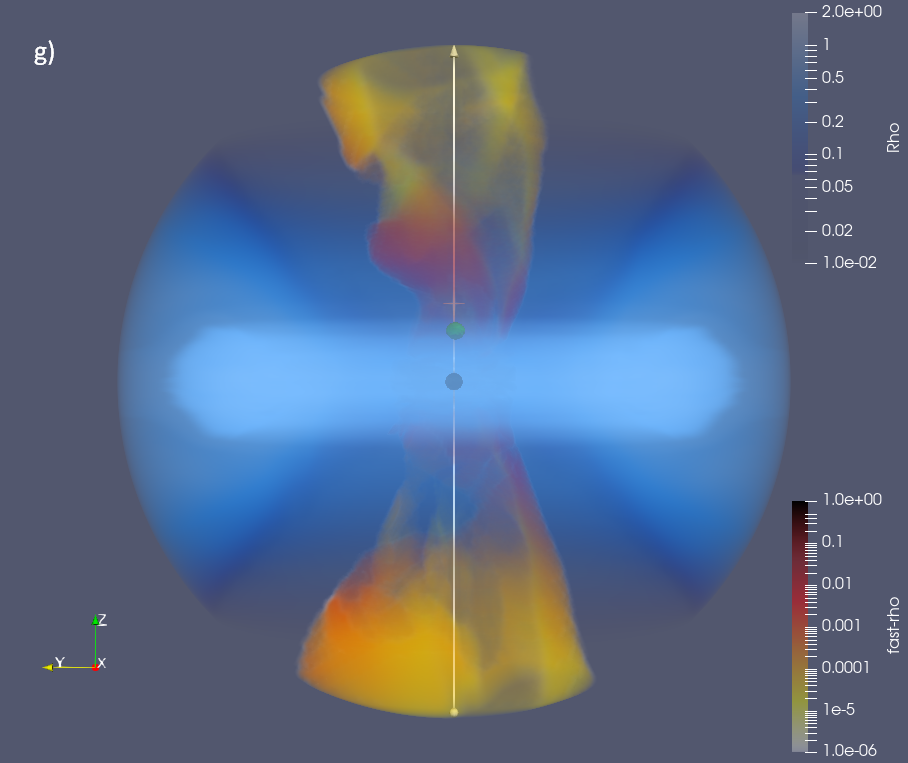}
    \includegraphics[width=0.49\textwidth]{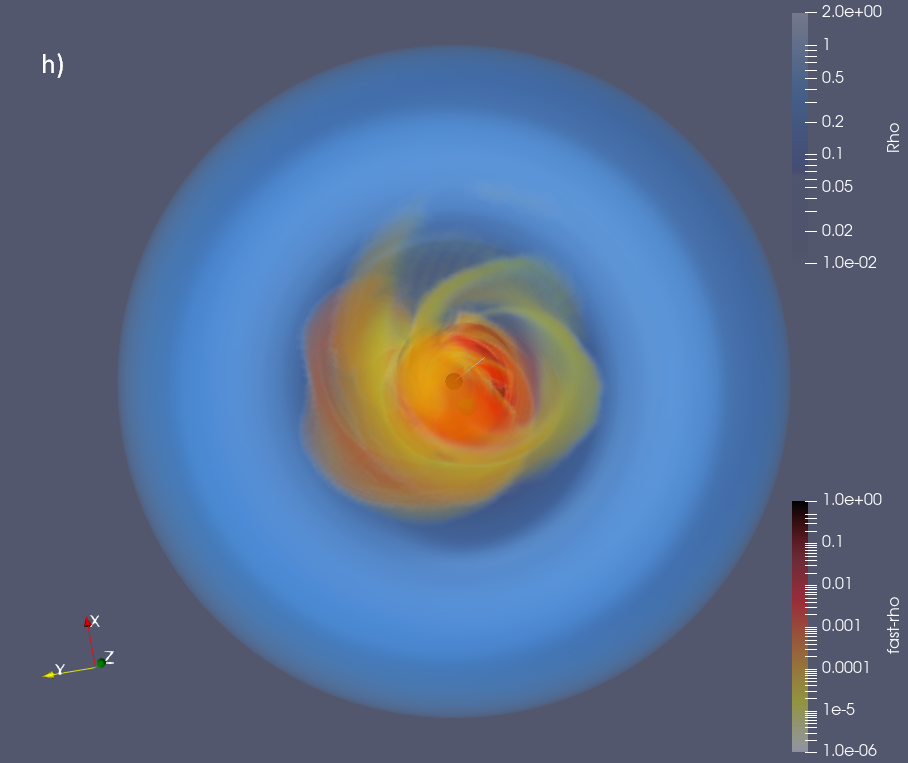}
    \caption{\textbf { \textbf{Run 16:} a)  accretion rate $\dot{\mathcal{M}}(t)$, b) outflow rate $\dot{\mathcal{M}}_{\rm out-up}(t)$ with $0.3c<v<0.4c$, c) outflow rate $\dot{\mathcal{M}}_{\rm out-up}(t)$ with $0.4c<v<0.5c$, d) - f) Lomb-Scargle periodogram of a) - c). The initial transient time $t_t =600M$ was omitted from the analysis. g) - h) snapshots from Run 16 showing the density of the flow.  The blue color scale shows the slowly moving matter, while the yellow-red color scale shows the fast-outflowing gas. The pertuber is shown by green color. This 3D movie is available at \href{https://youtu.be/fwgUEzGpApU}{https://youtu.be/fwgUEzGpApU}. }} 
    \label{fig_run16}

\end{figure}
% FIG+++FIG+++FIG+++FIG+++FIG+++FIG+++FIG+++FIG+++FIG+++FIG+++FIG+++FIG+++
\clearpage
% FIG+++FIG+++FIG+++FIG+++FIG+++FIG+++FIG+++FIG+++FIG+++FIG+++FIG+++FIG+++
%                   Extended Data Figure 5: XMM RGS spectrum
% FIG+++FIG+++FIG+++FIG+++FIG+++FIG+++FIG+++FIG+++FIG+++FIG+++FIG+++FIG+++
\begin{figure}[!htp]
    \centering
    \includegraphics[width=0.7\textwidth]{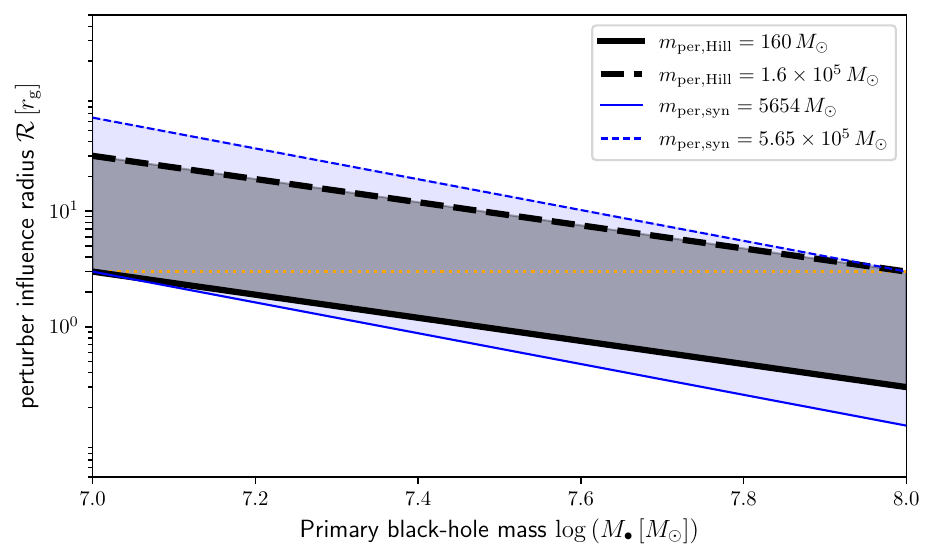}
    \includegraphics[width=0.7\textwidth]{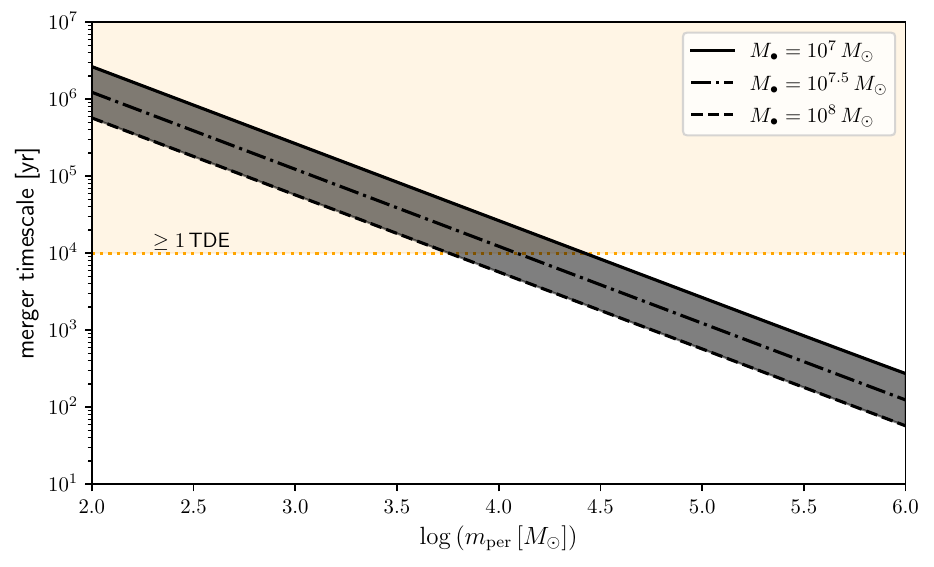}
    \caption{\textbf{Top: The influence radius of the perturber expressed as a function of the primary SMBH mass.} The calculations are performed for the limiting lower and upper masses using the tidal (Hill) radius relation, Eq.~\eqref{eq_hill_radius}, and the synchronization-radius relation, Eq.~\eqref{eq_synchronization_radius}, respectively, and the adopted rest-frame orbital period is $8.05$ days. The horizontal dotted orange line stands for the influence radius of $\mathcal{R}=3r_{\rm g}$, which is preferred based on the comparison of GRMHD simulations with the inferred outflow/inflow rate ratio during ODR minima. \textbf{Bottom: Merger timescale for the two-body system SMBH-IMBH with the rest-frame orbital period of $8.05$ days as a function of the IMBH mass.} We depict the merger timescales for the three cases of the primary SMBH mass, $10^7\,M_{\odot}$, $10^{7.5}\,M_{\odot}$, and $10^8\,M_{\odot}$, using the solid, dash-dotted, and the dashed lines, respectively. The horizontal orange dotted line marks the timescale of $10^4$ years, which is considered to be the typical minimum timescale for the tidal disruption event to take place per galaxy. The IMBH masses, for which the merger timescale is greater than this TDE timescale ($m_{\rm per}\lesssim 10^4\,M_{\odot}$), are favoured due to an increased chance of the simultaneous SMBH-IMBH inspiral and the TDE. }
    \label{fig_Hill_syn_radius}
\end{figure}

% FIG+++FIG+++FIG+++FIG+++FIG+++FIG+++FIG+++FIG+++FIG+++FIG+++FIG+++FIG+++
\clearpage
\begin{figure}[!htp]
    \centering
    \includegraphics[width=0.49\textwidth]{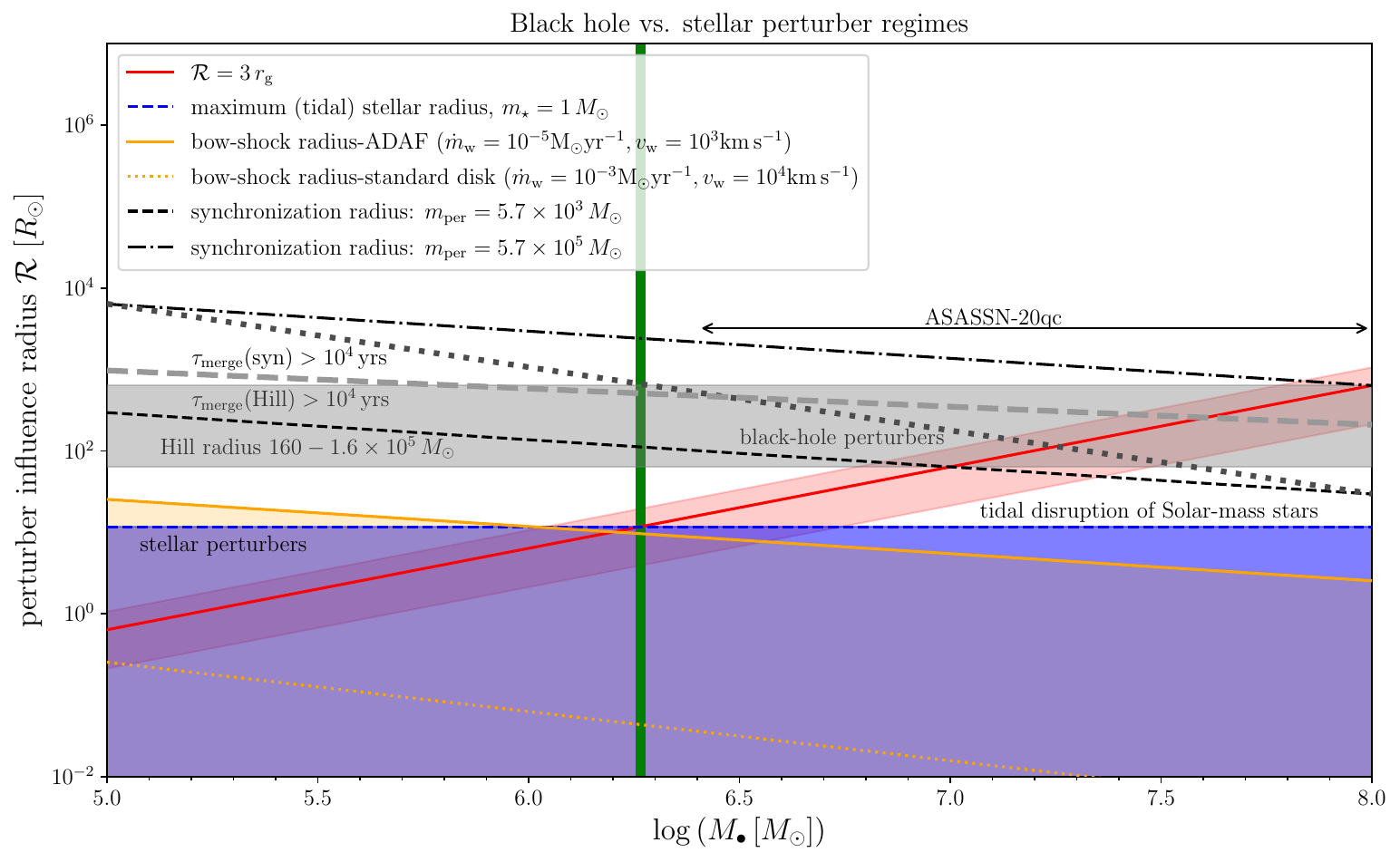}
    \includegraphics[width=0.5\textwidth]{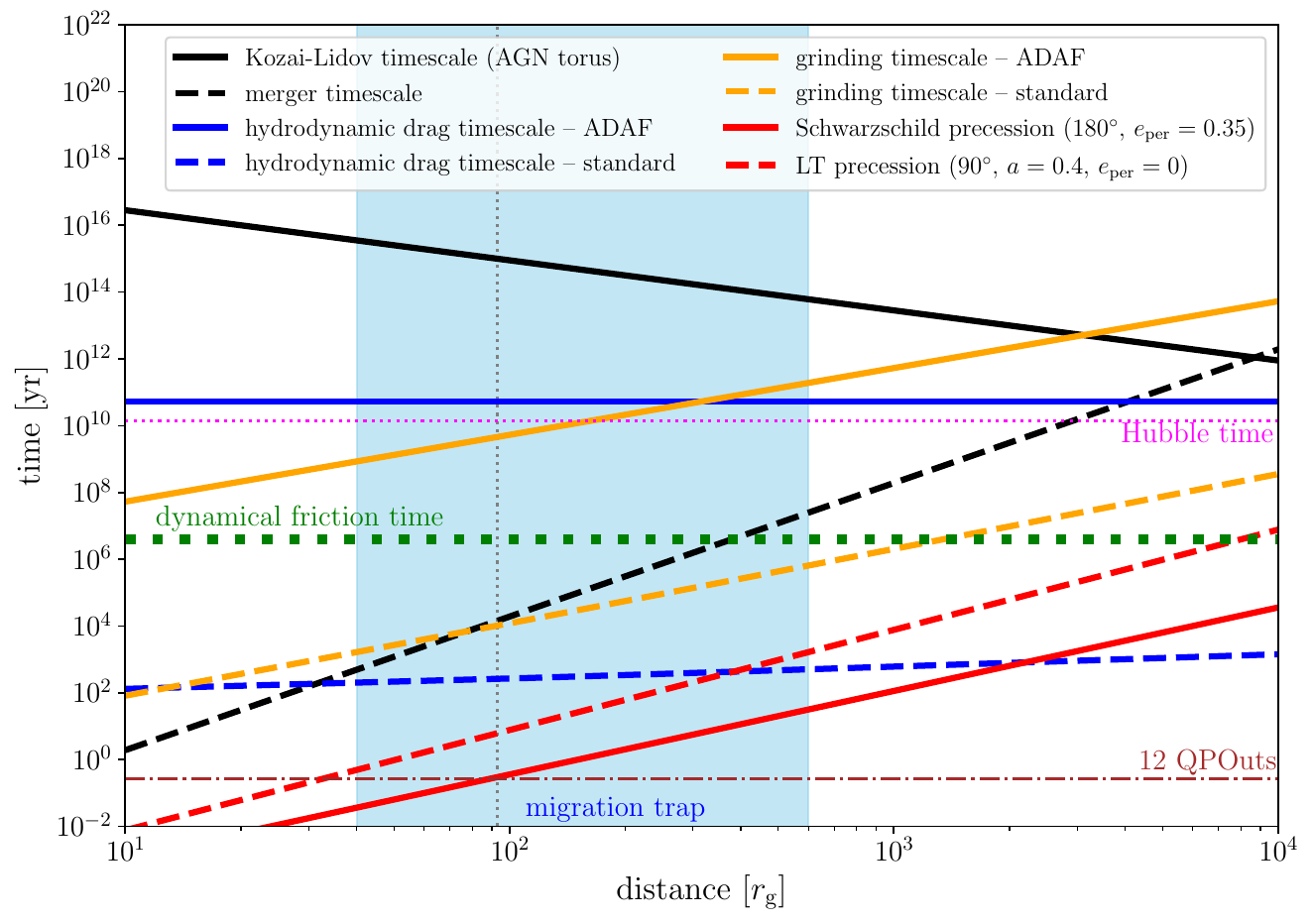}
    \caption{\textbf{Left: Influence radius of a disk perturber expressed in Solar radii as a function of the SMBH mass.} For lighter SMBHs of $\lesssim 10^{6.27}\,M_{\odot}$ (vertical green line), perturbers can be stars due to large enough relative cross-sections (physical or due to a strong stellar wind), while for heavier SMBHs, to which \target SMBH belongs, stellar-mass and intermediate-mass black holes will yield large enough spheres of gravitational influence while tidally stable stars will remain significantly below the limiting value of three gravitational radii (solid red line). The red shaded area expresses influence radii in the range of $\mathcal{R}=1-5\,r_{\rm g}$. The dotted black line marks the limit where $\tau_{\rm merge}=10^4$ years for the influence radius given by the synchronization radius, while the dashed gray line represents the same limit when the influence radius is calculated using the Hill expression. \textbf{Right: Dependency of dynamical timescales on the distance of the IMBH from the SMBH.} We show the radial dependency of the Kozai-Lidov timescale for the AGN torus perturbation, the merger timescale for the SMBH-IMBH pair, hydrodynamical drag and grinding timescales for both the ADAF and the standard disk cases; see the legend. The solid and dashed red lines represent general relativistic Schwarzschild and Lense-Thirring precession timescales, respectively. The shaded blue rectangle represents an approximate location of migration traps within the accretion disk ($\sim 40-600\,r_{\rm g}$ according to \cite{2016ApJ...819L..17B}). The horizontal dotted green line stands for the dynamical friction time as given by Eq.~\eqref{eq_dynamical_friction_time}. Dashed vertical gray line depicts the orbital distance of the considered IMBH perturber orbiting around $10^{7.4}\,M_{\odot}$ SMBH once in every 8.5 days. The dash-dotted brown horizontal line represents the duration of 12 QPOuts (96.6 days in the source frame). The dotted horizontal magenta line shows the Hubble time.  For all the timescale radial profiles, the IMBH mass is set to $10^4\,M_{\odot}$ and the SMBH mass to $10^{7.4}\,M_{\odot}$ when relevant.}
    \label{fig_perturber_regime}
\end{figure}

\clearpage
\textbf{Supplement Movie Caption S1: General-relativistic magnetohydrodynamical (GRMHD) simulation of repetitive stellar transits through the accretion flow onto a supermassive black hole with 10$^{7.4}$ M$_{\odot}$ and $a=0.4$.} The perturbing companion is consistent with an intermediate-mass black hole of $\gtrsim 100\,M_{\odot}$, which is moving along an eccentric geodesics in Kerr spacetime with r = 93 M. The influence radius of the perturber is $\mathcal{R}=3M$ and is shown to scale. The simulation was performed using the GRMHD code HARM, with the observed orbital period of the perturber set to $8.5$ days. The simulation is run in 2D, where the phi-coordinate of the star position is not taken into account. {\bf Top panels, from the left to the right:  (a) Spatial distribution of the logarithm of mass density.} The horizontal and the vertical axes are spatial coordinates expressed in gravitational radii. The white contours indicate the magnetic field configuration. The position and size of the perturber is shown by the black circle, while the grey line displays its trajectory in the 2D slice.  {\bf (b) Spatial distribution of the Lorentz factor of the gas bulk motion; } {\bf (c) Spatial distribution of the mass-outflow rate with $v>0.2c$. } The outflow rate is colour-coded using arbitrary units according to the colour-bar to the right. {\bf Bottom panel: Temporal profiles of the inflow rate (blue), the outflow rate through the upper funnel (purple), and the outflow rate through the lower funnel (green).} The inflow and the outflow rates are expressed in arbitrary units. The time is expressed in days in the observed frame measured from the moment, when the perturber was launched into the flow. The coloured points indicate the actual inflow and the outflow rates. YouTube link to the movie:  \href{https://youtu.be/WQmS8h3Zzeo}{https://youtu.be/WQmS8h3Zzeo}.
\vfill\eject

\clearpage

\nolinenumbers
\setcounter{page}{40}
\putbib
\nolinenumbers
\end{bibunit}

\end{document}